\documentclass[10pt,twocolumn,letterpaper]{article}

\usepackage[pagenumbers]{cvpr} 
\makeatletter
\@namedef{ver@everyshi.sty}{}
\makeatother

\usepackage{graphicx}
\usepackage{amsmath}
\usepackage{amssymb}
\usepackage{booktabs}
\usepackage{multirow}

\usepackage{mathtools}
\usepackage{bm}
\usepackage{empheq}
\usepackage{appendix}
\usepackage[most]{tcolorbox}
\usepackage{amsthm}
\usepackage[pagebackref,breaklinks,colorlinks]{hyperref}
\usepackage[ruled]{algorithm2e}
\usepackage{array}
\newcolumntype{C}[1]{>{\centering\arraybackslash}p{#1}}

\def\bal#1\eal{\begin{align}#1\end{align}} 

\renewcommand{\hm}[1]{\hat{\mathbf{#1}}}

\newcommand{\br}[1]{\left[#1\right]} 
\newcommand{\pr}[1]{\left(#1\right)} 
\DeclareMathOperator*{\argmin}{arg\,min} 
\DeclareMathOperator*{\argmax}{arg\,max} 

\def\transp{\mathsf{T}} 
\def\m{\mathbf}
\def\mc{\mathcal}
\def\R{\mathbb{R}}

\def\md{\mathbf{d}}
\def\ast{*}

\newcommand{\grad}[2]{\ensuremath{\nabla_{#2}#1}} 
\newcommand{\norm}[2]{\ensuremath{\left\|#1\right\|_{#2}}}

\newcommand{\abs}[1]{\ensuremath{\lvert #1\rvert}}
\newcommand {\bbmtx}{\begin{bmatrix}} 
\newcommand {\ebmtx}{\end{bmatrix}} 

\DeclareMathOperator*{\vctr}{vec} 
\newcommand{\vc}[1]{\vctr{\left(#1\right)}}
\newcommand{\deriv}[2]{\frac{\partial{#1}}{\partial{#2}}}

\DeclareMathOperator*{\diagonal}{diag} 
\newcommand{\diag}[1]{\diagonal\pr{#1}}

\DeclareMathOperator*{\sgn}{sgn} 

\newtcbox{\mymath}[1][]{%
	nobeforeafter, math upper, tcbox raise base,
	enhanced, colframe=red!100!white,
	colback=white!30, boxrule=1pt,
	#1}

\usepackage[capitalize]{cleveref}
\crefname{section}{Sec.}{Secs.}
\Crefname{section}{Section}{Sections}
\Crefname{table}{Table}{Tables}
\crefname{table}{Tab.}{Tabs.}

\date{\vspace{-0.2cm}}
\begin{document}
\title{DeepRLS: A Recurrent Network Architecture with Least Squares Implicit Layers for Non-blind Image Deconvolution}

\author{
{\hskip-.3cm\begin{tabular}{C{0.25\textwidth}C{0.3\textwidth}C{0.4\textwidth}}
Iaroslav Koshelev      & Daniil Selikhanovych      & Stamatios Lefkimmiatis \\
Skoltech, Moscow        & Skoltech, Moscow        & Huawei Noah's Ark Lab, Moscow   \\
\multicolumn{2}{c}{\tt\small \{Iaroslav.Koshelev, Daniil.Selihanovich\}@skoltech.ru} & \tt\small stamatios.lefkimmiatis@huawei.com
\end{tabular}}
}
\maketitle

\begin{abstract}
In this work, we study the problem of non-blind image deconvolution and propose a novel recurrent network architecture that leads to very competitive
restoration results of high image quality. Motivated by the computational efficiency and robustness of existing large scale linear solvers, we manage to express the solution to this problem as the solution of a series of adaptive non-negative least-squares problems. This gives rise to our proposed Recurrent Least Squares Deconvolution Network (RLSDN) architecture, which consists of an implicit layer that imposes a linear constraint between its input and output. By design, our network manages to serve two important purposes simultaneously. The first is that it implicitly models an effective image prior that can adequately characterize the set of natural images, while the second is that it recovers the corresponding maximum a posteriori (MAP) estimate.
Experiments on publicly available datasets, comparing recent state-of-the-art methods, show that our proposed RLSDN approach achieves the best reported performance both for grayscale and color images for all tested scenarios. Furthermore, we introduce a novel training strategy that can be adopted by any network architecture that involves the solution of linear systems as part of its pipeline. Our strategy eliminates completely the need to unroll the  iterations required by the linear solver and, thus, it reduces significantly the memory footprint during training.  Consequently, this enables the training of deeper network architectures which can further improve the reconstruction results.
\end{abstract}


\begin{figure}[t]
\centering
\begin{tabular}{@{} c @{ } c @{ }}
  \includegraphics[width=.45\linewidth]{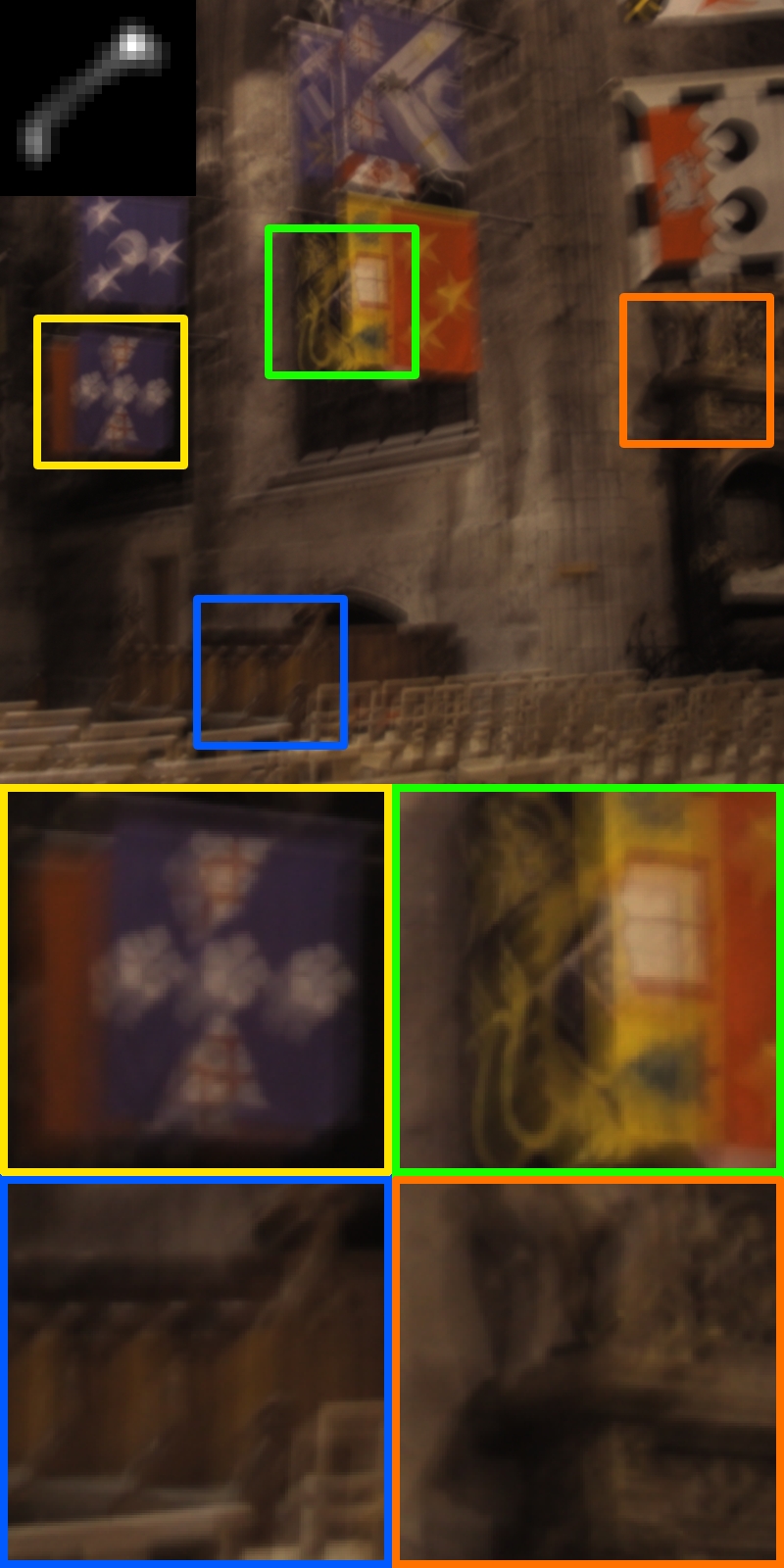}&
  \includegraphics[width=.45\linewidth]{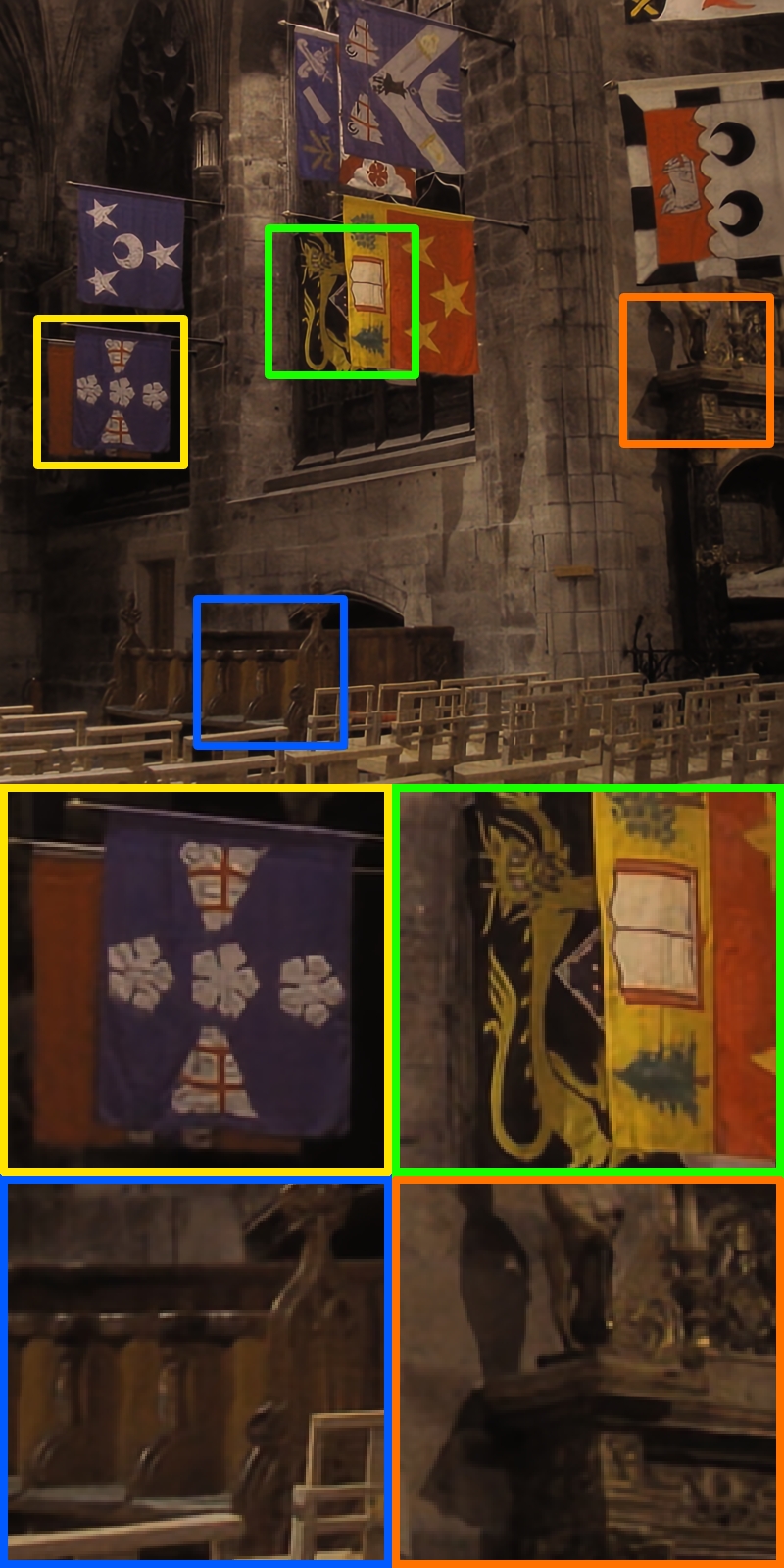}\\
  Input / Estimated blur kernel & RLSDN (ours)
\end{tabular}\vspace{-0.2cm}
  \caption{Real image deblurring result by the proposed RLSDN network. The blurred input was obtained from \cite{Kohler2012}, while the blur kernel was estimated using the method of~\cite{Pan2016}.}
  \label{fig:RealColorExample}
  \vspace{-0.6cm}
\end{figure}

\section{Introduction}
\label{sec:intro}
Image deconvolution belongs to the category of inverse imaging problems~\cite{Bertero1998} and it appears in a host of applications ranging from computational photography and biomicroscopy to remote sensing and astronomical imaging. The goal of deconvolution is to recover the sharp latent image from a blurry and noisy captured version. The blurring effect is typically modeled as the convolution of the underlying image with the point spread function (psf) of the imaging system. It can be caused by several factors including motion during image acquisition, which can be due to camera shake or due to moving objects in the captured scene when long exposure times are used, out-of-focus optics, scattered light distortion in confocal microscopy, atmospheric turbulence in astronomy, \etc~\cite{Hansen2006}.

Image deconvolution is a very challenging problem and a plethora of methods have been proposed in the literature to address it. We can classify all these methods into two categories, those that deal with the estimation of both the underlying sharp image and the blur kernel (psf)~\cite{Cho2009, Ren2020, Xu2017, Dong2017, Tran2021} and those that work under the assumption that the blur kernel is given and aim to estimate only the underlying image~\cite{Kruse2017, Zhang2017, Gong2020, Dong2021, Dong2021b}. The methods in the first class are referred to as blind deconvolution methods while those in the second as non-blind deconvolution methods. In general, though, when dealing with blind deconvolution problems the common strategy is to follow a two phase approach. In the first phase an accurate estimate of the blur kernel is obtained, while in the second phase a non-blind deconvolution method is applied to recover the underlying image  (for a comprehensive review of blind methods see \cite{Lai2016} and references therein). Therefore, the development of efficient and effective non-blind image deconvolution methods is still of high significance, and is the main focus of this work.

The presence of blur and noise usually leads to an acquired image that has suffered a significant loss of information. As a result the recovery of the sharp underlying image is unattainable without further taking into account additional available information. There are several ways one can incorporate such information and exploit it during the image restoration process. One way to achieve this is using model-based methods that adopt certain image priors, which are able to encode statistical or physical properties of the latent image~\cite{Lefkimmiatis2012J, Whyte2014, Dong2017}. Then, the image deconvolution is re-casted to a constrained optimization problem. However, the most recent paradigm, which has shown great potentials, involves deep-learning methods which are able both to implicitly encode prior image information and obtain the deblurred result relying on specific network architectures~\cite{Kruse2017, Ren2020, Dong2021, Dong2021b}. 

In this work, we follow an approach that combines ideas both from model-based and deep-learning methods. In particular, motivated by model-based methods and large-scale efficient optimization techniques we manage to express the solution of the deblurring problem as the solution of a series of adaptive non-negative least-squares (NNLS) problems. Then, based on this result we design a recurrent deconvolution network which can solve convincingly the deblurring problem and lead to state-of-the-art results. Moreover, we adopt a novel network training strategy, which is not specific to our proposed architecture but it applies to any network that involves the solution of a linear system. Our strategy eliminates completely the need to unroll the iterations of the linear solver and as a result it allow us to reduce significantly the memory requirements during the training stage. This makes it possible to use a larger number of network iterations, which lead to improved reconstruction quality.


\section{Problem Formulation}
\label{sec:PF}

To deal with the problem of non-blind image deconvolution we first need to consider the observation (forward)  model, which relates the observed blurry and noisy image with
the latent image that we aim to restore. In this work we adopt the following linear observation model
\bal
\m y = \m H\m x + \m n,
\label{eq:fwd_model}
\eal
which is the most widely used in the literature and usually can serve as an adequate approximation of the image acquisition process. In the above formula 
$\m y\in\R^M$ and $\m x\in\R^N$ represent the vectorized forms of the observed and underlying images, respectively, assuming that they have been raster-scanned using a lexicographical order.  Under this notation, $\m H\in\R^{M\times N}$, with $M < N$,  is the degradation matrix that models the spatial response of the imaging device, which is responsible for the presence of blur in the observed image. Note that according to the model of \cref{eq:fwd_model},  the observed image $\m y$ has smaller spatial dimensions than the underlying image $\m x$,  which translates to $\m H$ being a Toeplitz convolution matrix. This is a more realistic assumption than the alternative and frequently used one of considering $\m H$ to be a square circulant convolution matrix.  Apart from the blur degradation, the image measurements are also perturbed by noise, which hereafter we will assume it to be zero mean i.i.d Gaussian noise of variance $\sigma^2$, \ie
$\m n\sim\mc{N}\pr{0, \sigma^2}$.

The recovery of $\m x$ from the distorted measurements $\m y$ belongs to the broad class of linear inverse problems~\cite{Bertero1998}. Despite the linear nature of the acquisition process, the image restoration is far from a trivial task. This is due to the presence of noise, whose exact realization is unknown, and the fact that the blurring operator $\m H$ in practice is singular. These two factors turn image deblurring to a highly ill-posed problem~\cite{Hansen2006}. This has the implication that a unique solution to the problem does not exist and therefore we cannot solely rely on the image evidence but we further need to take into account \emph{a priori} information about the solution. 

One way to move forward is to adopt a Bayesian approach and seek for the Maximum A Posteriori (MAP) estimate~\cite{Kay1993}
\bal
\m x^\ast 
&= \argmax_{\m x} \log\pr{p\pr{\m y |\m x}} + \log\pr{p\pr{\m x}},
\label{eq:MAP}
\eal
where $\log\pr{p\pr{\m y| \m x}}$ corresponds to the log-likelihood of the observation $\m y$ and $\log\pr{p\pr{ \m x}}$ is the log-prior of $\m{x}$. Given our initial assumption that the noise perturbing the measurements is i.i.d Gaussian, the problem in~\eqref{eq:MAP} can be equivalently reformulated as the minimization problem
\begin{equation}
\label{eq:var}
{\m x}^\star = \argmin_{\m x}\pr{ \mc{E}\pr{\m x;\m y, \m H}\equiv\frac{1}{2\sigma^2} \norm{\m y-\m H \m x}{2}^2 + r\pr{\m x}},
\end{equation}
where the first term of the objective function $\mc{E}\pr{\cdot}$ corresponds to the negative log-likelihood and the second term corresponds to the negative log-prior. This problem formulation has direct links to variational methods where the first term of the objective can be interpreted as the data-fidelity that quantifies the proximity of the solution to the observation, while the second term, $r\pr{\m x}$, amounts to the regularizer, whose role is to promote solutions that exhibit certain favorable image properties. 

Under this framework, it becomes apparent that the selection of a proper regularizer (image prior) is of utmost importance and it relates directly to the quality of the reconstruction. This has led to a wide research interest for developing novel ways to effectively model key image properties that can subsequently lead to improved reconstruction results. The majority of the existing regularizers in the literature can be expressed in the following generic form 
\bal
r\pr{\m x} = \phi\pr{\abs{\m G\m x}},
\label{eq:reg}
\eal 
where $\m G:\R^N\mapsto\R^{F\cdot D}$ is a linear operator that acts on the latent image $\m x$ and maps it to a linear space of $F$ features of $D$ dimensions each, which is also referred to as the regularization operator, $\phi:\R^{F\cdot D}_+\mapsto\R_+$ is a non-decreasing potential function which penalizes the response of the operator $\m G$ on $\m x$, while the operation $\abs{\cdot}$ is meant to act element-wise.

In the recent past, very popular choices for the regularization operator have been first and second order differential operators such as the gradient~\cite{Rudin1992}, the structure tensor~\cite{Lefkimmiatis2015J}, the Laplacian and the Hessian~\cite{Lefkimmiatis2012J, Lefkimmiatis2013J}, wavelet-like operators such as wavelets, curvelets and ridgelets~(see \cite{Figueiredo2007} and references therein), and learned convolution operators~\cite{Roth2009}, while for the potential function the predominant choice had been the squared $\ell_2$ norm, which leads to the well known Tikhonov regularization strategy~\cite{Hansen2006}. The reason behind the strong preference in using the squared $\ell_2$ norm has been that in this case the entire objective function is quadratic and thus computationally efficient linear solvers can be employed to obtain the solution. Indeed, the minimizer of a quadratic objective function can be derived as the solution of the corresponding normal equations.

However, a significant drawback of using this potential function in the regularizer is the extensive over-smoothing that the resulting reconstructed images typically exhibit. Nowadays, it is widely acknowledged that employing different and more expressive potential functions, such as $\ell_p$ norms or pseudo-norms with $0 \le p < 1$ or the logarithm, 
can lead to sharper and higher-quality results~\cite{Fergus2006, Krishnan2011, Babacan2012, Xu2013}. Nevertheless, one great challenge that arises with the use of alternative potential functions is that the solution cannot anymore be obtained by simply solving a system of linear equations and more advanced optimization techniques are needed. Indeed, there is a variety of existing strategies to deal with the resulting objective functions, such as FISTA~\cite{Beck2009}, Split Bregman~\cite{Goldstein2009}, Alternating Method of Multipliers~\cite{Boyd2011}, HQS~\cite{Nikolova2005} just to name a few. The common underlying idea behind all these optimization strategies is that in order to find a minimizer which corresponds to the reconstructed image, instead of directly dealing with the original minimization problem of \cref{eq:var}, we consider several easier to solve problems. 

From the previous discussion it becomes clear that in order to be in position of obtaining a satisfactory solution to the image deblurring problem, first we have to address two important issues. The first one is the selection of an appropriate regularizer, by wisely choosing the regularization operator and the potential function. This will allow us to promote meaningful solutions that exhibit key properties  adequately describing the set of natural images. The second challenge is to come up with an optimization strategy that can find such solutions in a computationally efficient way.

In \cref{sec:optim} we focus on the design of an optimization strategy that can efficiently deal with objective functions of the form provided in \cref{eq:var}, while in \cref{sec:DRLS} we describe how we avoid to specify the exact form of the image regularizer and instead implicitly model it using a novel network architecture. 

\section{Image Restoration via Fixed Point Iteration}
\label{sec:optim}
There are two key difficulties in the minimization of the objective function in \cref{eq:var}. The first one is the coupling that exists between the singular convolution degradation operator $\m H$ and the latent image $\m x$. The second one is that the regularizer $r\pr{\m x}$, as defined in \cref{eq:reg}, has typically a non-quadratic form. These two factors prevent us from aiming for a direct solution. Thus, we can only opt for an iterative-based minimization strategy. Now, if we assume that the potential function $\phi$ is smooth and $\m x$ doesn't belong to the null space of the regularization operator $\m G$, then we can compute the gradient of the regularizer as:
\bal
\grad{r}{}\pr{\m x} &= \m G^\transp\diag{\sgn\pr{\m G\m x}}\grad{\phi}{}\pr{\abs{\m G\m x}}\nonumber\\
&=\m G^\transp\diag{\grad{\phi}{}\pr{\abs{\m G\m x}}}\diag{\abs{\m G\m x}}^{-1}\m G\m x\nonumber\\
&=\m G^\transp \m W\pr{\m G\m x}\m G\m x,
\label{eq:reg_grad}
\eal
where we use the notation $\m W\pr{\m G\m x}$ to denote the diagonal and positive semi-definite matrix\footnote{Note that since the potential functional $\phi$ is non-decreasing, it's gradient will consist of non-negative values, which in turn implies that the diagonal matrix $\m W\pr{\m G\m x}$ will be positive semi-definite.} that has a direct dependency on $\m G\m x$. 

Next, it is straightforward to show that $\m x^\ast$ is a minimizer (stationary point) of the objective function  $\mc{E}\pr{\m x;\m y, \m H}$  if it holds:
\bal
\frac{1}{\sigma^2}\m H^\transp\pr{\m H\m x^\ast - \m y} + \m G^\transp \m W\pr{\m G\m x^\ast}\m G\m x^\ast = \m 0.
\label{eq:nonlin}
\eal
\noindent
Therefore, in order to find the solution to our problem, it is sufficient to solve a system of non-linear equations as shown in \cref{eq:nonlin}. By carefully inspecting the above system of equations, we observe that its non-linear nature stems exclusively from the dependency of the matrix $\m W$ on $\m x^*$. This suggests that we can use the following fixed-point iteration strategy:
\bal
\frac{1}{\sigma^2}\m H^\transp\pr{\m H\m x^{k+1} - \m y} + \m G^\transp \m W^k\m G\m x^{k+1} = \m 0,
\label{eq:fixed_point}
\eal
with $\m W^k\equiv\m W\pr{\m G\m x^k}$, 
which in turn implies that the solution can be obtained through a sequence of updates of the form:
\bal
\m x^{k+1} &= \argmin_{\m x} \frac{1}{\sigma^2} \norm{\m y-\m H \m x}{2}^2 + \norm{\m G\m x}{\m W^k}^2\nonumber\\
&= \argmin_{\m x} \frac{1}{\sigma^2} \norm{\m y-\m H \m x}{2}^2 + \norm{{\m W^k}^{\frac{1}{2}}\m G\m x}{2}^2,
\label{eq:irls}
\eal
where $\norm{\m x}{\m A}^2 = \m x^\transp\m A\m x$, with $\m A\succeq \m 0$. The update rule provided in \cref{eq:irls} is reminiscent of the classical Tikhonov-regularized solution~\cite{Hansen2006}, with the regularization operator selected to be of the form $\m W^{\frac{1}{2}}\m G$. The only difference is that in our case the regularization operator is not fixed but it changes in every iteration according to the solution of the previous iteration.  An additional improvement to the previous update rule, which leads to a more stable and robust iterative strategy, is to include an extra term that enforces the solution of the current iteration to be not too  far from the previous one. According to this reasoning, we modify 
\cref{eq:irls} to be of the form:
\bal
\m x^{k+1} \!\!=\! \argmin_{\m x}\!\frac{1}{\sigma^2}\!\norm{\m y-\m H \m x}{2}^2 \!+\! \norm{\m G\m x}{\m W^k}^2 \!+\!\alpha\!\norm{\m x -\m x^k}{2}^2,
\eal
where $\alpha$ is a positive constant. We note that the addition of the last term doesn't affect the final solution, since upon convergence of the algorithm to a fixed point, it will hold that $\m x^{k+1} = \m x^k$ and thus the extra term will become zero.

Finally, based on all the above we end up with an iterative optimization strategy that allows us to solve the original minimization problem of interest by solving a sequence of non-negative least squares (NNLS) problems, whose solutions can be computed as
\bal
\m x^{k+1} &= \pr{\m S^k\equiv\frac{1}{\sigma^2} \m H^\transp\m H +  \m G^\transp \m W^k\m G + \alpha\m I}^{-1}\tilde{\m y}^k,\!\!
\label{eq:robust_irls}
\eal
with $\tilde{\m y}^k  = \frac{1}{\sigma^2}\m H^\transp\m y + \alpha\m x^k$.
Compared to other alternative minimization strategies, which also attack the problem by splitting it in a series of simpler sub-problems, our proposed approach has an important advantage, since it relies solely on existing efficient and fast matrix-free linear solvers designed for large scale problems. More specifically, given that the corresponding system matrix is symmetric and positive definite we can readily use the Conjugate Gradient (CG) method~\cite{Shewchuk1994} and its variants, which are specifically designed for this task and have been shown to be robust and very efficient. It is also worth noting that our strategy has close ties to the Iterative Reweighted Least Squares (IRLS) method~\cite{Daubechies2010}, which has been the minimization strategy of choice in the field of compressive sensing~\cite{Donoho2006, Candes2008} and matrix completion~\cite{Keshavan2010}. In fact, our strategy can be considered as an extension of IRLS to the case of regularizers with more generic form than $\ell_p$ norms.

\section{Deep Recurrent Least Squares Network}
\label{sec:DRLS}

\begin{figure}[!t]
\centering
   \includegraphics[width=1\columnwidth]{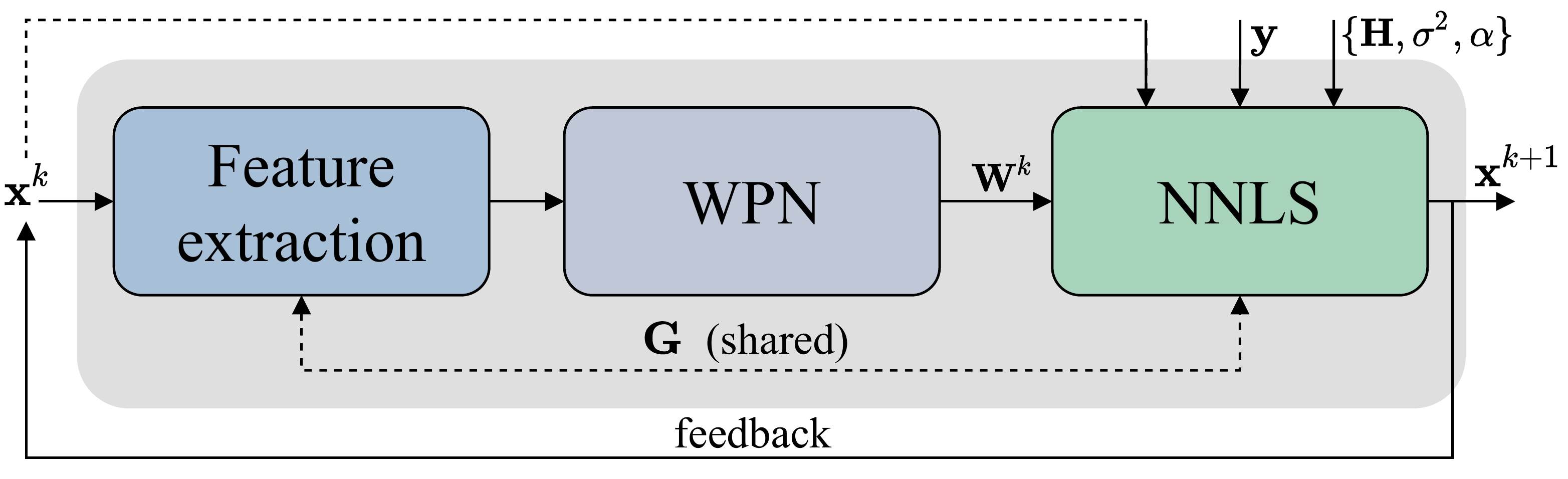}
   \caption{Architecture of the proposed Recurrent Least Squares Deconvolution Network. The proposed network consists of three main components: (a) The linear feature extraction network, (b) the weight prediction network, and (c) the non-negative least squares implicit layer. Since the network is recurrent, the output of the current iteration along with the blurry input $\m y$ serve as the inputs of the network in the next iteration.}
   \label{fig:DRLS_net}
  \vspace{-0.5cm}
\end{figure}

Until now we have refrained entirely from discussing specific choices for the linear operator $\m G$ and the potential function $\phi$, which constitute the image regularizer and as mentioned earlier play a crucial role in the reconstruction quality. In fact, the main idea that we explore in this work is that instead of manually selecting these parameters, we consider them as parameters of a network and learn them during training.

To design such a network, we rely on the update rule of \cref{eq:robust_irls} and we end up with the recurrent architecture depicted in \cref{fig:DRLS_net}. Our proposed network consists of three main components: (\textbf{a}) The linear feature extraction module, which initially accepts as input either some   estimate of the reconstructed image or the blurry image itself, while in the subsequent iterations it accepts as input the output of the network from the previous iteration. This linear sub-network essentially constitutes a parametrization of the regularization operator $\m G$. (\textbf{b}) The weight prediction network (WPN) which acts on the output of the feature extraction module. The role of this sub-network is to predict the diagonal positive semi-definite matrix $\m W^k$, which as described in \cref{eq:fixed_point} has a direct dependency on $\m x^k$, and according to \cref{eq:irls} is used to weight the regularization operator $\m G$. WPN is implemented as an image-to-image network with the additional constraint that its output is non-negative, so as to ensure the positive semi-definite nature of $\m W^k$. We note that WPN models the potential function $\phi$ in an implicit way. Indeed from \cref{eq:reg_grad} we see that $\m W^k$ is defined as $\m W^k = \diag{\frac{\grad{\phi}{}\pr{\abs{\m G\m x^k}}}{\abs{\m G\m x^k}}}$, with the division being applied element-wise, and involves the gradient of the potential function. (\textbf{c}) The NNLS layer, whose role is to refine the current estimate of the reconstructed image by solving a  NNLS problem according to \cref{eq:robust_irls}. 

One important point regarding the NNLS layer is that in real applications the inversion of the system matrix $\m S^k$ in \cref{eq:robust_irls} is practically infeasible. Therefore we need to rely on a matrix-free large-scale linear solver such as CG. However, since there is a variety of existing linear solvers that we could use, we implement our layer as an implicit one~\cite{Bai2019}. The key difference of implicit layers over conventional explicit layers, which are usually employed in deep learning, is that instead of having to explicitly specify a set of operations that the layer needs to perform to its input in order to produce the output, we only need to specify a set of constraints that the input and the output should satisfy. Then we are free to use any algorithm among the available ones that can enforce the desired constraints imposed by the layer. This is a rather different approach and leads to more flexible layers that offer one additional level of abstraction. In our case, the constraint imposed by the NNLS layer can be expressed as: 
\bal
g\pr{\m x^{k+1}, \m x^k, \m y} = \m S^k\m x^{k+1} - \frac{1}{\sigma^2}\m H^\transp\m y - \alpha\m x^k = \m 0,
\eal 
where $\m y$ and $\m x^k$ are the inputs and $\m x^{k+1}$ is the output of the layer, while $\m S^k$ is a short-hand notation for the system matrix that appears in \cref{eq:robust_irls}.

\begin{algorithm}[!t]
\small
 \SetAlgoCaptionSeparator{\unskip:}
\SetKwInput{Np}{Layer's parameters}
\SetKwBlock{Fwd}{Forward Pass}{}
\SetKwBlock{Bwd}{Backward Pass}{}
\Np{$\bm w$}
\Fwd{
Compute $\m x^\ast$ as the solution of the linear system: \vspace{-0.2cm}$$\m A\pr{\bm w}\m x = \m b\pr{\bm w}.\vspace{-0.2cm}$$}
\Bwd{
\begin{enumerate}
\item Use $\m x^\ast$ as the input to an auxiliary residual layer\\ with parameters $\bm w$ and compute its output as:
\vspace{-0.2cm}$$\bm r = \m b\pr{\bm w} -\m A\pr{\bm w}\m x^\ast.\vspace{-0.35cm}$$
\item Compute $\bm g$ by solving the linear system \vspace{-0.2cm}$$\m A^\transp\pr{\bm w}\bm g = \bm \rho,\vspace{-0.2cm}$$
where $\bm\rho =\grad{\mc L}{\bm x^\ast}$ and $\mc L$ is the
training loss function.
\item Obtain the gradient $\grad{\mc L}{\bm w}$ by computing the \\product $\grad{\bm r}{\bm w}\cdot\bm g$ using any of the existing auto-\\grad libraries (pytorch, tensorflow, \etc).
\item Use $\grad{\mc L}{\bm w}$ to update the layer's parameters $\bm w$.
\end{enumerate}
}
 \caption{Back-propagation for an implicit layer whose output is the solution of a linear system.}
  \label{alg:IBP}
\end{algorithm}

\section{Network Training}
\label{sec:net_train}
\vspace{-.2cm}
As we have discussed earlier, the output of our network is obtained by finding the minimizers of a sequence of NNLS problems, which boils down to computing the solutions of 
a sequence of linear problems. A strategy that is commonly adopted in cases of recurrent networks like ours, is to unroll the network using a fixed number of iterations and update the network parameters either by means of back-propagation through time (BPTT) or by its truncated version (TBPTT)~\cite{Robinson2008, Kokkinos2019}. Unfortunately, our proposed architecture is not entirely compatible to such a training strategy. The reason is that apart from the external iterations, our network also involves the internal iterations required by the linear solver in order to compute the solutions of each one of the related linear problems. This means that if we had to unroll both the external and the internal iterations of the network, then we would naturally end up with a very deep architecture. However, unrolling such a deep network would be prohibitive since the overall depth of the network is bounded by the available GPU memory or RAM that is required during its training. 

\subsection{Implicit Back-Propagation}\vspace{-0.2cm}
Fortunately, it is possible to come around this problem and completely avoid the need of unrolling the iterations of each linear solver. Indeed, if we denote by $\bm w$ the set of the network parameters, and consider the generic linear system:
\vspace{-0.5cm}
\bal
\m A\pr{\bm w}\m x = \m b\pr{\bm w},
\vspace{-.5cm}
\eal
where we specifically indicate the dependency of the system matrix $\m A$ and the rhs vector $\m b$ on $\bm w$, then we can compute the gradient of 
some loss function $\mc L$ w.r.t the network parameters $\bm w$, following the steps described in \cref{alg:IBP}. In the supplementary material we provide a formal proof that motivates
the proposed algorithm. The important implication of using \cref{alg:IBP} as part of our network training strategy, is that now we can use any linear solver running for any required number of iterations without having to worry about saving any intermediate results that would result in high memory utilization during training.  Indeed, the only intermediate results we are required to save during the forward pass of the network are the final solutions of each one of the linear systems that we encounter. The cost we pay for this reduction of the memory footprint is that during the backward pass, for each one of the external network iterations we have to compute the forward pass of an auxiliary residual network and solve another linear system. While this can increase the training time, it lifts the memory constraints and allow us to use a larger number of external and internal iterations during training.
\begin{figure*}[!t]
\centering
\begin{tabular}{@{} c @{ } c @{ } c @{ } c @{ } c @{ } c @{ }}
    \includegraphics[width=.16\linewidth]{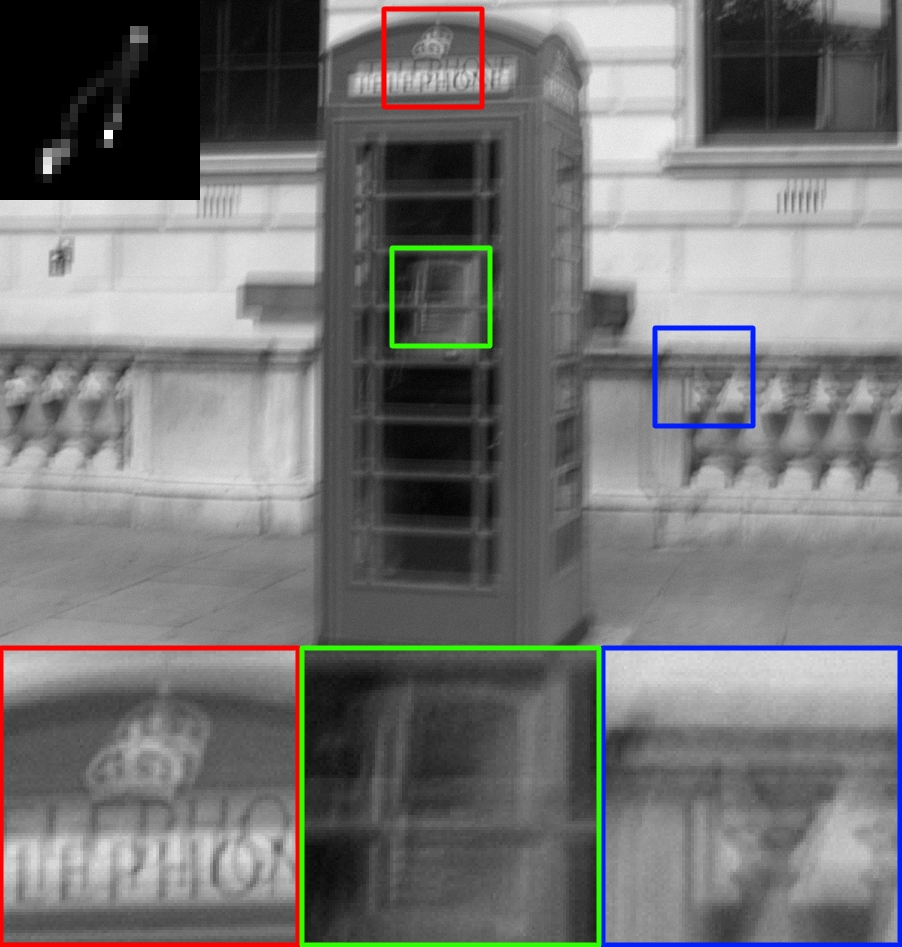}&
    \includegraphics[width=.165\linewidth]{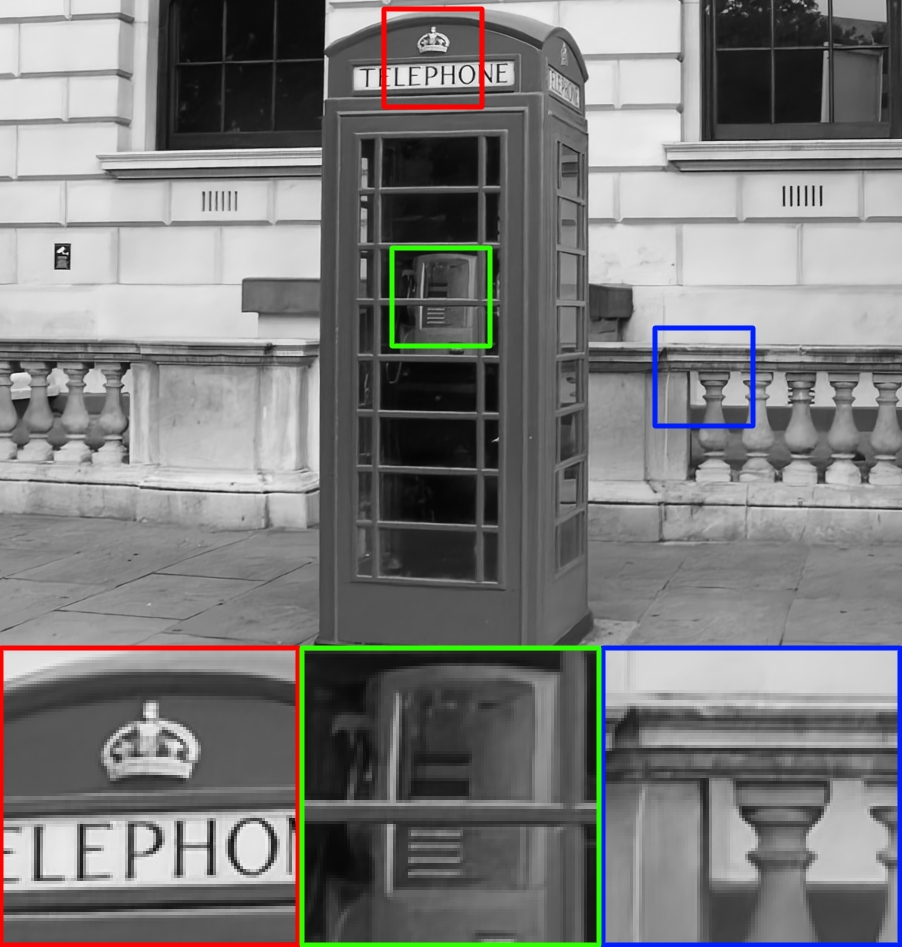}&
    \includegraphics[width=.165\linewidth]{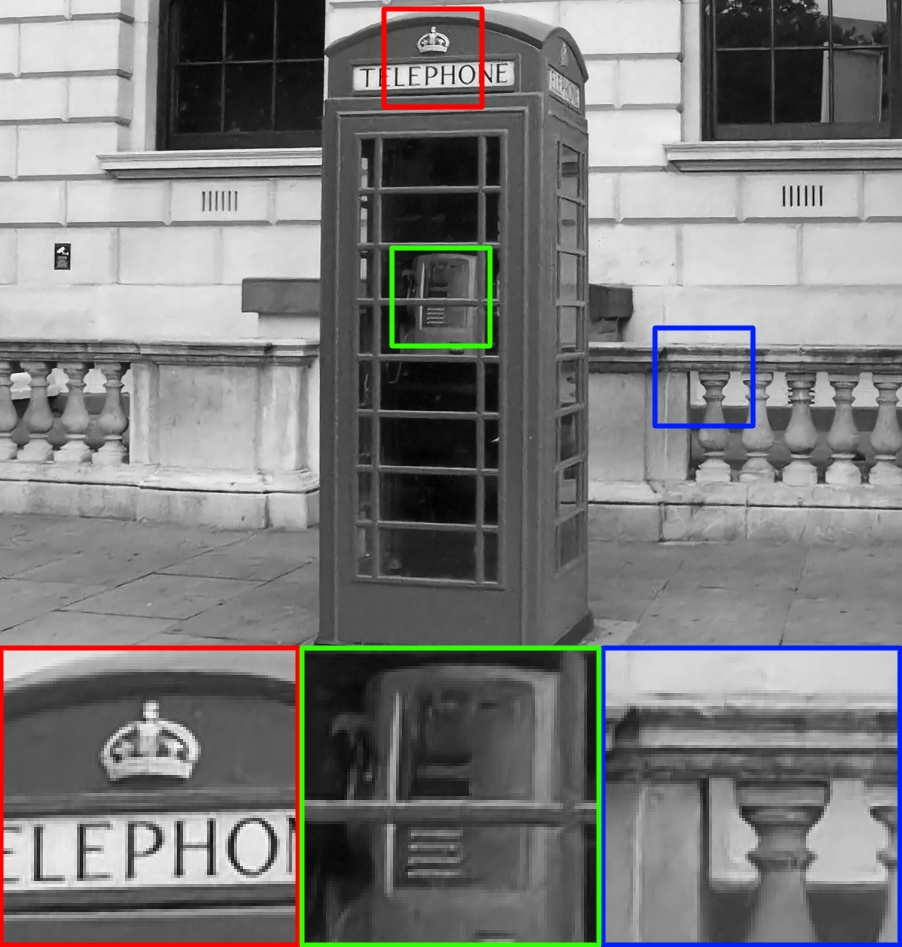}&
    \includegraphics[width=.165\linewidth]{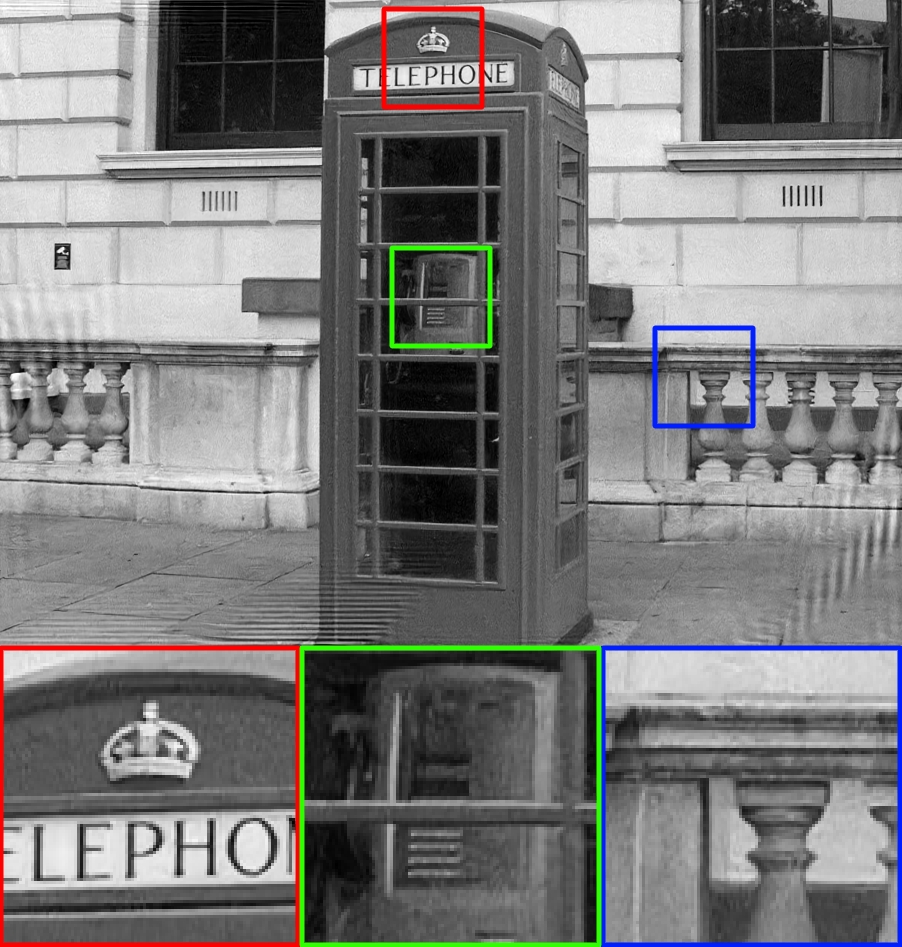}&
    \includegraphics[width=.165\linewidth]{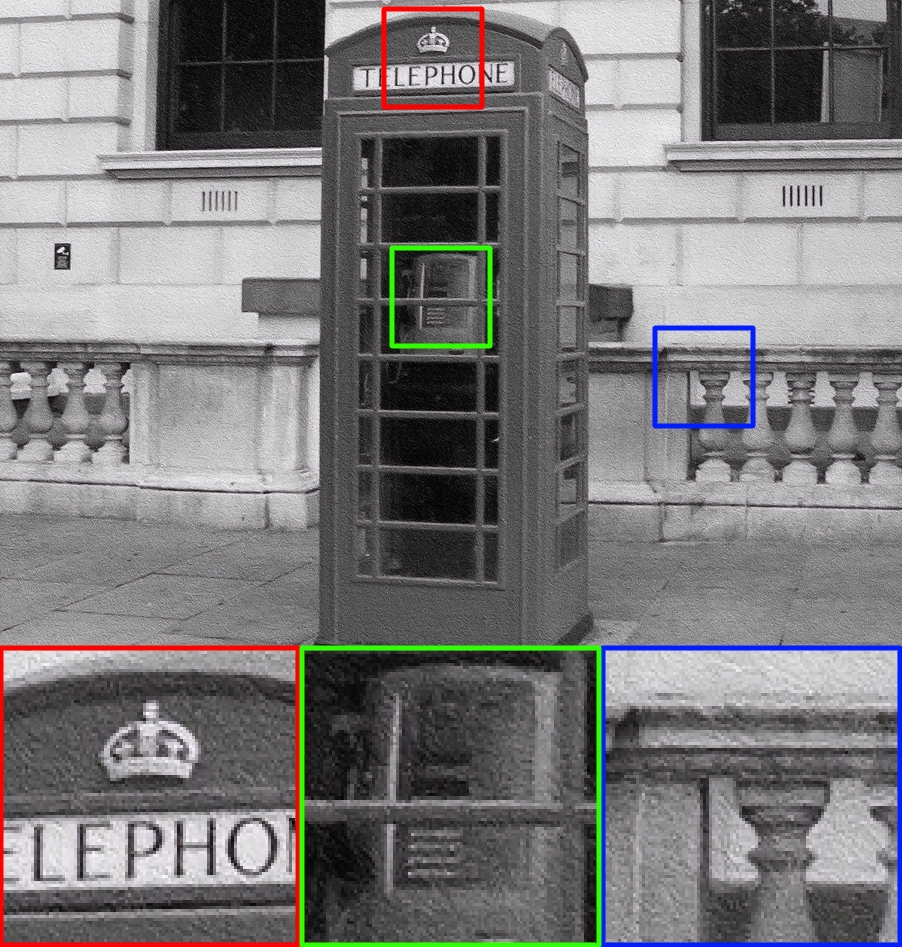}&
    \includegraphics[width=.165\linewidth]{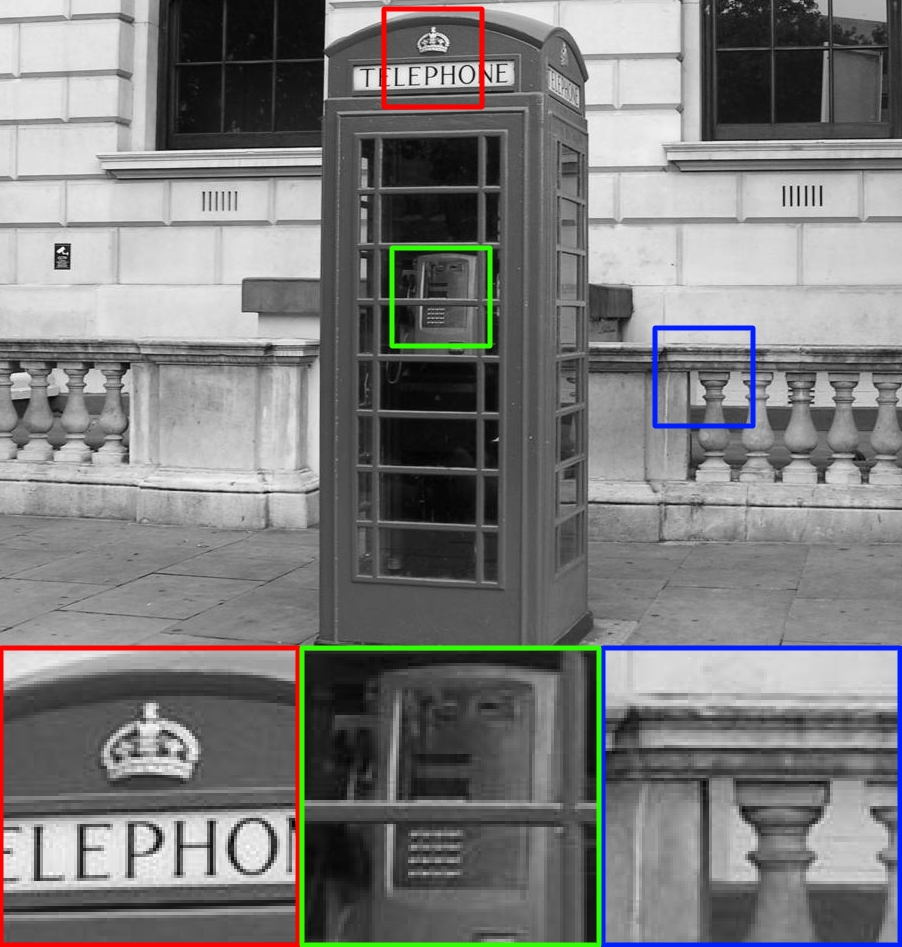}\\
    \small{Input / 20.13} & \small{RLSDN (ours) / 34.91} & \small{FDN \cite{Kruse2017} / 33.48} & \small{IRCNN \cite{Zhang2017b} / 29.82} & \small{RGDN \cite{Gong2020} / 26.57} & \small{Ground-truth}\\
    
   \includegraphics[width=.165\linewidth]{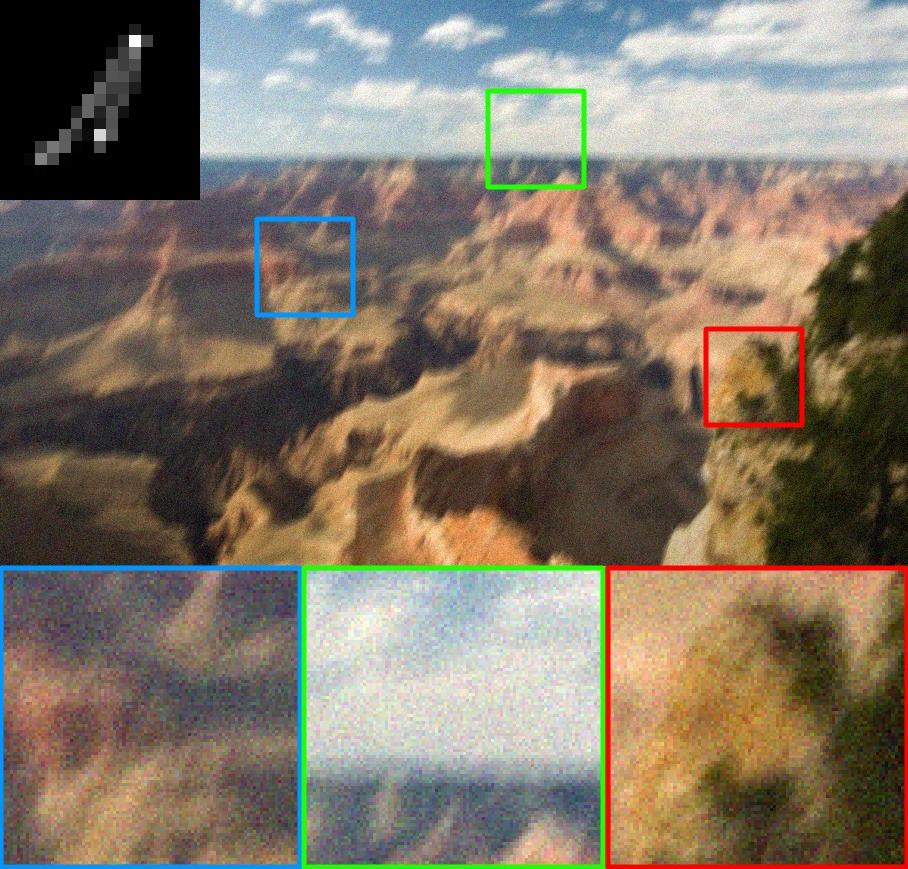}&
   \includegraphics[width=.165\linewidth]{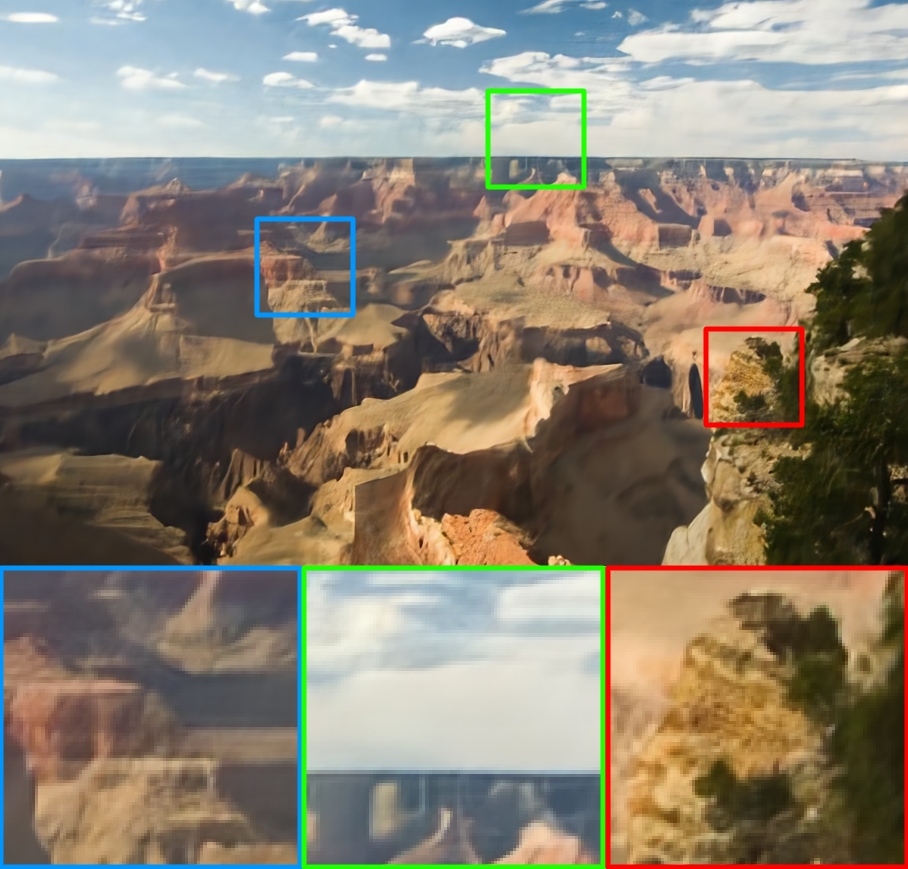}&
   \includegraphics[width=.165\linewidth]{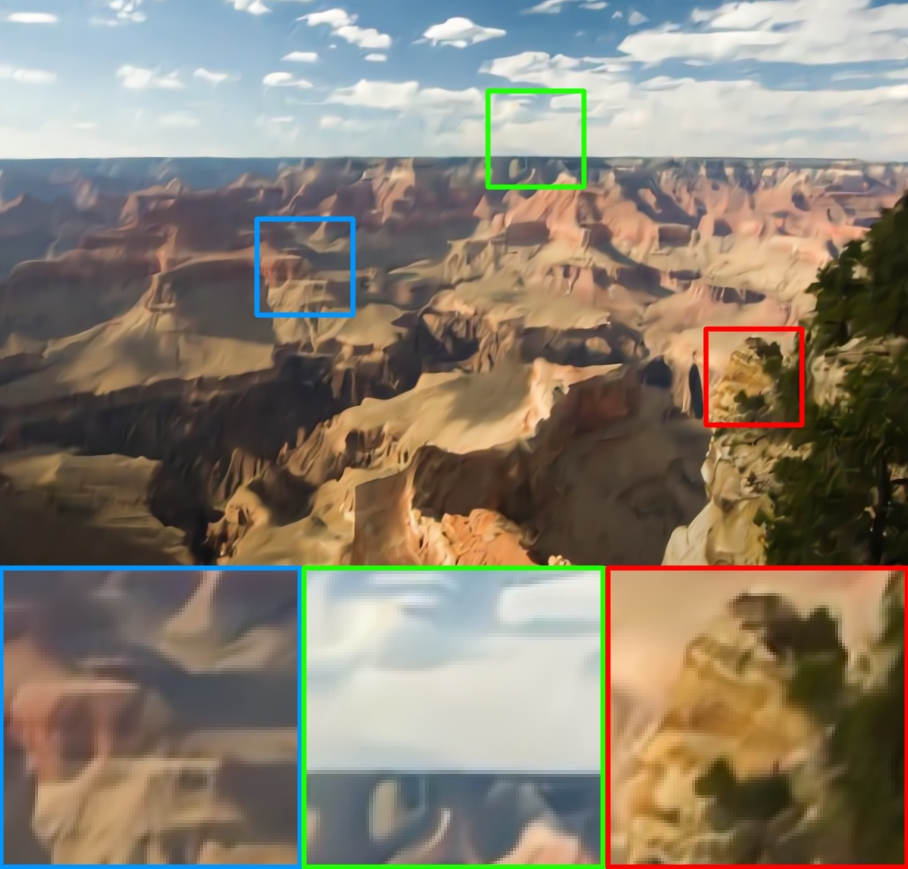}&
   \includegraphics[width=.165\linewidth]{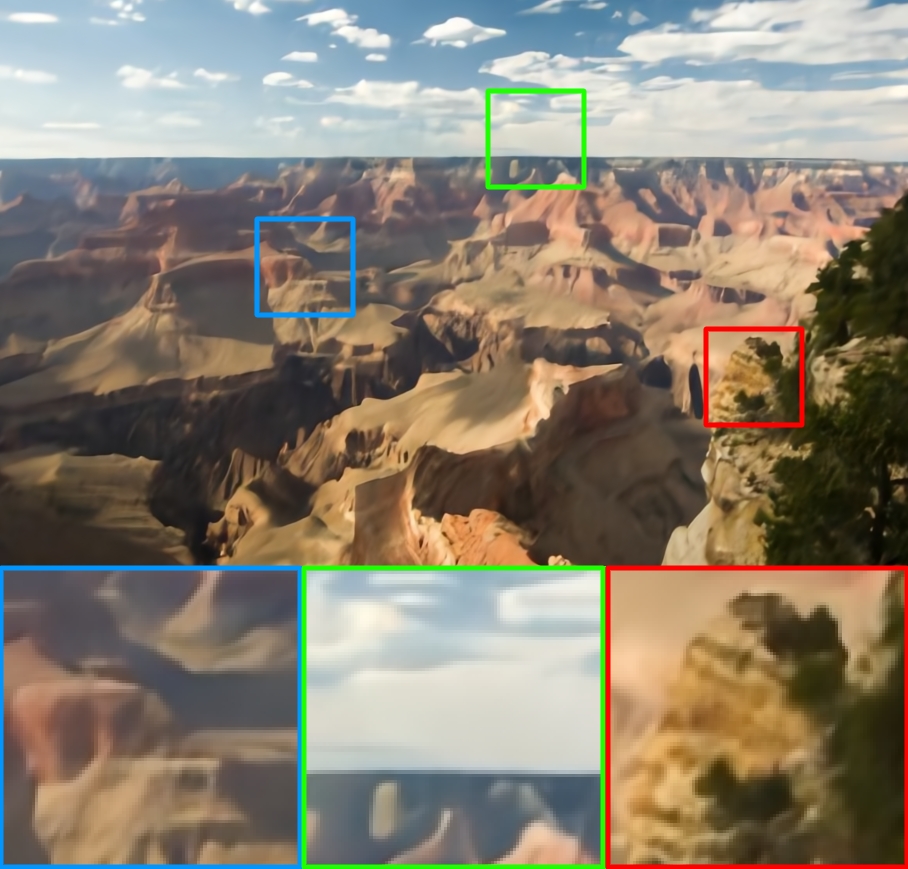}&
   \includegraphics[width=.165\linewidth]{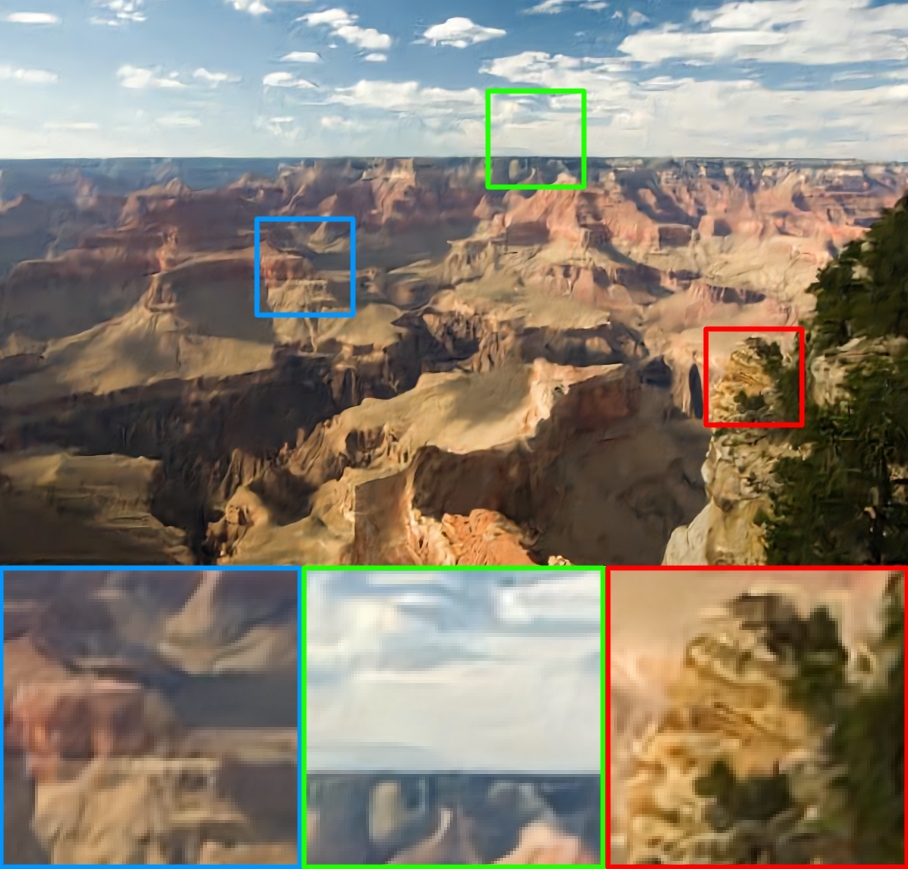}&
   \includegraphics[width=.165\linewidth]{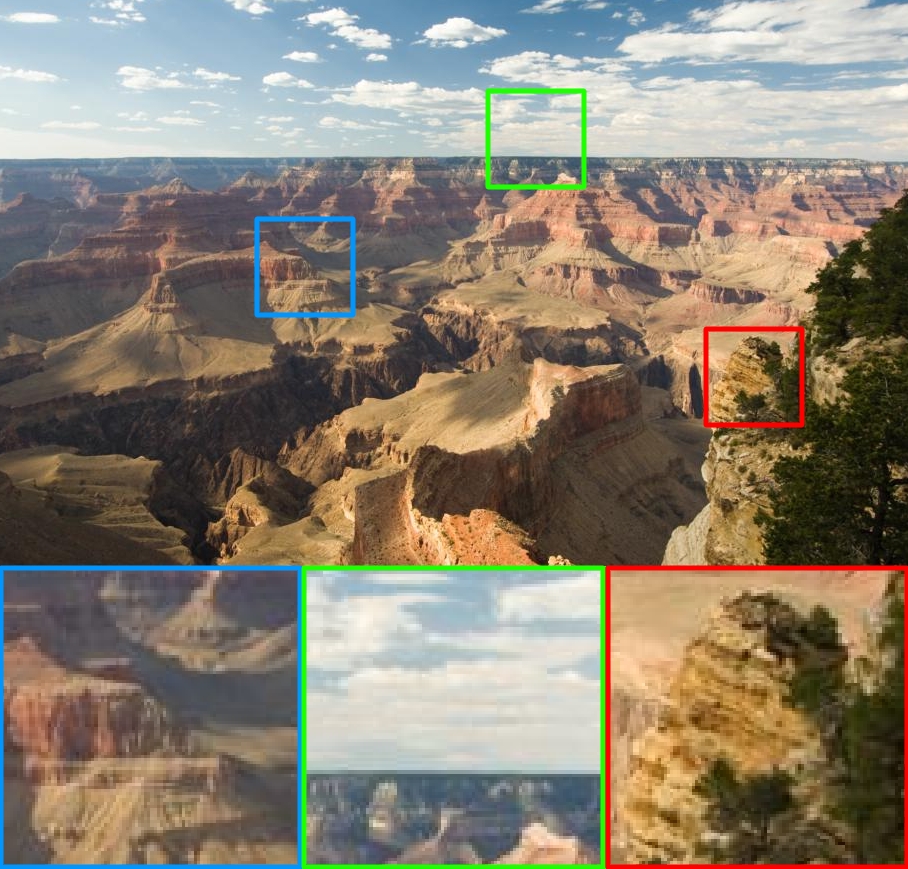}\\
    \small{Input / 22.72} & \small{RLSDN (ours) / 29.31} & \small{SVMAP \cite{Dong2021} / 28.94} & \small{DWDN \cite{Dong2021b} / 29.04} & \small{IRCNN \cite{Zhang2017b} / 29.00} & \small{Ground-truth}\vspace{-0.2cm}
\end{tabular}
   \caption{Visual comparisons with state-of-the art methods on synthetically blurred images from the Sun \etal \cite{Sun2013} dataset. The top row refers to grayscale deblurring results with $1\%$ noise, while the bottom row to color deblurring with $5\%$ noise. For each image its PSNR value is provided in dB.}
   \label{fig:SynthColorComp}
\end{figure*}

\begin{figure*}[t]
\centering
\begin{tabular}{@{} c @{ } c @{ } c @{ } c @{ } c @{ }}
   \includegraphics[width=.19\linewidth]{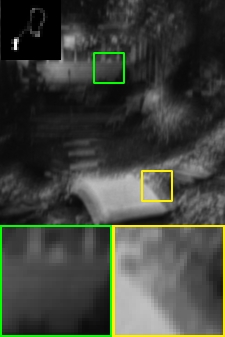}&
   \includegraphics[width=.19\linewidth]{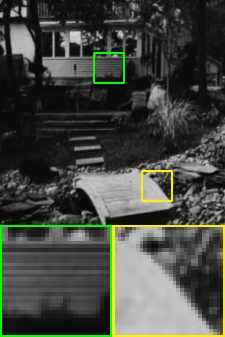}&
   \includegraphics[width=.19\linewidth]{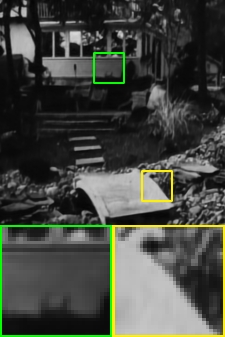}&
   \includegraphics[width=.19\linewidth]{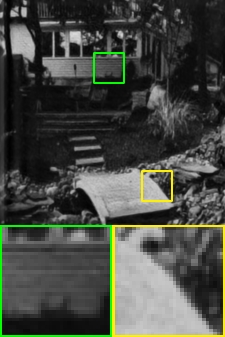}&
   \includegraphics[width=.19\linewidth]{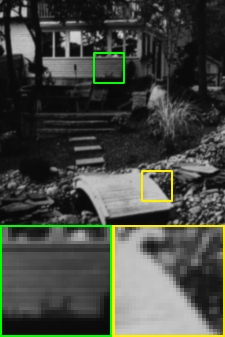}\\
    \small{Input / 22.11} & \small{RLSDN (ours) / 36.78} & \small{FDN \cite{Kruse2017} / 36.02} & \small{RGDN \cite{Dong2021b} / 31.95} & \small{Ground-truth}
\end{tabular}
    \vspace{-.2cm}
   \caption{Visual comparisons with state-of-the art methods on a real blurred example from the Levin \etal \cite{Levin2009} dataset. For each image its PSNR value is provided in dB.}
   \label{fig:RealGrayscaleComp}
   \vspace{-.4cm}
\end{figure*}

\begin{table}[t]
\centering
\caption{Comparison of RLSDN with state-of-the-art methods on the Levin \etal \cite{Levin2009} real grayscale benchmark.}
\begin{tabular}{lll}
\hline
Method  & PSNR (dB)  & SSIM   \\ \hline
IRCNN \cite{Zhang2017b}   & 27.05 & 0.8200 \\ 
RGDN \cite{Gong2020}     & 33.57 & 0.9648 \\ 
FDN \cite{Kruse2017}     & 36.20 & 0.9811 \\ 
RLSDN (ours)& \bf{36.92} & \bf{0.9842}
\end{tabular}
\label{tab:levin}
\vspace{-.3cm}
\end{table}

\begin{table*}[]
\centering
\caption{Comparison of RLSDN with state-of-the-art methods on the Sun \etal \cite{Sun2013} synthetic benchmark. With -- we indicate the methods whose inference code is publicly available only for color images and underperform when applied on grayscale images.}
\begin{tabular}{lllllllll}
\hline
                           & Noise          & Metrics      & IRCNN \cite{Zhang2017b}  & RGDN \cite{Gong2020}   & FDN \cite{Kruse2017}   & DWDN \cite{Dong2021b}   & SVMAP \cite{Dong2021}  & RLSDN (ours) \\ \hline
\multirow{4}{*}{Grayscale} & \multirow{2}{*}{1\%} & PSNR & 29.31  & 26.90  & 32.63  & -- & --  & \bf{33.07}   \\
                           &                      & SSIM      & 0.8047 & 0.6216 & 0.8894 & -- & -- & \bf{0.9006}  \\ \cline{2-9} 
                           & \multirow{2}{*}{5\%} & PSNR & 27.60  & 14.06  & 27.72  & --  & --  & \bf{28.13}   \\
                           &                      & SSIM      & 0.7444 &  0.1267   & 0.7348 & -- & -- & \bf{0.7573}  \\ \hline
\multirow{4}{*}{Color}     & \multirow{2}{*}{1\%} & PSNR & 29.68  & 30.98  & 32.51  & 34.09  & 34.36  & \bf{34.64}   \\
                           &                      & SSIM      & 0.8311 & 0.8840 & 0.8857 & 0.9197 & 0.9249 & \bf{0.9287}  \\ \cline{2-9} 
                           & \multirow{2}{*}{5\%} & PSNR & 28.85  & 26.93    & 27.66  & 29.16  & 29.15  & \bf{29.44}   \\
                           &                      & SSIM      & 0.7961 & 0.7121   & 0.7292 & 0.7905 & 0.7925 & \bf{0.8077} 
\end{tabular}
\label{tab:sun}
\vspace{-.3cm}
\end{table*}

\begin{table}[]
\centering
\caption{Comparison of RLSDN with state-of-the-art methods on images with saturated pixels.}
\begin{tabular}{lll}
\hline
Method  & PSNR (dB)  & SSIM   \\ \hline
Whyte \etal~\cite{Whyte2014} & 25.30 & 0.6626 \\
Cho \etal~\cite{Cho2011} & 32.20 & 0.8936 \\
SVMAP \cite{Dong2021} & 30.30 & 0.8191 \\ 
RLSDN (ours) & \bf{32.76} & \bf{0.9230}
\end{tabular}
\label{tab:saturated}
\vspace{-.5cm}
\end{table}

\vspace{-.1cm}
\section{Experiments and Results}
\label{sec:experimental}\vspace{-0.2cm}
In this section, we discuss the implementation details of our network and report the performed comparisons against recent state-of-the-art deblurring methods on several publicly available datasets. Additional results are provided in the supplementary material.

\vspace{-.2cm}
\subsection{Train and test data}\vspace{-.2cm}
For training purposes, we have combined DIV2K \cite{Agustsson2017}, and Flickr2K \cite{Wang2018} datasets, which allowed us to use 3450 high-resolution source images in total. We used blur kernels of support sizes lying in the range between $13 \times 13$ to $35 \times 35$ pixels, randomly sampled from a Brownian motion model using the procedure proposed in \cite{Boracchi2012}. Ground truth crops of size $128 \times 128$ pixels were selected based on responses of the Laplacian filter in order for each sample to contain high-frequency details. For all target samples, we follow the degradation model in \cref{eq:fwd_model} to obtain the corresponding blurred inputs. We have trained two separate models considering small and high noise level scenarios with standard deviations lying within the range of $[1.0, 3.0]$ for the former and $[11.75, 13.75]$ for the latter.

To show the general applicability of our network, we have conducted a separate experiment considering deblurring of saturated images. For this purpose, we have used a similar procedure as the one proposed in \cite{Dong2021} to obtain train samples. Although the degradation model for such cases departs from the adopted model of \cref{eq:fwd_model}, we have used the same RLSDN architecture and retrained the network to handle saturated cases with noise levels of standard deviation lying in the range of $[1.0, 3.0]$. By construction, none of our training data intersects with any data we have used for validation and testing purposes.

To validate the performance of our network and compare it against other methods we consider color and grayscale deblurring benchmarks consisting of both synthetic and real images. For synthetically blurred images, a standard dataset has been proposed by Sun \etal \cite{Sun2013}. This dataset consists only of grayscale images. In order to be able to compare both for grayscale and color cases, we have used the \textit{original} color images from \cite{Sun2012} and the same blur kernels as those considered in \cite{Sun2013}. Then, we created the test dataset by applying a valid convolution as the blurring operation and two different noise levels corresponding to standard deviations of 1\% and 5\% of the peak image intensities. To validate all the methods, we follow the protocol proposed by \cite{Sun2013} and we compute all the metrics by considering the central part of the \textit{original} image (we discard 50 pixel from each border), so as to avoid the influence of boundary artifacts in the SSIM and PSNR scores. 

Several synthetic benchmarks were previously proposed to evaluate deblurring methods on large set of images with saturated pixels~\cite{Dong2021}. Unfortunately, none of them contain neither publicly released data, nor concrete instructions on how to reproduce them. For this reason we collected a set of 21 images with saturated areas and used those to produce synthetically blurred results using the same kernels as those used above. All synthetic data used for evaluation can be downloaded from \href{https://drive.google.com/drive/folders/1G31zQ1H-00IJ-w_uKbRdXkZCouo_Q6nm?usp=sharing}{this link}.

To compare our approach on real data with uniform blur, we have used a benchmark dataset provided by Levin \etal~\cite{Levin2009}, which consists of optically blurred images and accurate blur kernels. Since the degraded images and corresponding ground truths are not well aligned, we compute the scores using the original Matlab code provided by the authors in \cite{Levin2011B}, which internally performs image alignment. 

\subsection{Model Specification and Training}
\vspace{-.2cm}
The concrete implementation of the proposed RLSDN model requires us to parameterize several ``free" variables and in particular the regularization operator $\m G$, the weight prediction network and the regularization constant $\alpha$, as well as other hyper-parameters. In all our experiments, we parameterize $\m G$ with a valid convolution layer that consists of $128$ output channels and filters of size $13 \times 13$. As it was proposed in \cite{Laurent2018} we further overparameterize this operation by a deep linear network (i.e. a composition of convolution layers with smaller filter sizes) to accelerate the training. We apply spectral normalization to each layer of the deep linear network to stabilize the training and resolve the ambiguity that can be introduced due to the interplay between WPN and $\m G$. To parameterize the WPN, we use the RRDB backbone \cite{Wang2018} with $128$ input and output channels, and we add a final ReLU activation layer to ensure non-negativity of its output. Finally, the regularization constant $\alpha$ is modeled as $\alpha = e^{\beta}$, where $\beta$ is learned without any restrictions on its values.

We have trained our network by performing 5 steps, with the first one involving a separately learned Wiener filter. Its output $\m x^{1} = \pr{\m H^\transp\m H + {\sigma^2}\m G_\textrm{w}^\transp \m G_\textrm{w}}^{-1}{\m H^\transp\m y}$, provides a good initial result that we further refine in the following four adaptive steps. At each iteration, we limit the amount of CG steps to $250$ during forward and $500$ during backward passes. We also employ an early exit strategy if the relative tolerance of the residual lies below $10^{-3}$ for all elements of the batch. During the forward pass, the output from the previous RLSDN recurrent step is used as the initial solution of the linear solver to further reduce the amount of iterations required for convergence. 

All our models were learned using back-propagation by minimizing the sum of mean squared errors between ground truth and outputs of each recurrent step. We used Adam optimizer \cite{Kingma2014} with a learning rate $2 \cdot 10^{-4}$ decreasing by a factor of $0.98$ after each epoch and a warm-up strategy for the first two epochs. We set the batch size to $32$ and trained our models for $100$ epochs, where each epoch consists of $2000$ batch passes. 

\begin{figure*}[!t]
\centering
\begin{tabular}{@{} c @{ } c @{ } c @{ } c @{ } c @{ }}
   \includegraphics[width=.2\linewidth]{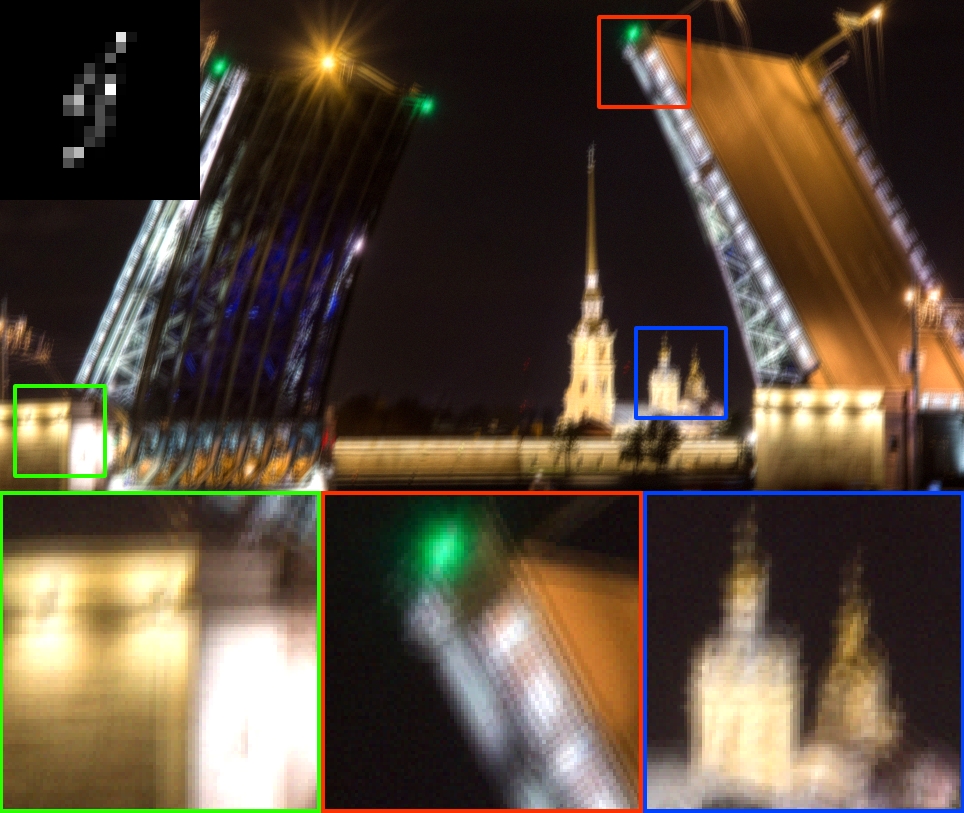}&
   \includegraphics[width=.2\linewidth]{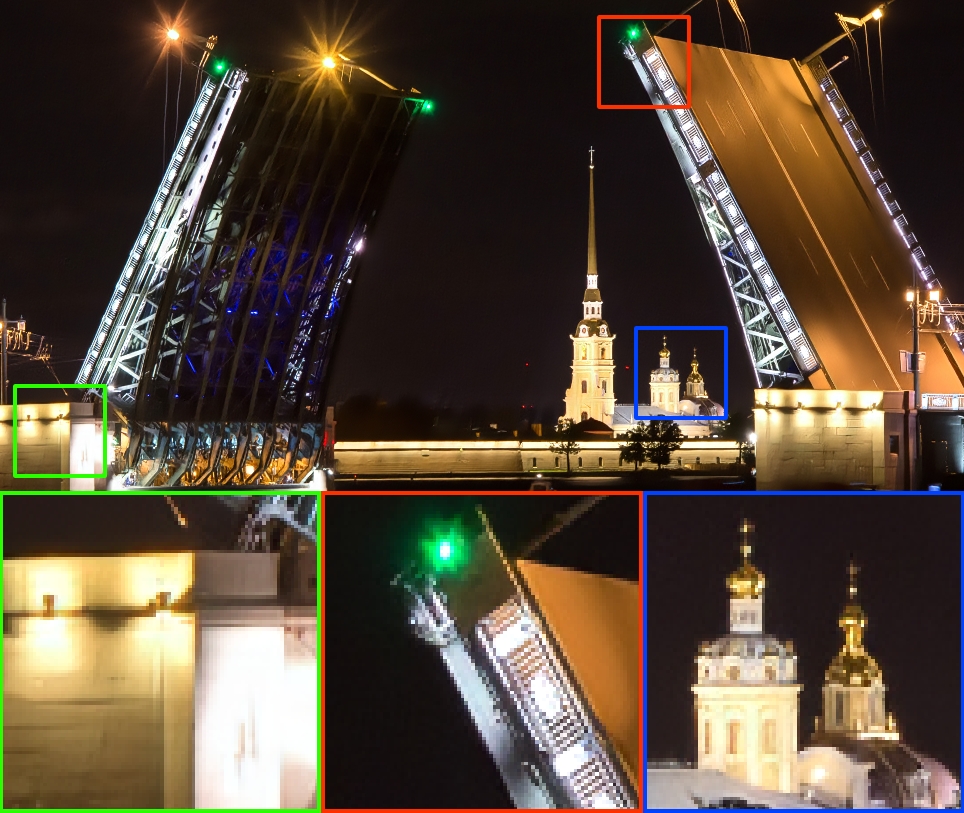}&
   \includegraphics[width=.2\linewidth]{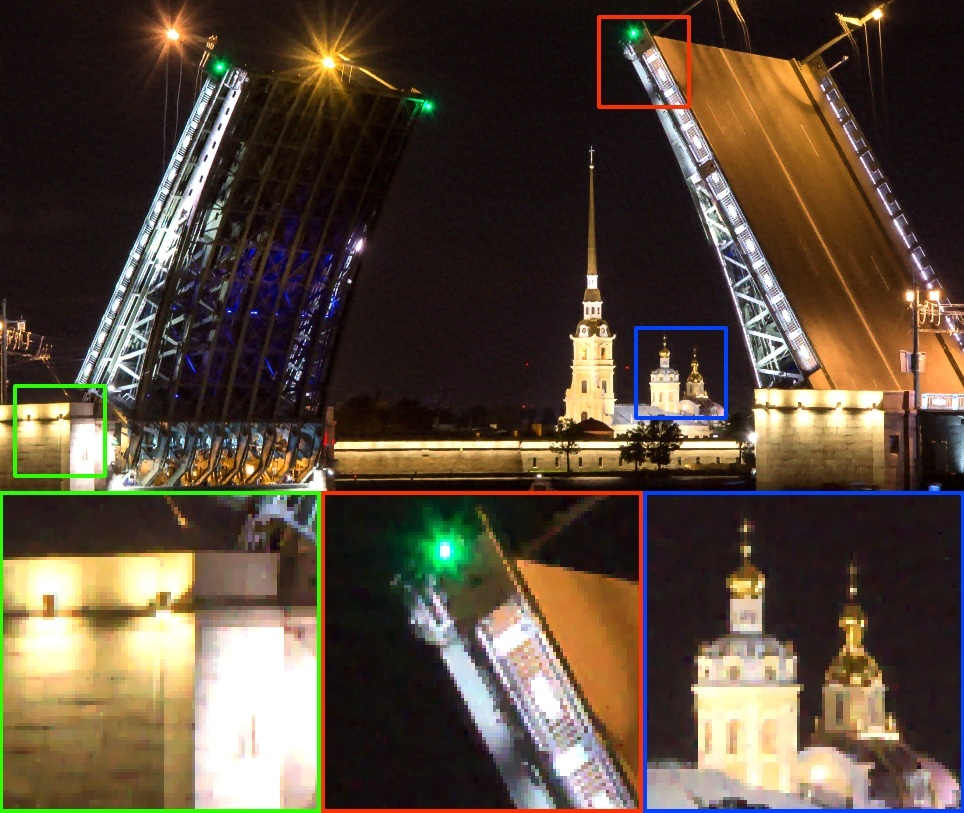}&
   \includegraphics[width=.2\linewidth]{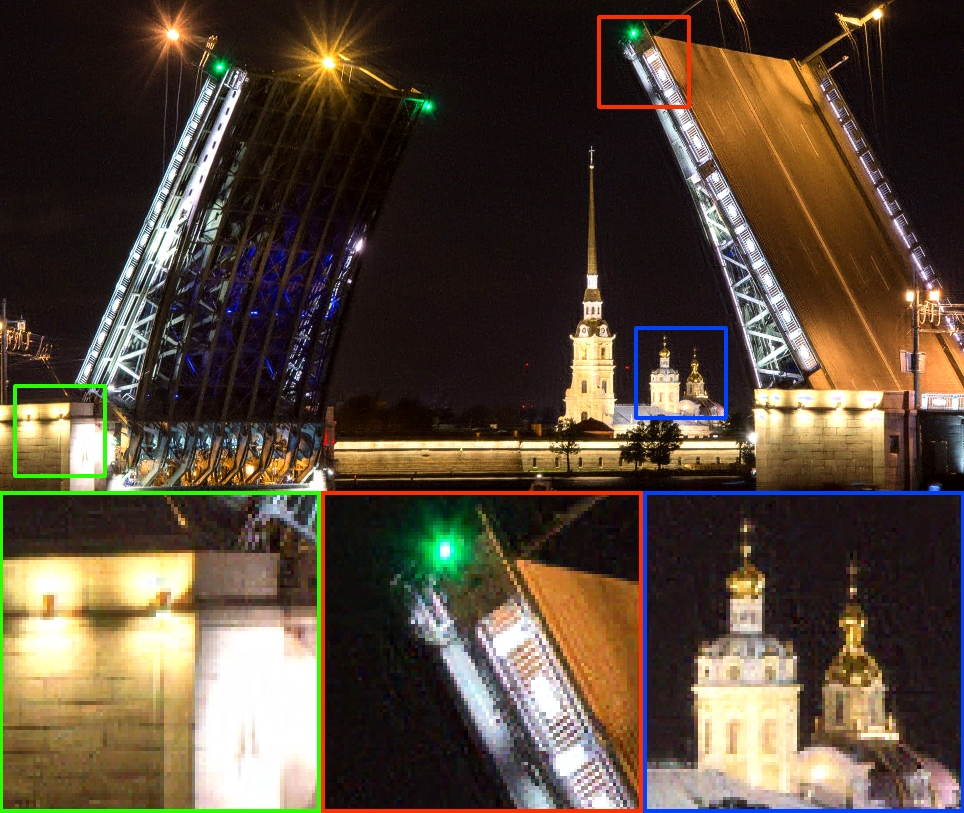}&
   \includegraphics[width=.2\linewidth]{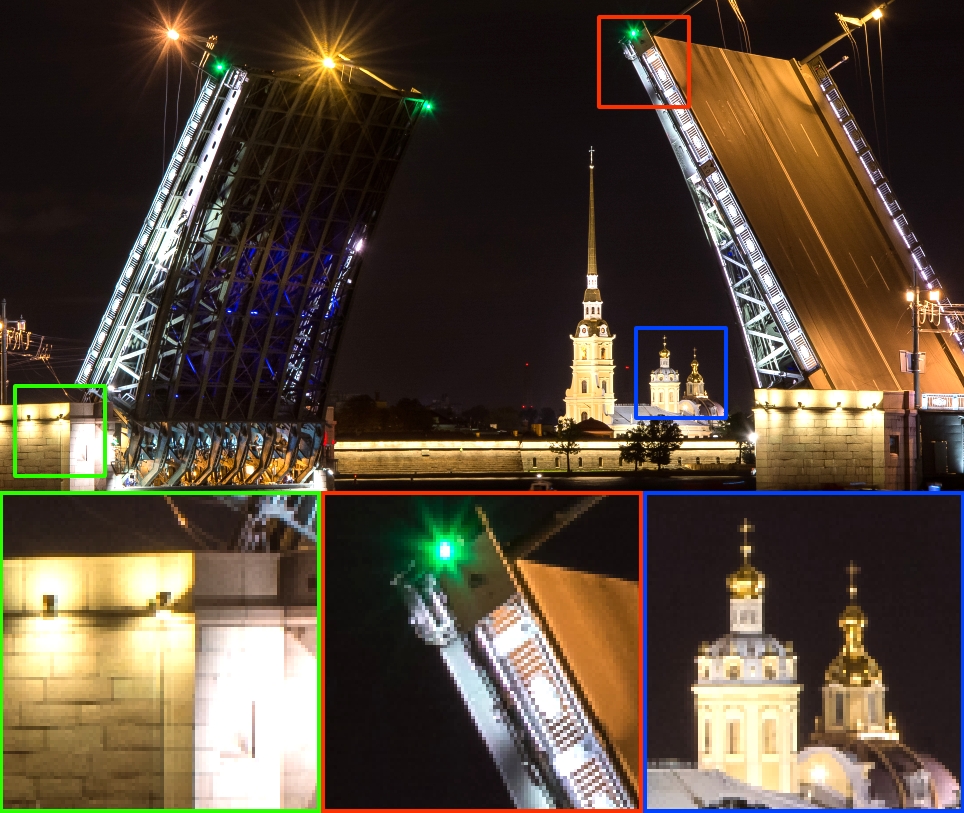}\\
    \small{Input / 20.69} & \small{RLSDN (ours) / 31.14} & \small{Cho \etal \cite{Cho2011} / 29.49} & \small{SVMAP \cite{Dong2021} / 29.04} & \small{Ground-truth}\vspace{-0.2cm}
\end{tabular}
   \caption{Visual comparisons with state-of-the-art methods on synthetically blurred images with saturated pixels. For each image its PSNR value is provided in dB. Original image taken from \url{https://flic.kr/p/ZMVcuC}.}
   \label{fig:SynthSatColorComp}
   \vspace{-.3cm}
\end{figure*}

\subsection{Results}
\vspace{-.2cm}
{\bf Synthetic blur.} In \cref{tab:sun} we provide quantitative comparisons with state-of-the-art methods for two different noise levels. From the reported results it is clear, that our proposed network outperforms all competing methods by a good margin in both color and grayscale deblurring on small and large noise levels. In particular, for the case of color deblurring our method leads to improved results of almost $0.3$dB for both $1\%$ and $5\%$ Gaussian noise compared to the recent best performing network~\cite{Dong2021}. The improved performance is even more pronounced in the case of grayscale image deblurring, where our method outperforms the second best FDN network \cite{Kruse2017} by nearly $0.4$dB in all tested cases. Representative grayscale and color deblurring results that demonstrate visually the restoration quality of the proposed models are shown in \cref{fig:SynthColorComp}.

Comparisons among methods specifically designed for image deblurring of images with saturated pixels are presented in \cref{tab:saturated} and follow a similar trend as the previous results. In particular, our network outperforms the second best method~\cite{Cho2009} by $0.56$dB, SVMAP~\cite{Dong2021} by $2.46$dB and the method of ~\cite{Whyte2014} by more than $7$dB. 
It is noteworthy to mention that while SVMAP's network architecture is also inspired by IRLS, it performs significantly worse than our network. Our improved performance can be attributed both to our specific network architecture and our training strategy, which allow us to employ more recurrent steps and linear solver iterations. Indeed, SVMAP consists of two external iterations with five CG iterations each, which typically are not enough for large-scale restoration problems. A visual comparison of the different restored results obtained by three of the methods under consideration is possible by referring to \cref{fig:SynthSatColorComp}. According to these results we observe that our network leads to a visually better restored image, without the presence of strong artifacts or noise residual. This is not the case both for SVMAP~\cite{Dong2021} and the method of~\cite{Cho2009}, where the restored image by the former method suffers from strong noise while the image recovered by the latter method is significantly over-smoothed and as a result fine details of the image are failed to be recovered entirely.

{\bf Real blur.} In \cref{tab:levin} we report results for real grayscale image deblurring. Our comparisons involve only methods that work well on grayscale images and we refrain from reporting results by other methods appearing in \cref{tab:sun} which are designed for color image deblurring and are not producing competitive enough results. \cref{tab:levin} clearly demonstrates the superiority of our approach and its ability to handle real motion blur. Similar to the synthetic case, our network outperforms FDN \cite{Kruse2017}, which is the current state-of-the-art grayscale deblurring method, by $0.72$dB. An important observation is that our network, while it is exclusively trained on synthetic examples,  performs very well on real images when accurate blur kernels are provided. For a visual inspection of the restoration performance of our grayscale model we refer to \cref{fig:RealGrayscaleComp}. For the most  practical cases when an accurate motion trajectory is not available, a third party kernel estimation method such as~\cite{Pan2016} can be used to complement our network. As it can be seen from \cref{fig:RealColorExample}, even in cases where only an estimate of the blur kernel is available, our network is still able to restore the image adequately and produce results of good visual quality.

\vspace{-0.1cm}
\section{Conclusions and Future Work}
\vspace{-.1cm}
In this work we introduced a novel recurrent network architecture that we deploy to deal with the task of non-blind image deconvolution. The design of the proposed network is inspired by a large-scale fixed-point iteration method that we developed and it amounts to solving a series of adaptive non-negative least squares problems. This approach coupled with a new training strategy that eliminates the need to unroll the involved iterations of the linear solver leads to a very effective deconvolution network that produces state-of-the-art results under different scenarios, including noise of low and high levels and images with saturated pixels.

While fixed-point iteration methods are frequently met in optimization, unless they satisfy certain conditions they might diverge. Since our proposed network architecture implements a particular fixed-point iteration strategy, not all of its possible configurations are guaranteed to always converge, irrespectively of the input data. Based on this, an interesting future research direction that we plan to pursue is investigate possible ways to ensure that the learned network parameters will lead to a configuration with provable convergence guarantees. This is a very important property for a network to possess and can be extremely useful in practical applications. Another possible direction is to investigate the applicability of our network architecture to other inverse imaging problems such as demosaicking and image super-resolution.

\appendix
\addcontentsline{toc}{section}{Appendices}
\section*{Appendix}
\section{Implicit Back-Propagation for Linear Systems}

Let $\m A\in\R^{n\times n}$ be a non-singular matrix and $\m b\in\R^n$ a vector. Both $\m A$ and $\m b$ depend on a parameter vector $\bm w\in\R^m$ and form the following linear system:
\bal
\m A\pr{\bm w}\m x = \m b\pr{\bm w}.
\label{eq:normal_eqs}
\eal
\
Let us also consider an implicit network layer whose parameters correspond to the latent vector $\bm w$ and it's output is given as:
\bal
\hm x = \m A^{-1}\pr{\bm w}\m b\pr{\bm w}.
\label{eq:qpl_output}
\eal
In order to learn the layer's parameters using back-propagation, we need to be able to compute the gradient of its output w.r.t $\bm w$.
To do so, first we compute the differential of $\hm x$ as:
\bal
&\md\hm x = \pr{\md\m A^{-1}\pr{\bm w}}\m b\pr{\bm w} + \m A^{-1}\pr{\bm w}\md\m b\pr{\bm w}\notag\\
&=-\m A^{-1}\pr{\bm w}\pr{\md\m A\pr{\bm w}}\m A^{-1}\pr{\bm w}\m b\pr{\bm w}+\m A^{-1}\pr{\bm w}\md\m b\pr{\bm w}\notag\\
&\overset{\eqref{eq:qpl_output}}{=}-\m A^{-1}\pr{\bm w}\pr{\md\m A\pr{\bm w}}\hm x +\m A^{-1}\pr{\bm w}\md\m b\pr{\bm w}\notag\\
&=-\pr{{\hm x}^\transp\otimes \m A^{-1}\pr{\bm w}}\vc{\md \m A\pr{\bm w}}+\m A^{-1}\pr{\bm w}\md\m b\pr{\bm w}\notag\\
&=-\pr{{\hm x}^\transp\otimes \m A^{-1}\pr{\bm w}}\deriv{\vc{\m A\pr{\bm w}}}{\bm w}\md\bm w\notag\\
&+\m A^{-1}\pr{\bm w}\deriv{\m b\pr{\bm w}}{\bm w}\md\bm w,
\label{eq:diff_hx}
\eal
where we have used the property
$\md\m A^{-1} = -\m A^{-1}\pr{\md\m A}\m A^{-1}$~\cite{Magnus1999}
and the $\vctr$ identity $\vc{\m A\m B\m C}=\pr{\m C^\transp\otimes \m A}\vc{\m B}$ with $\m A, \m B, \m C$ being proper-sized matrices and $\otimes$ denoting the Kronecker product.

From Eq.~\eqref{eq:diff_hx} and using the identification theorem~\cite{Magnus1999} we can compute the gradient of $\hm x$ w.r.t $\bm w$ as:
\bal
&\grad{\hm x}{\bm w} = -\grad{\vc{\m A\pr{\bm w}}}{\bm w}\pr{\hm x\otimes \m A^{-\transp}\pr{\bm w}}\notag\\
&+\grad{\m b\pr{\bm w}}{\bm w}\m A^{-\transp}\pr{\bm w}
\label{eq:grad_hx_w}.
\eal
Next, let us compute the gradient of a scalar loss function $\mc L:\R^n\mapsto\R$ w.r.t the parameter vector $\bm w$ given that $\grad{\mc L}{\hm x}=\bm\rho\in\R^n$. Using the chain rule and Eq.~\eqref{eq:grad_hx_w}, we get:
\bal
&\grad{\mc L}{\bm w} = \grad{\hm x}{\bm w}\grad{\mc L}{\hm x} = \br{-\grad{\vc{\m A\pr{\bm w}}}{\bm w}\pr{\hm x\otimes \m A^{-\transp}\pr{\bm w}} \right. \notag\\
&\left. + \grad{\m b\pr{\bm w}}{\bm w}\m A^{-\transp}\pr{\bm w}}\bm\rho\notag\\
&=-\grad{\vc{\m A\pr{\bm w}}}{\bm w}\vc{\m A^{-\transp}\pr{\bm w}\bm\rho{\hm x}^{\transp}}\notag\\ 
&+ \grad{\m b\pr{\bm w}}{\bm w}\m A^{-\transp}\pr{\bm w}\bm\rho\notag\\
&=-\grad{\vc{\m A\pr{\bm w}}}{\bm w}\vc{\bm g{\hm x}^{\transp}}+\grad{\m b\pr{\bm w}}{\bm w}\bm g,
\label{eq:grad_L_w}
\eal
where $\bm g = \m A^{-\transp}\pr{\bm w}\bm\rho$ and can be efficiently computed by using a ``matrix-free" linear solver.

According to Eq.~\eqref{eq:grad_L_w}, in order to compute the exact expression of $\grad{\mc L}{\bm w}$ we first need to be able to compute the gradients of $\vc{\m A\pr{\bm w}}$  and $\m b\pr{\bm w}$ w.r.t $\bm w$. In certain cases, these gradients can be challenging to be derived in analytical form. This mostly depends on the specific structure of $\m A$ and $\m b$ and their dependency on $\bm w$. Moreover, in most of the cases we don't have access to the individual elements of the matrix $\m A$, but rather to the product of $\m A$ with a vector. Fortunately, as we describe next, there is a workaround which allows us to avoid the explicit computation of these gradients and instead rely on their computation using an auto-grad library. Specifically, let us consider the residual vector
\bal
\bm r = \m b\pr{\bm w} - \m A\pr{\bm w}\hm x 
\eal
and let us compute its differential as:
\bal
&\md\bm r = \md \m b\pr{\bm w} - \md\pr{\m A\pr{\bm w}}\hm x\notag\\ 
&= \md \m b\pr{\bm w} - \pr{{\hm x}^\transp\otimes \m I_n}\vc{\md\pr{\m A\pr{\bm w}}}\notag\\ 
&= \deriv{\m b\pr{\bm w}}{\bm w}\md\bm w - \pr{{\hm x}^{\transp}\otimes \m I_n}\deriv{\vc{\pr{\m A\pr{\bm w}}}}{\bm w}\md\bm w.\
\label{eq:residual_diff}
\eal
Now, from Eq.~\eqref{eq:residual_diff} and the identification theorem we can compute the gradient of $\bm r$ w.r.t $\bm w $ as:
\bal
\grad{\bm r}{\bm w} = \grad{\m b\pr{\bm w}}{\bm w}- \grad{\vc{\m A\pr{\bm w}}}{\bm w}\pr{\hm x\otimes\m I_n}.
\label{eq:grad_r_w}
\eal
Using Eq.~\eqref{eq:grad_r_w} we can show that:
\bal
&\pr{\grad{\bm r}{\bm w}}\bm g = \grad{\m b\pr{\bm w}}{\bm w}\bm g- \grad{\vc{\m A\pr{\bm w}}}{\bm w}\pr{\hm x\otimes\m I_n}\bm g\notag\\
&= \grad{\m b\pr{\bm w}}{\bm w}\bm g- \grad{\vc{\m A\pr{\bm w}}}{\bm w}\vc{\bm g {\hm x}^\transp}.
\label{eq:product_grad_r_g}
\eal
Comparing Eqs.~\eqref{eq:grad_L_w} and \eqref{eq:product_grad_r_g} we finally have that:
\bal
\grad{\mc L}{\bm w} = \pr{\grad{\bm r}{\bm w}}\bm g = \pr{\grad{\bm r}{\bm w}}\m A^{-\transp}\pr{\bm w}\grad{\mc L}{\hm x}.
\eal
The above equality suggests the following algorithmic steps for computing $\grad{\mc L}{\bm w}$, given $\bm\rho = \grad{\mc L}{\hm x}$:
\begin{enumerate}
	\item[1] \textbf{Forward Pass}\\
	Compute the solution $\hm x$ by solving the linear system
	$\m A\pr{\bm w}\m x = \m b\pr{\bm w}$.
	\item[2] \textbf{Backward Pass} 
	\subitem(a) Use $\hm x$ as the input of an auxiliary network that produces the output $$\bm r = \m b\pr{\bm w} - \m A\pr{\bm w}\hm x.$$
	\subitem(b) Compute $\bm g$ by solving the linear system $\m A^\transp\pr{\bm w}\bm g = \bm\rho.$
	\subitem(c) Obtain the gradient $\grad{\mc L}{\bm w}$ by computing the product $\pr{\grad{\bm r}{\bm w}}\bm g$. This last product can be computed automatically by any of the existing auto-grad libraries with $\bm g$ being the incoming gradient in the auxiliary network that produces the output $\bm r$.
\end{enumerate}

\section{Ablation Study}
\subsection{Convergence to a fixed point}
\vspace{-0.1cm}
In this section we provide some empirical evidence that our trained models have learned a mapping that converges to a fixed point. In order a fixed point iteration method to enjoy theoretical convergence guarantees, it is sufficient to show that the mapping of the input to the output is contractive, for any input. We note that in our case we do not impose any constraints on the learned parameters that can enforce our network to have this property. Nevertheless, from the results we report below we observe that in practice our trained networks indeed reach a fixed point in all cases and do not diverge.

\begin{figure}[!h]
\centering
   \includegraphics[width=\linewidth]{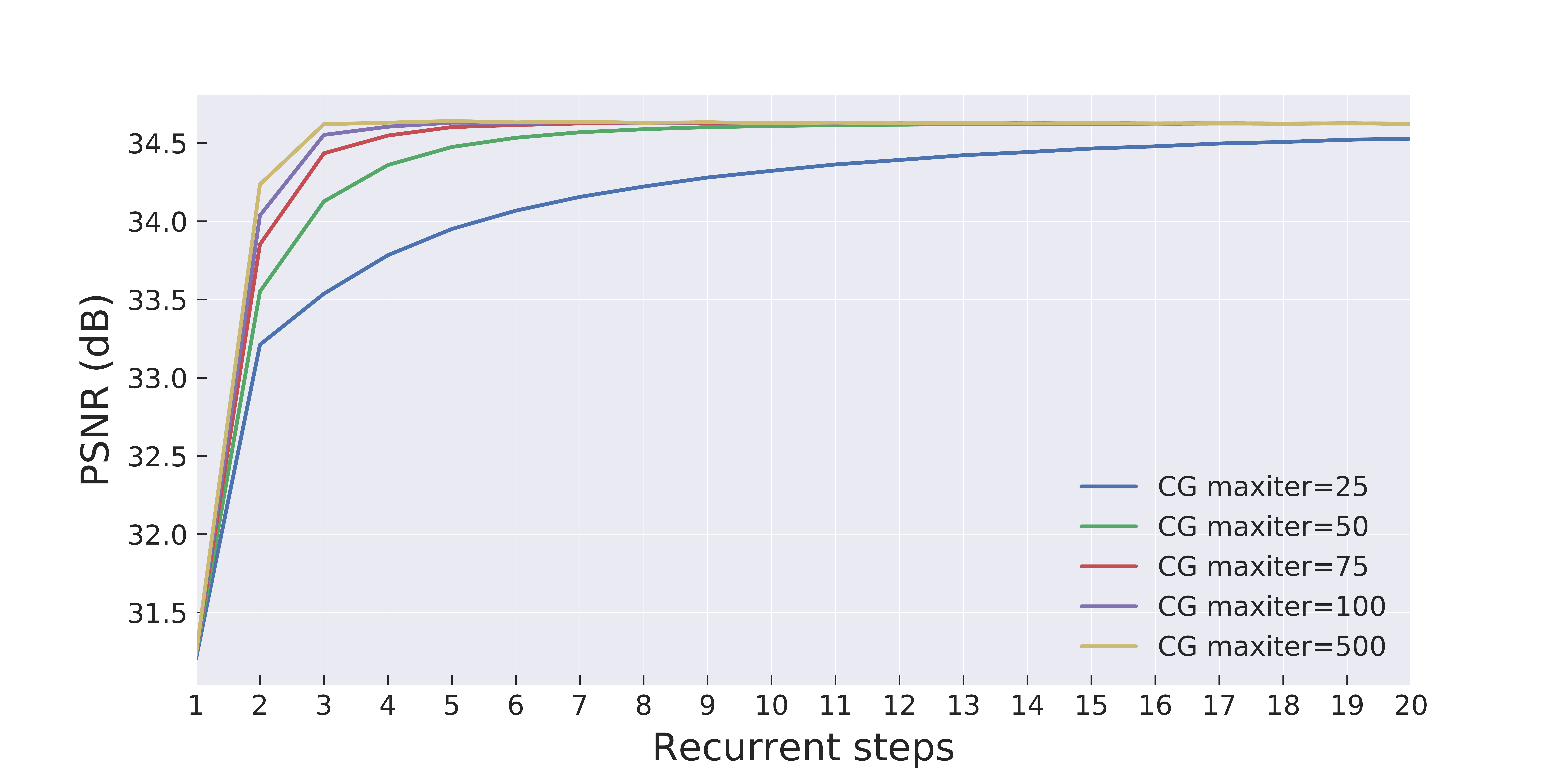}
   \caption{Convergence of RLSDN to a fixed point for different number of employed CG iterations per recurrent step.}
   \label{fig:ConvergencePlots}
   \vspace{-0.2cm}
\end{figure}

\begin{figure*}[!t]
\centering
\begin{tabular}{@{} c @{ } c @{ } c @{ } c @{ } c @{ } c @{ } c @{ } c @{ }}

    \small{Input} & \small{Step 1} & \small{Step 2} & \small{Step 3} & \small{Step 4} & \small{Step 5} & \small{Step 10} & \small{Step 20}\\
    \includegraphics[width=.125\linewidth]{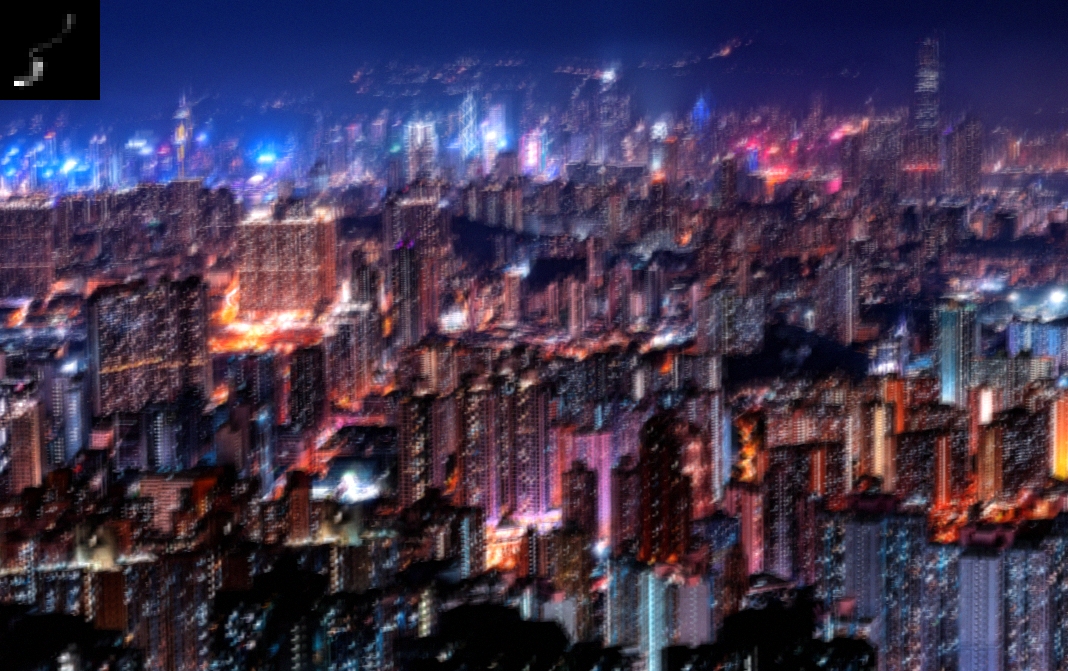}&
    \includegraphics[width=.125\linewidth]{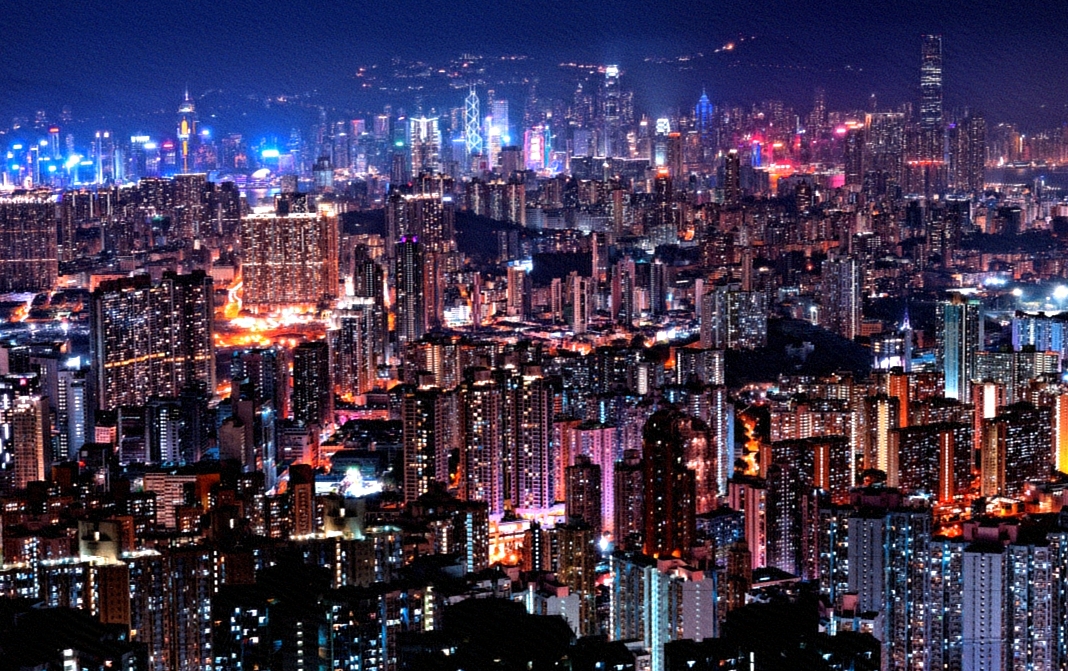}&
    \includegraphics[width=.125\linewidth]{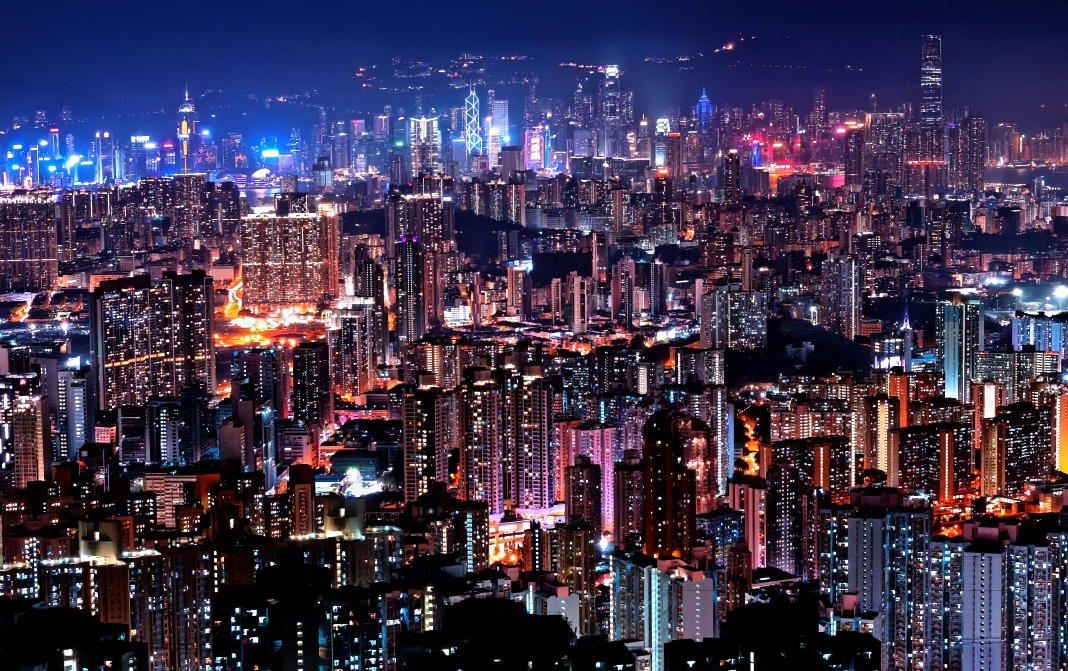}&
    \includegraphics[width=.125\linewidth]{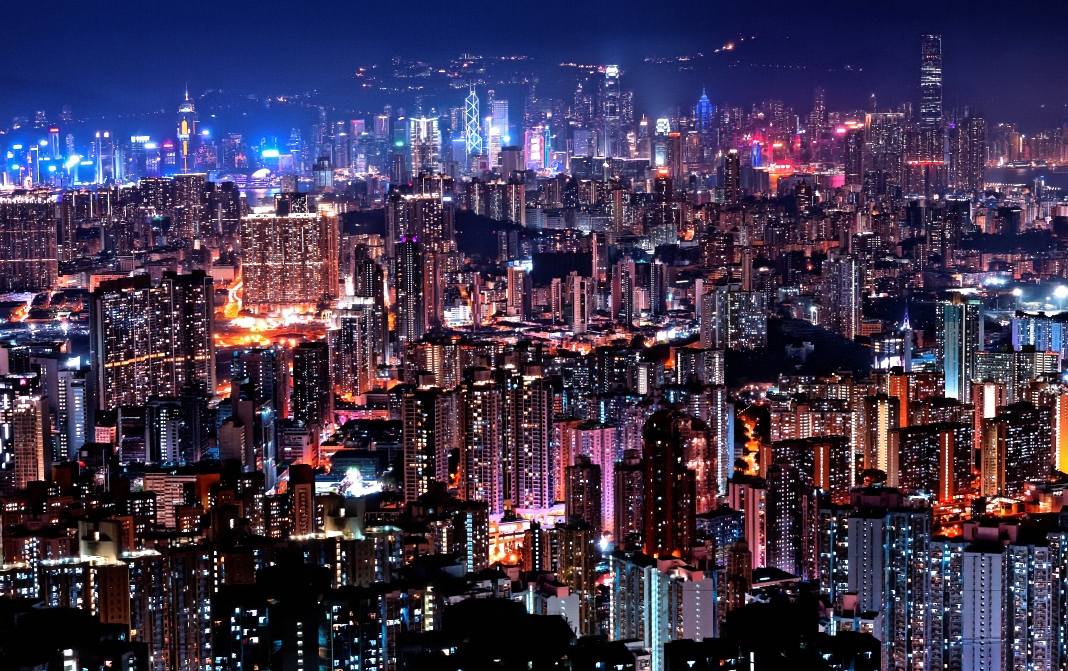}&
    \includegraphics[width=.125\linewidth]{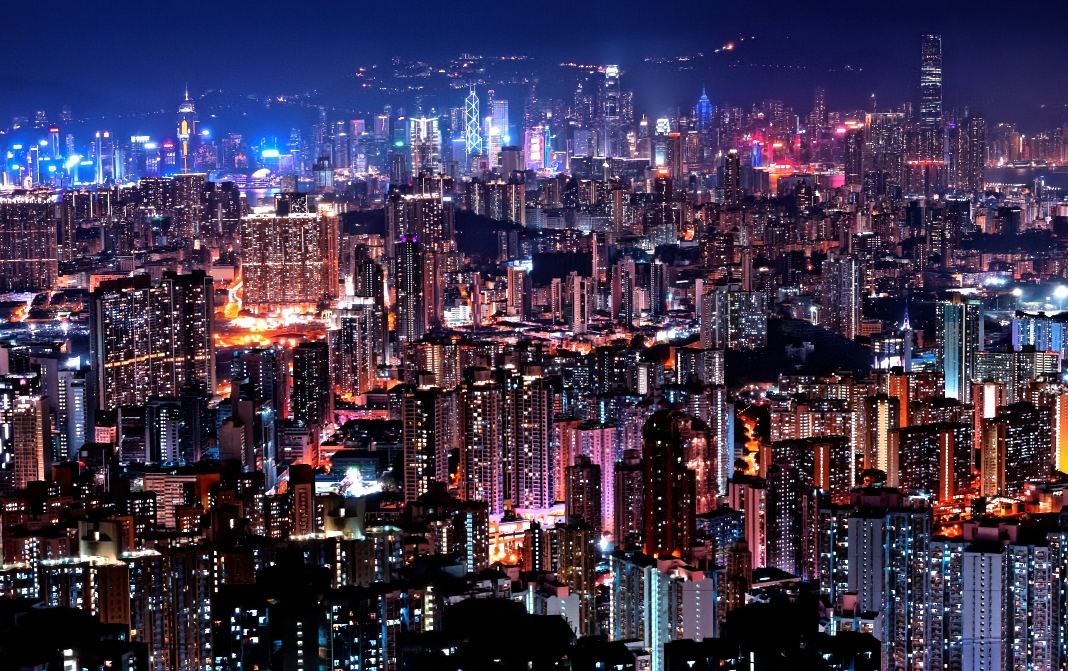}&
    \includegraphics[width=.125\linewidth]{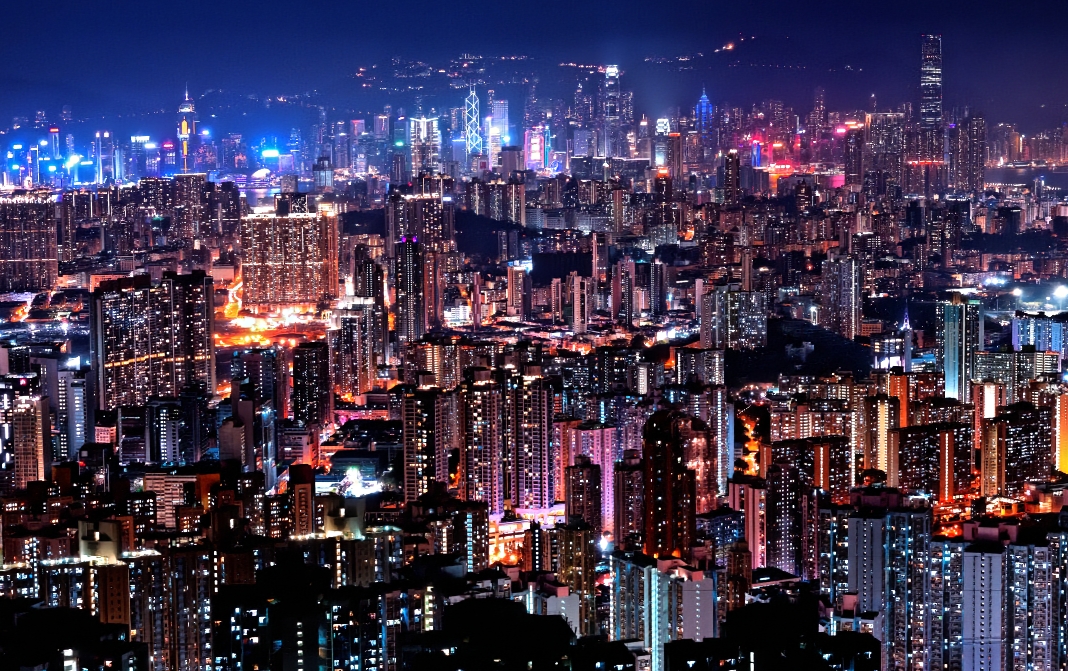}&
    \includegraphics[width=.125\linewidth]{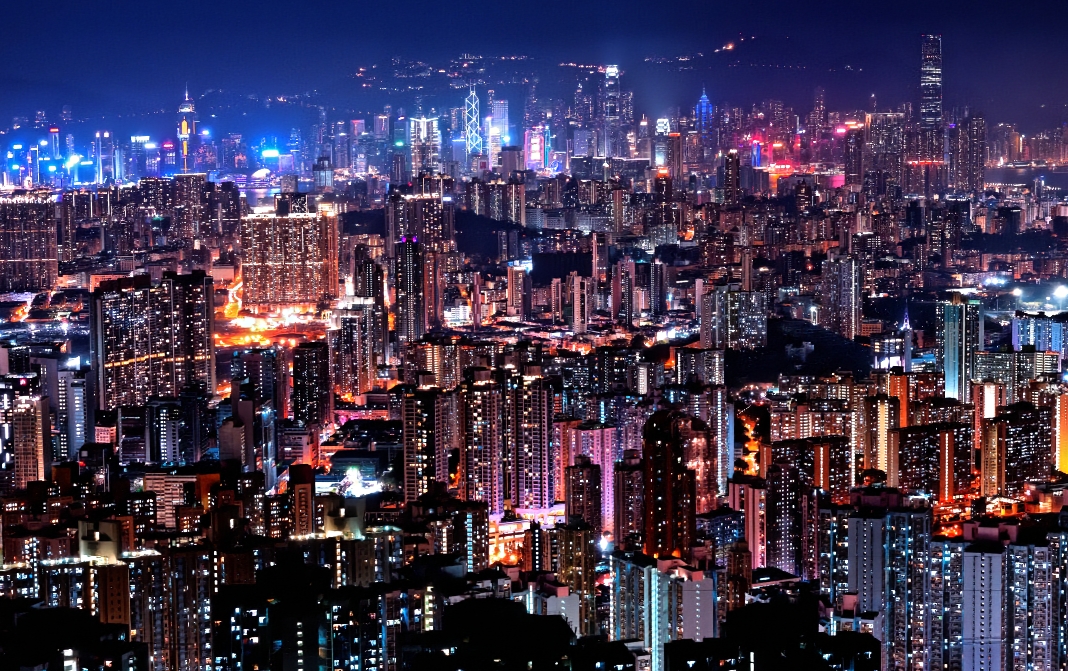}&
    \includegraphics[width=.125\linewidth]{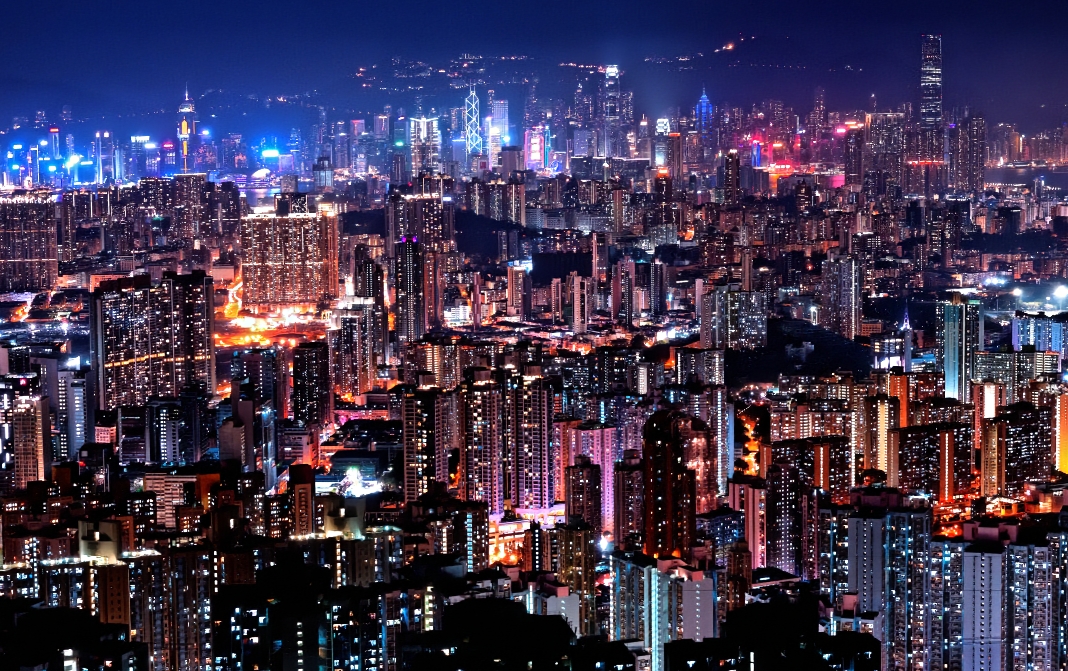}\\
    \small{PSNR=14.25dB} & \small{PSNR=25.39dB} & \small{PSNR=26.13dB} & \small{PSNR=29.52dB} & \small{PSNR=29.62dB} & \small{PSNR=29.62dB} & \small{PSNR=29.60dB} & \small{PSNR=29.60dB}\\
    &
    \includegraphics[width=.125\linewidth]{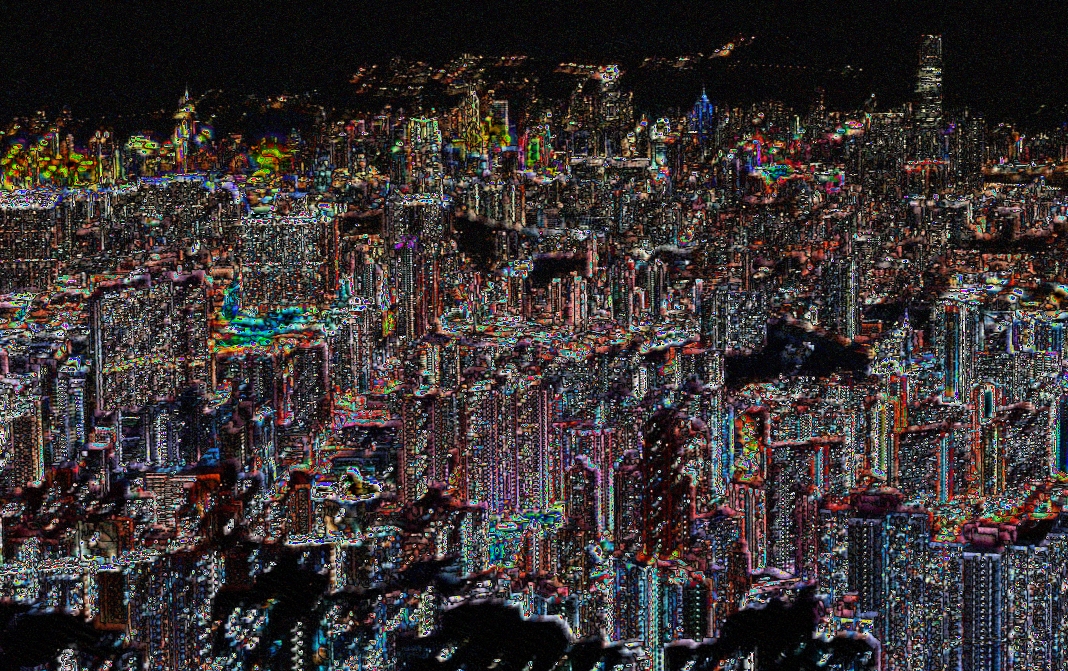}&
    \includegraphics[width=.125\linewidth]{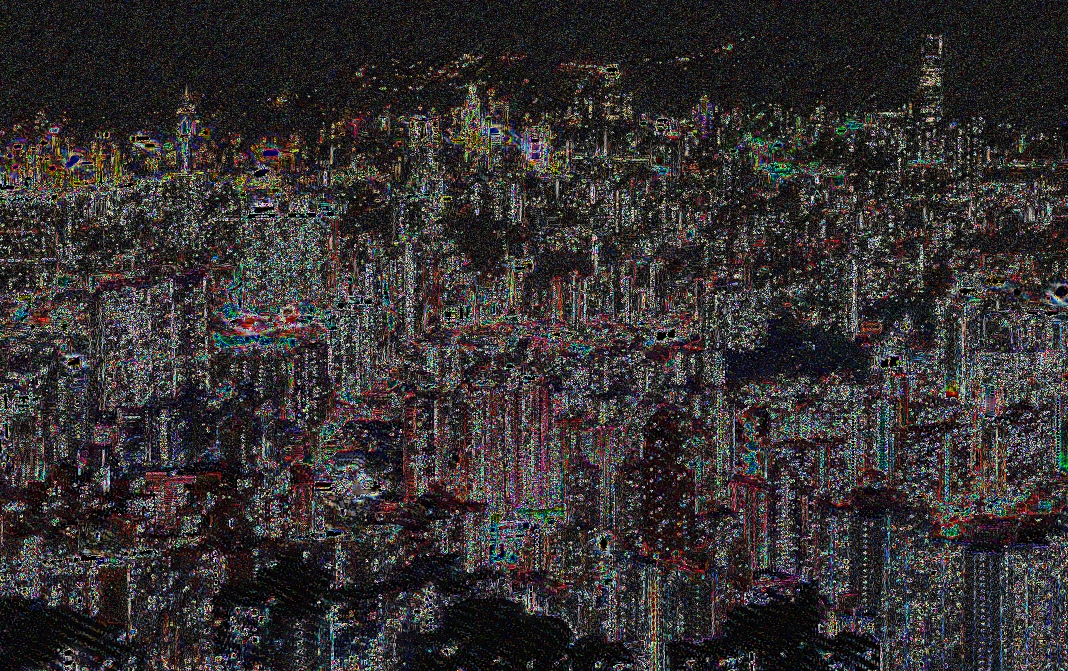}&
    \includegraphics[width=.125\linewidth]{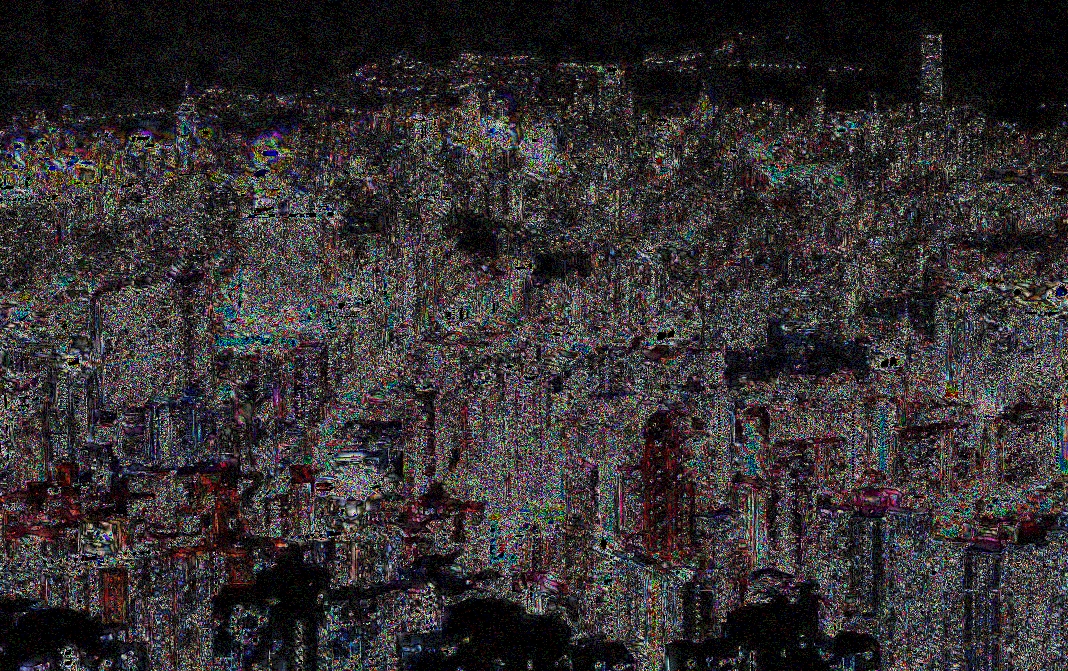}&
    \includegraphics[width=.125\linewidth]{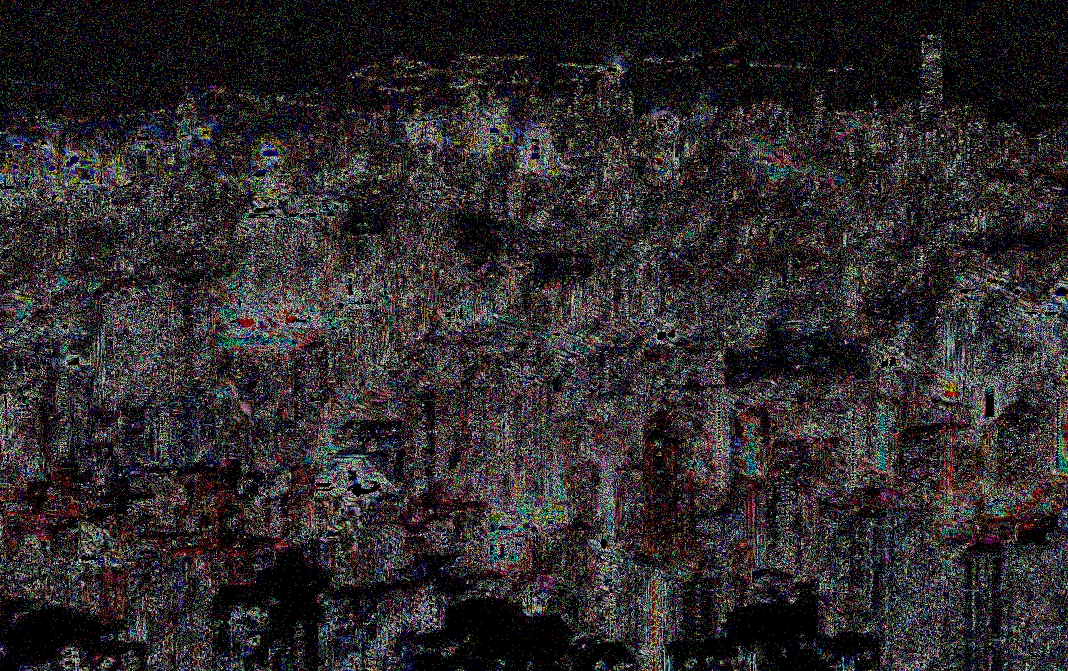}&
    \includegraphics[width=.125\linewidth]{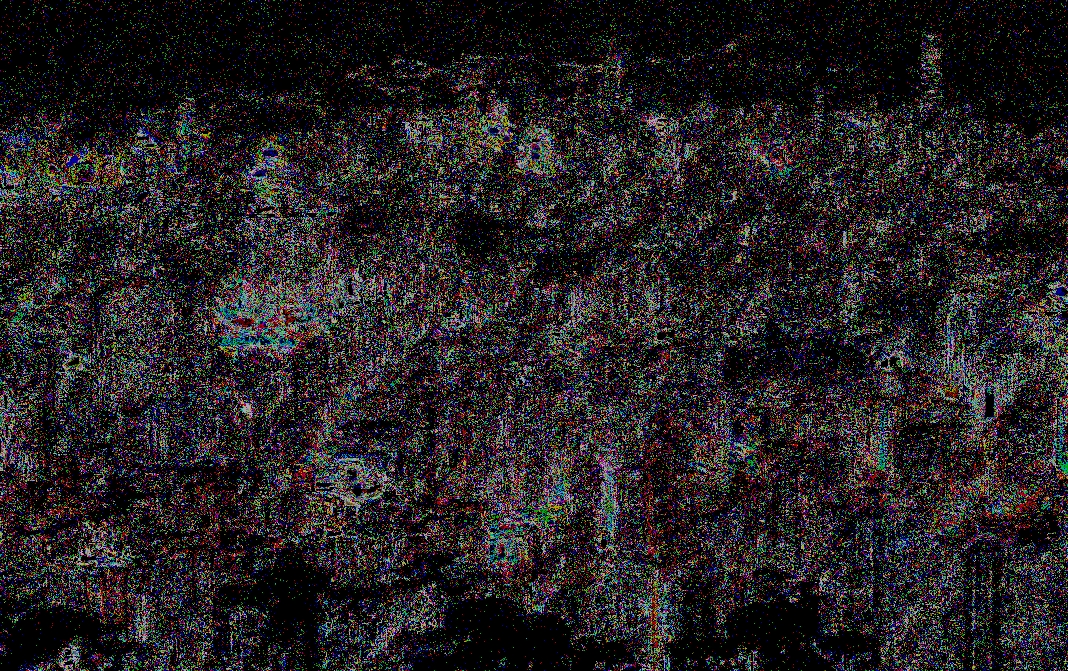}&
    \includegraphics[width=.125\linewidth]{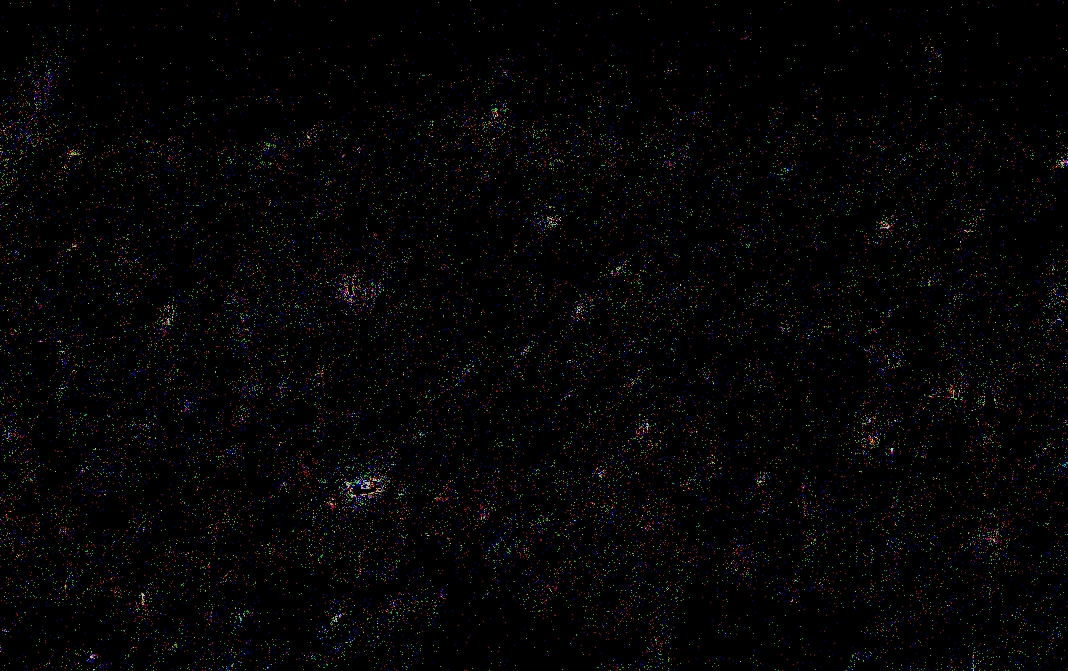}&
    \includegraphics[width=.125\linewidth]{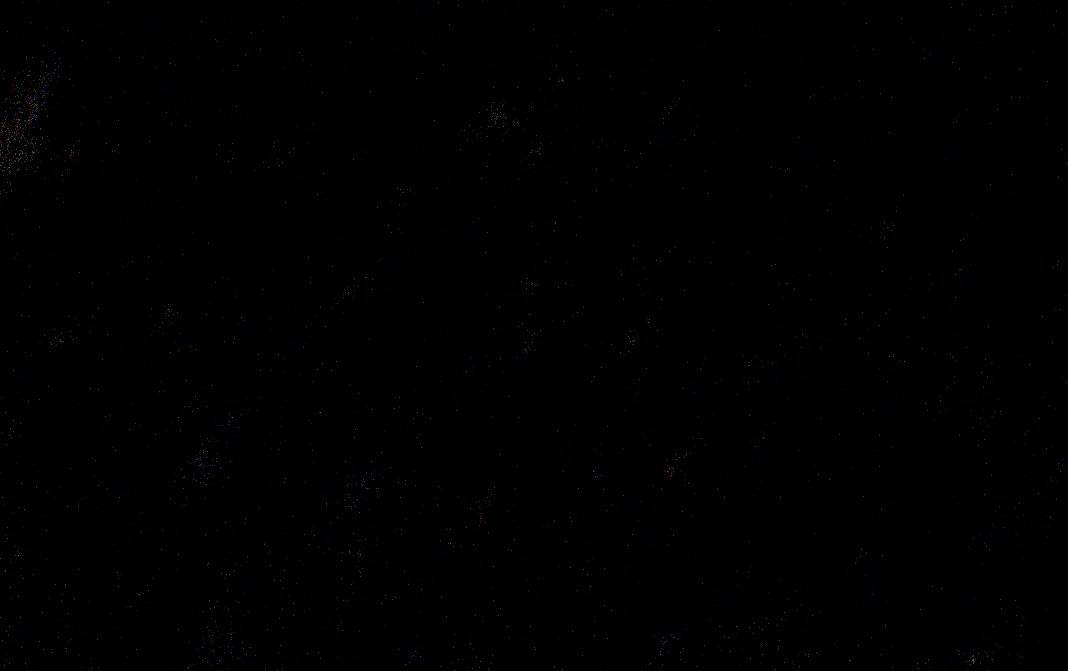}\\
    & \small{tol=$5.00\cdot 10^{-1}$} & \small{tol=$1.25\cdot 10^{-1}$} & \small{tol=$2.66\cdot 10^{-2}$} & \small{tol=$1.03\cdot 10^{-2}$} & \small{tol=$5.80\cdot 10^{-3}$} & \small{tol=$1.41\cdot 10^{-3}$} & \small{tol=$4.65\cdot 10^{-4}$}
\end{tabular}
   \caption{Visualization of convergence to a fixed point on an image from our saturated images dataset with 1\% noise. Top row refers to the per step deblurring results, while bottom row to the relative error, i.e. the difference between the current latent estimate and the one from the previous step. The corresponding PSNR values and relative errors are provided for each image. For visualization purposes the images in the even rows are normalized by their corresponding relative error.}
   \label{fig:PerStep}
   \vspace{-0.5cm}
\end{figure*}

To investigate, how the network's performance evolves with the amount of recurrent steps performed, we have run our model for 20 steps and evaluated the PSNR of the output from each step on the whole Sun \etal color dataset with noise level of 1\%. Moreover, we performed several runs where we used different settings for the linear solver. In particular, we investigated the achieved performance when the maximum allowed CG iterations vary between 25 to 500. The convergence behavior of all the different network configurations is depicted in \cref{fig:ConvergencePlots}.
From this figure it is clearly observed that in all the different cases the proposed RLSDN network manages to reach an equilibrium (the final output is close to a fixed point of the network). The exact amount of recurrent steps needed for convergence depends on the maximum allowed number of CG iterations. The higher the CG iterations the smaller the recurrent steps for which the network converges. It is also important to note that according to the reported results, by limiting the amount of CG iterations to 75 the performance of our network at the 5th step and beyond does not degrade comparing to the case where 500 CG iterations are used. This is an important empirical evidence which suggests that we can speed-up the inference without suffering any significant loss in reconstruction quality. Note that all the results we reported in the paper correspond to those obtained by using 5 recurrent steps with maximum 250 CG iterations per step.
\begin{figure}[!t]
\centering
\begin{tabular}{@{} c @{ } c @{ }}
    \includegraphics[width=.5\linewidth]{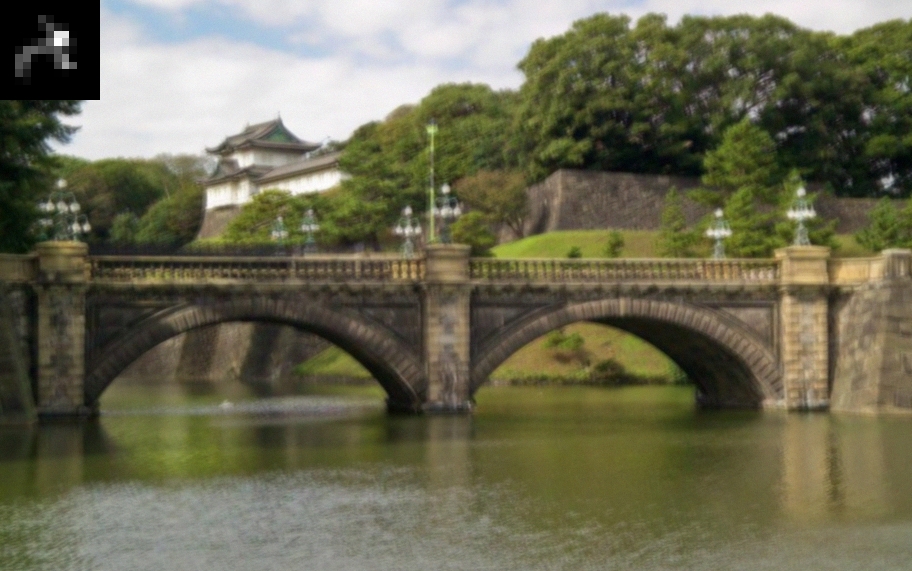}&
    \includegraphics[width=.5\linewidth]{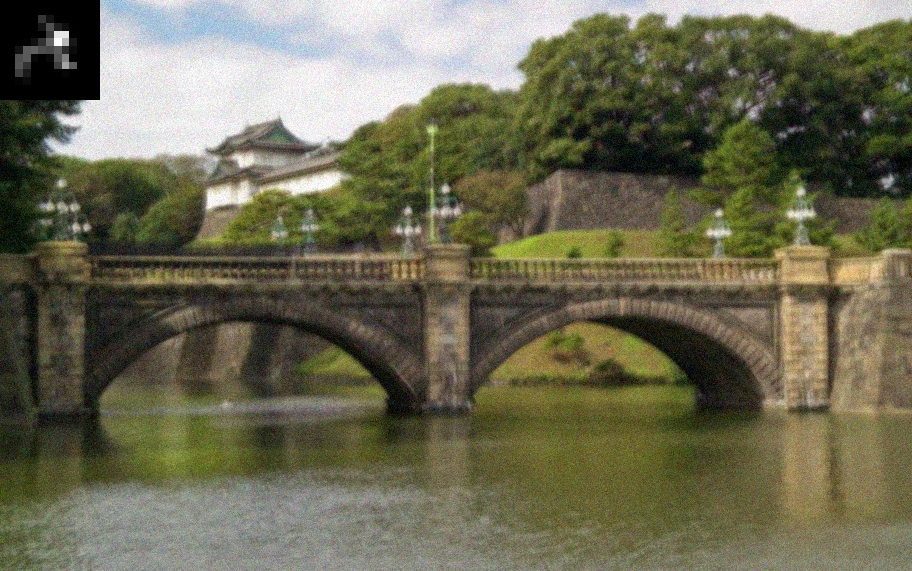}\\
    Input, 1\% noise & Input, 5\% noise\\
    \multicolumn{2}{l}{
    \includegraphics[width=1\linewidth]{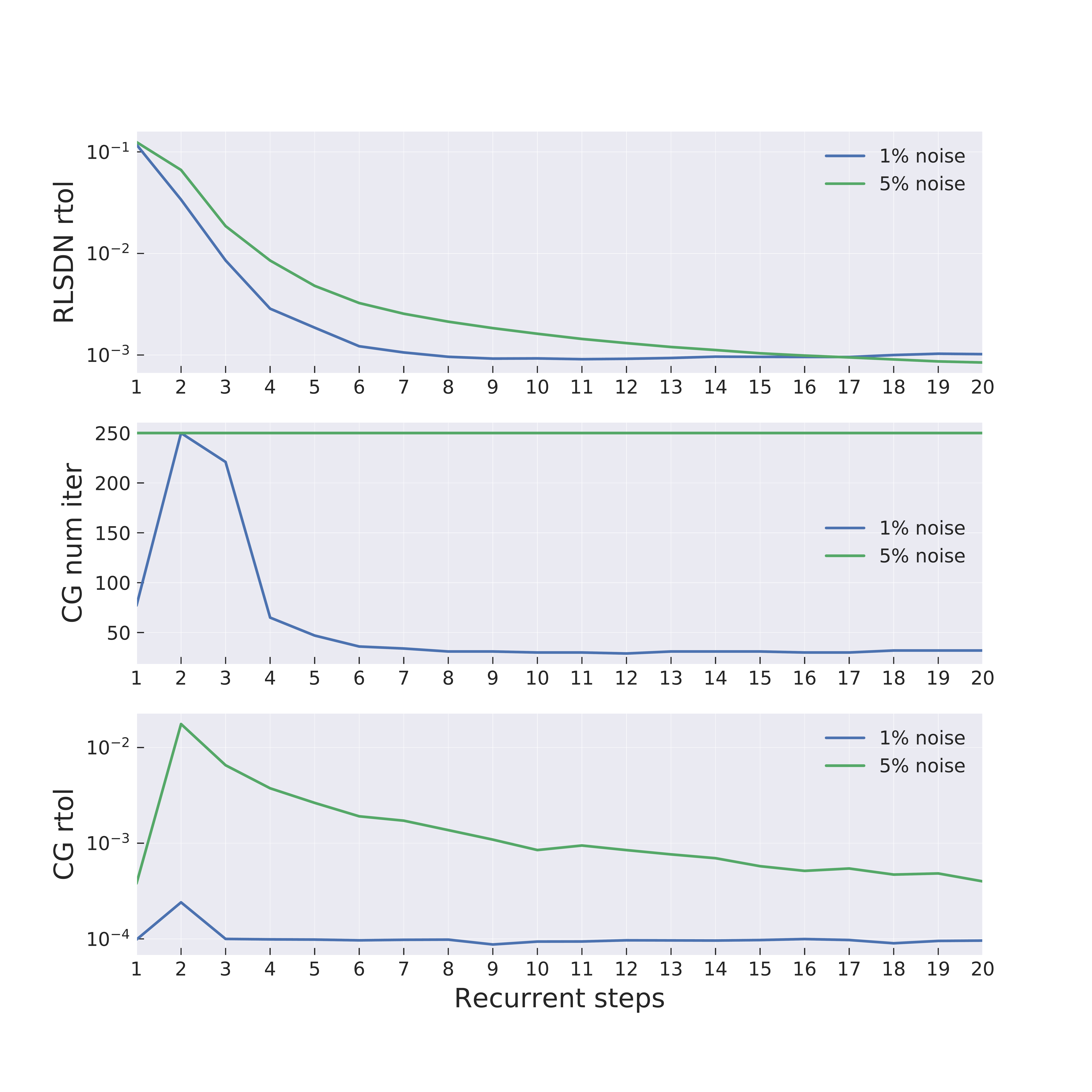}
    }
\end{tabular}
  \caption{Convergence plots of RLSDN for different levels of distortion.}
  \label{fig:EasyHard}
  \vspace{-0.5cm}
\end{figure}
We also provide a visual example in \cref{fig:PerStep} that further supports our claims regarding convergence to a fixed point. While we found the presented results valid for all our trained networks, in this figure we depict the specific example from our saturated images dataset and per step PSNR evolution of the RLSDN network trained for scenario of color saturated image deblurring with 1\% noise. We also provide the per step relative error, which is computed as $\textrm{tol} = ||x_{i} - x_{i-1}||_2 / ||x_i||_2$. From \cref{fig:PerStep} we notice the same behaviour as in \cref{fig:ConvergencePlots}, where the output of our network stabilizes at around the 4th step, and then approximately reaches an equilibrium. Finally, we observe that the relative error decreases monotonically up to the 20th step. 

\subsection{Adaptive complexity of RLSDN}
\vspace{-0.1cm}
A natural advantage of iterative algorithms is that they adapt their complexity according to the type and level of distortions of the input. Specifically, iterative methods tend to converge in less iterations when the degradation of the input is small, while they need more iterations when the degradation is excessive. To investigate whether this is also the case for the RLSDN network, we have used an image from the Sun \etal dataset, distorted it with 1\% noise and 5\% noise and studied the amount of CG iterations needed for convergence. We have used an early exit strategy when the relative error during CG iterations falls below $1\cdot10^{-4}$ and limit the maximum amount of iterations to 250. We report our findings regarding convergence in  \cref{fig:EasyHard}. From this figure it can be seen that for the case of 1\% noise RLSDN indeed converges faster, performing less amount of internal CG iterations than for the 5\% noise case. The faster convergence is clearly visible from the first plot of \cref{fig:EasyHard}, as the RLSDN relative error for the 5 \% case lies above the 1\% case and catches up only at the 17th step. The same can be seen from the bottom plot, which depicts the CG relative error. For the 5\% noise case the relative error is an order of magnitude larger than for the 1\% case. Indeed, as it can be seen from the central plot, for the case of 1\% noise CG performs an early exit at almost every RLSDN step, while for the 5\% noise case 250 iterations seems not to be enough to converge at any RLSDN step. From this study we can conclude that for the images presented in \cref{fig:EasyHard} in order for RLDN to converge to a relative tolerance of $1\cdot 10^{-3}$, it requires to perform 8 steps with 761 CG iterations in total for the 1\% noise case, while for the 5\% noise case it converges in 16 steps and 4000 CG iterations in total.

It is important to note, that the adaptive behaviour of RLSDN stems directly from our proposed architecture and our adopted training strategy which employs implicit back-propagation. In turn, this leads to a significant boost in restoration quality, which is impossible to achieve with traditional feed-forward convolution neural networks (CNN). Indeed, under the classical deep learning training strategy, when unrolled, our model consists of a sequence of CG iterations roughly represented by the following blocks: convolution layer, point-wise multiplication, transpose convolution layer, skip-connection. As discussed in the previous paragraph, in order to converge for a hard case scenario, our network requires 4000 CG iterations, which corresponds to 8000 convolution layers in total. Compared to a large ResNet-150 architecture\cite{He2016} with 150 convolution layers in total, where also a block of two convolution layers is followed by a skip connection, our approach represents 50 times deeper neural network for a hard case scenario, and 5 times deeper network for an easy case. Clearly, due to memory constraints, it is extremely difficult to train a ResNet-8000, while our network can be learned in an end-to-end fashion without any problems. 

\section{Real Color Image Deblurring Comparisons}
\begin{figure*}[!t]
\centering
\begin{tabular}{@{} c @{ } c @{ } c @{ } c @{ } c @{ } c @{ } c @{ }}
    \includegraphics[width=.142\linewidth]{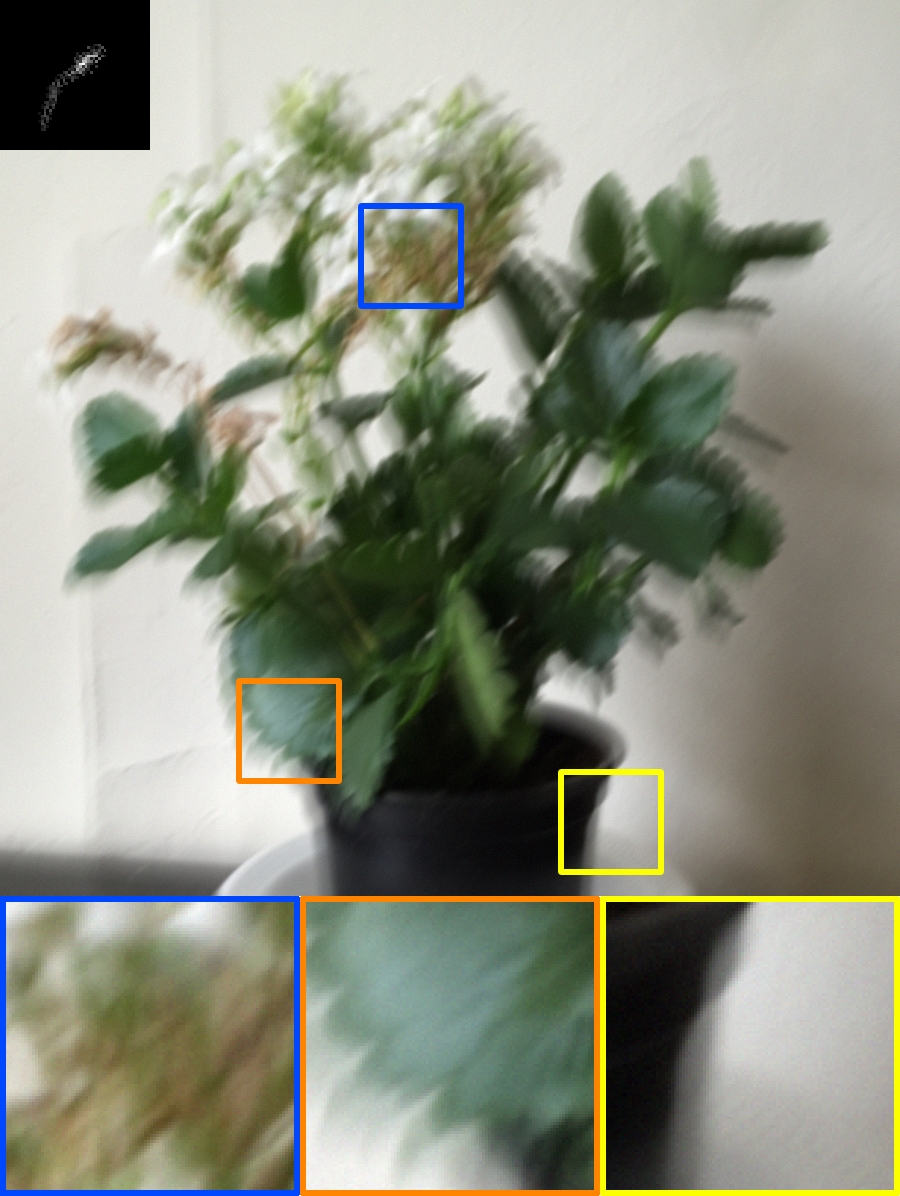}&
    \includegraphics[width=.142\linewidth]{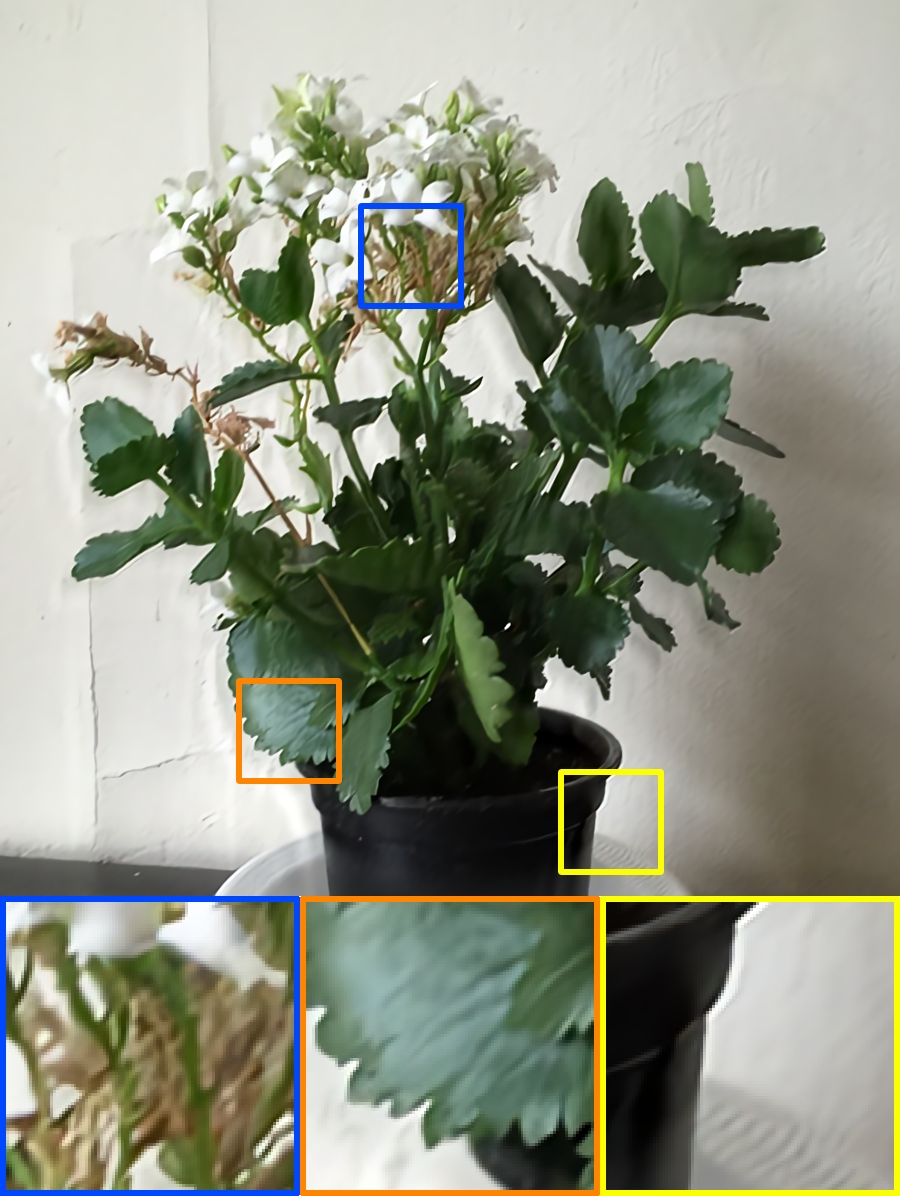}&
    \includegraphics[width=.142\linewidth]{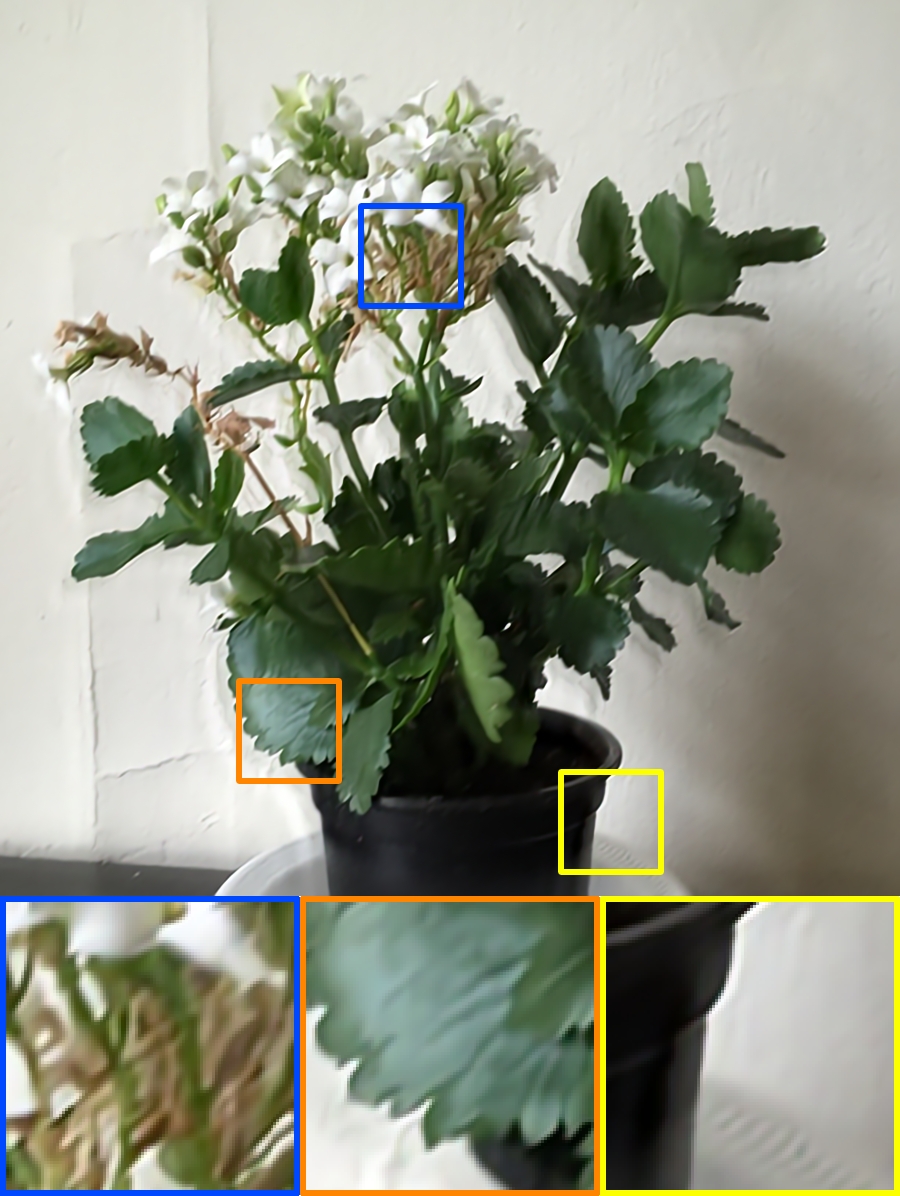}&
    \includegraphics[width=.142\linewidth]{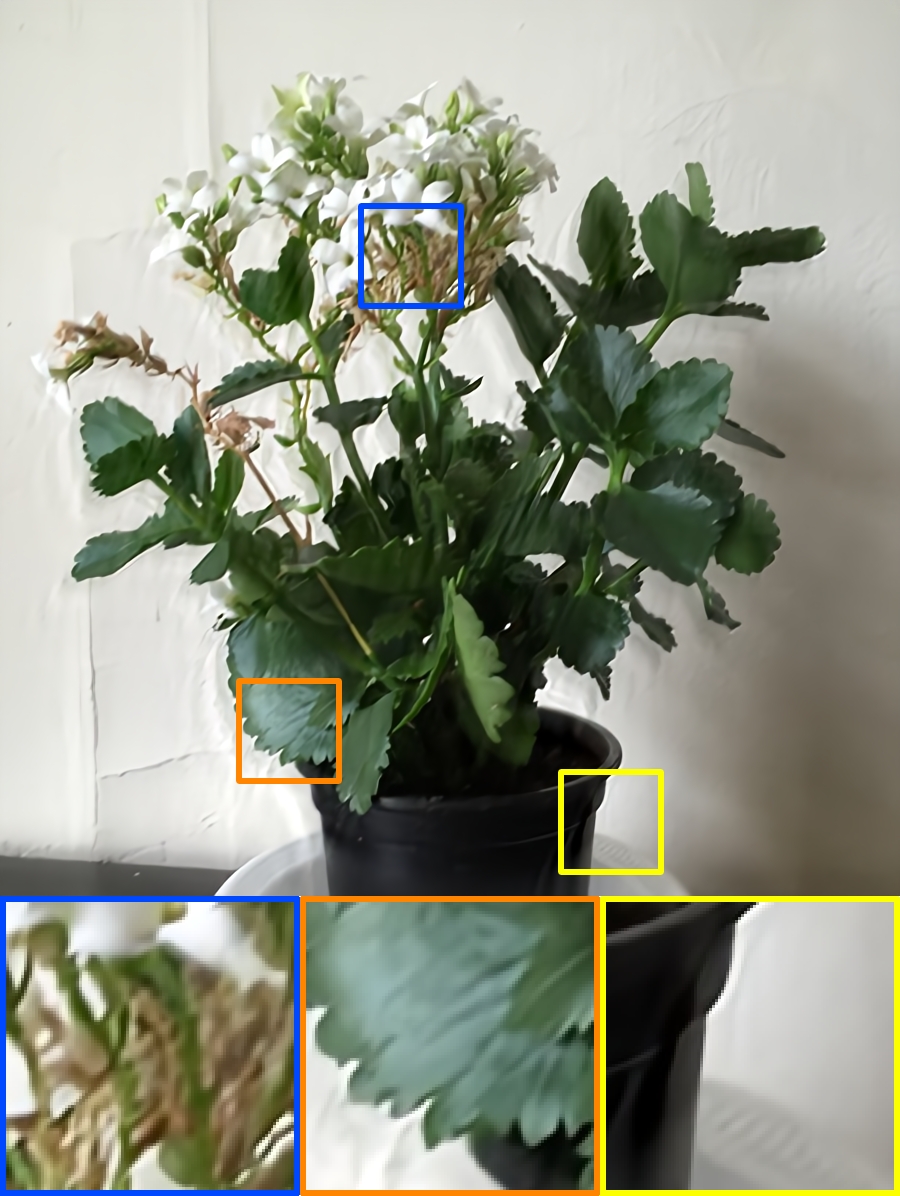}&
    \includegraphics[width=.142\linewidth]{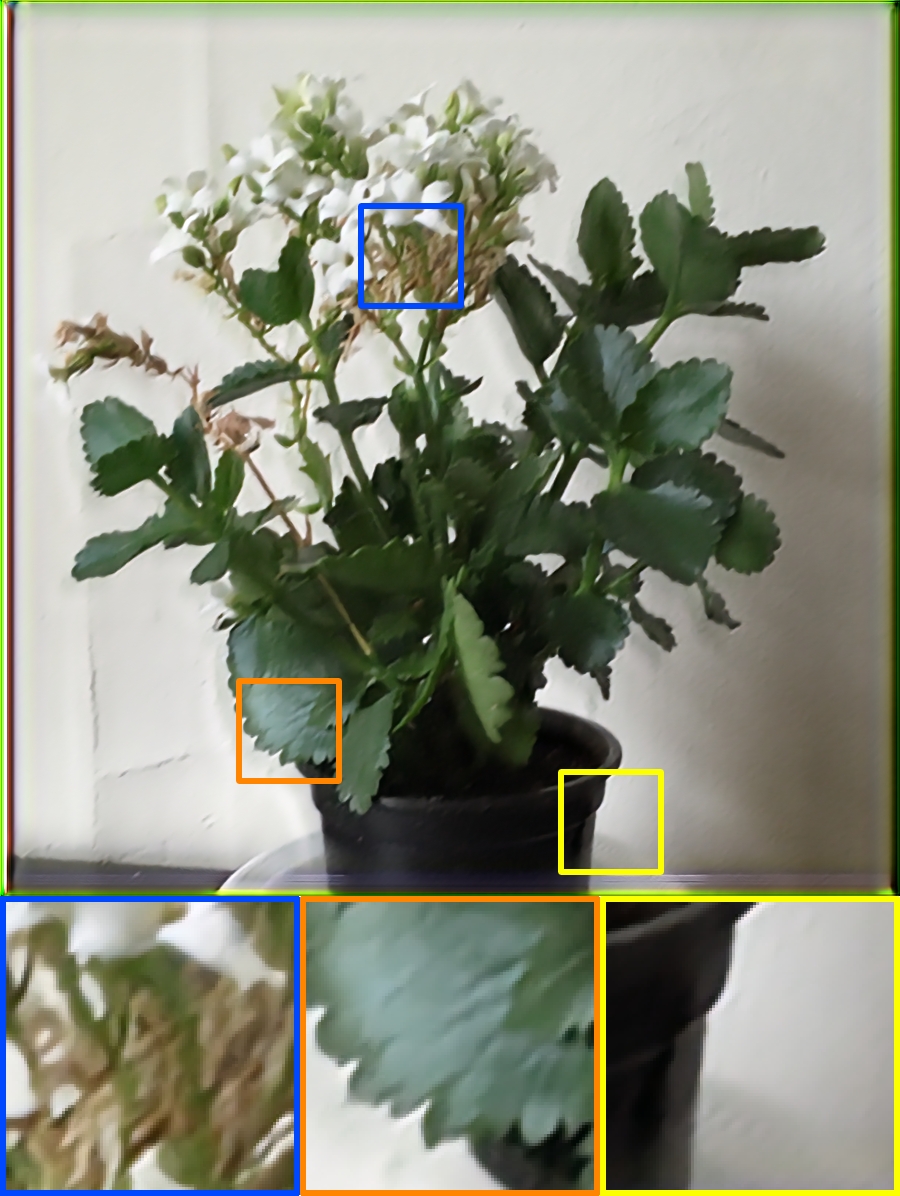}&
    \includegraphics[width=.142\linewidth]{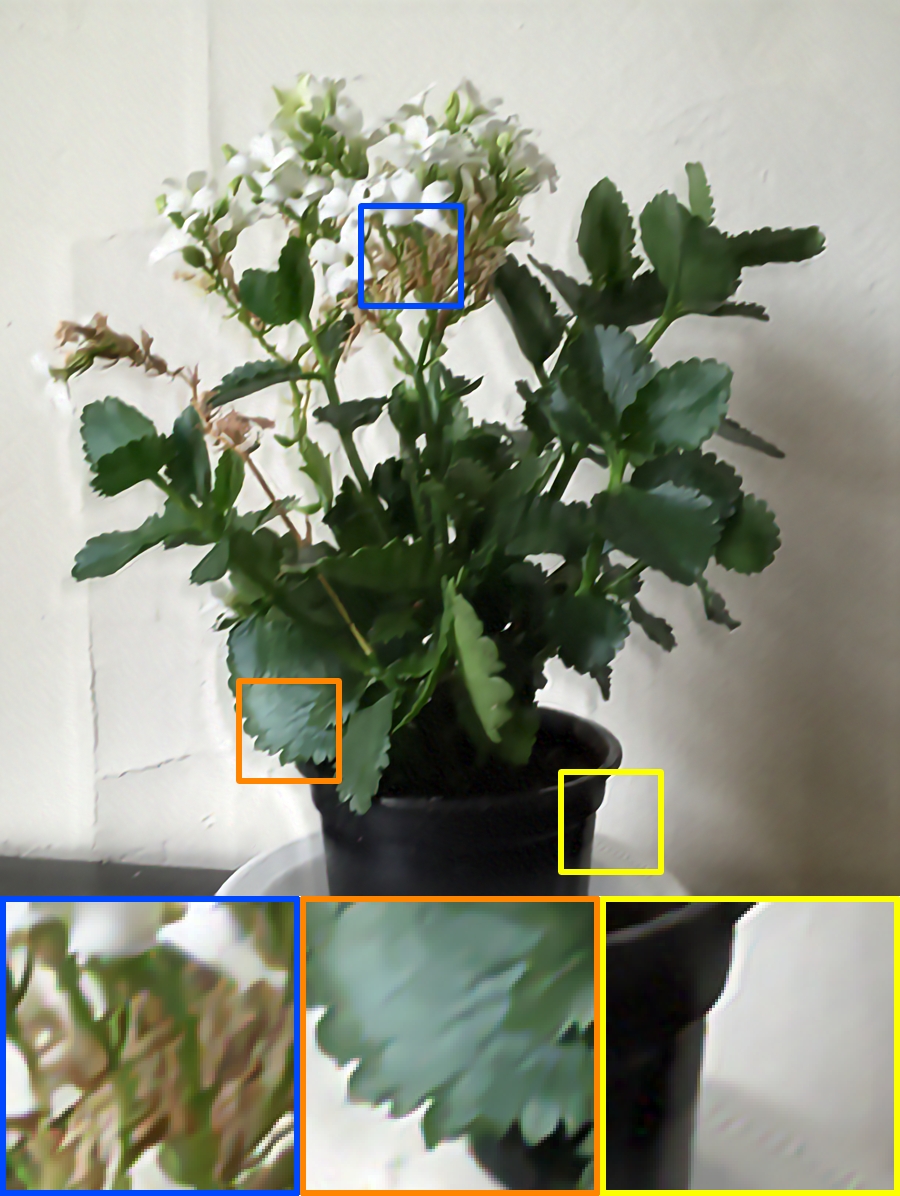}&
    \includegraphics[width=.142\linewidth]{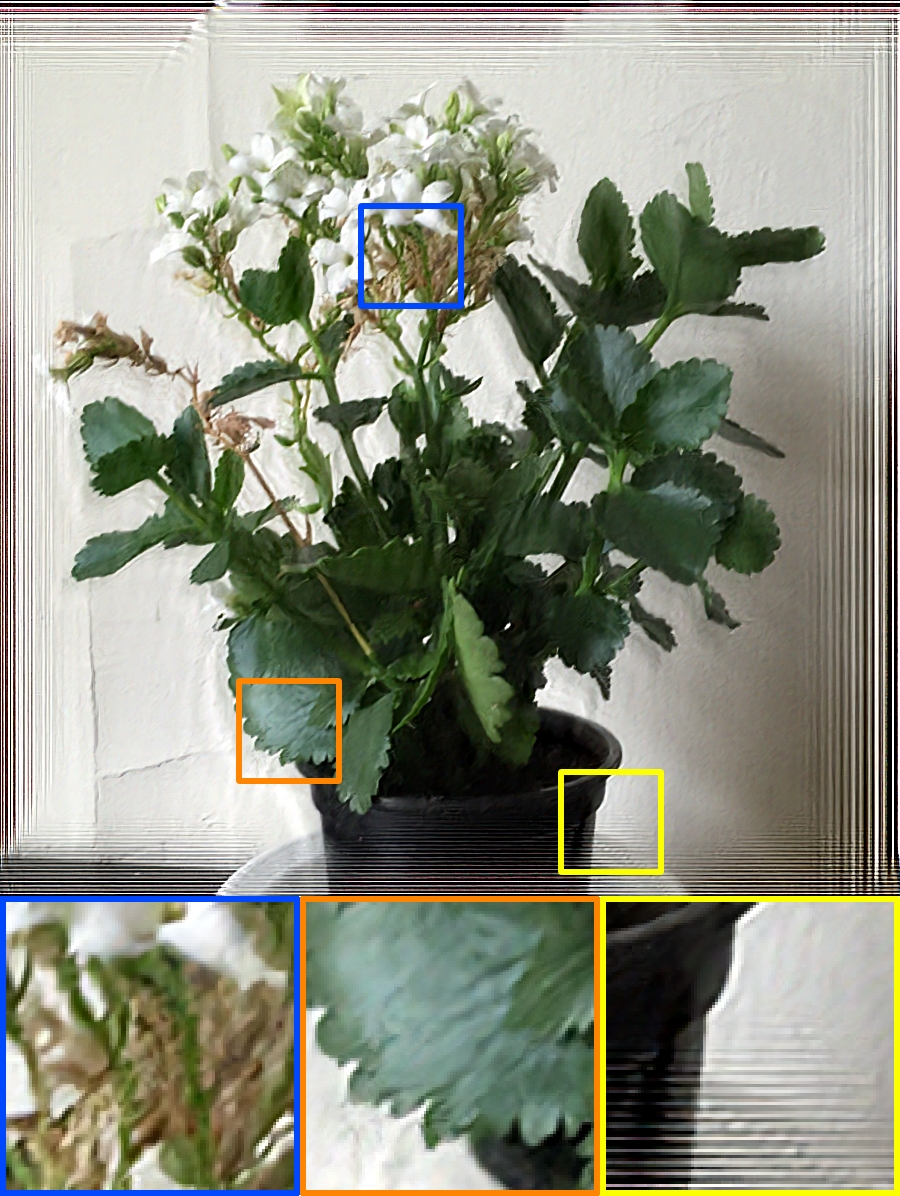}\\
    
    \includegraphics[width=.142\linewidth]{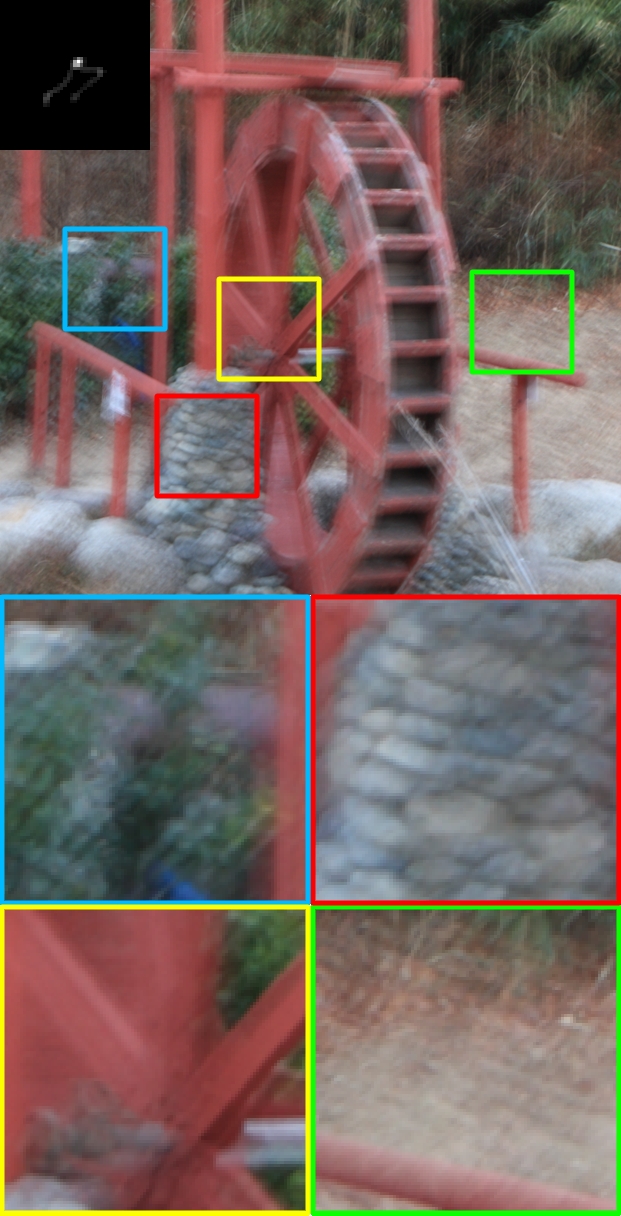}&
    \includegraphics[width=.142\linewidth]{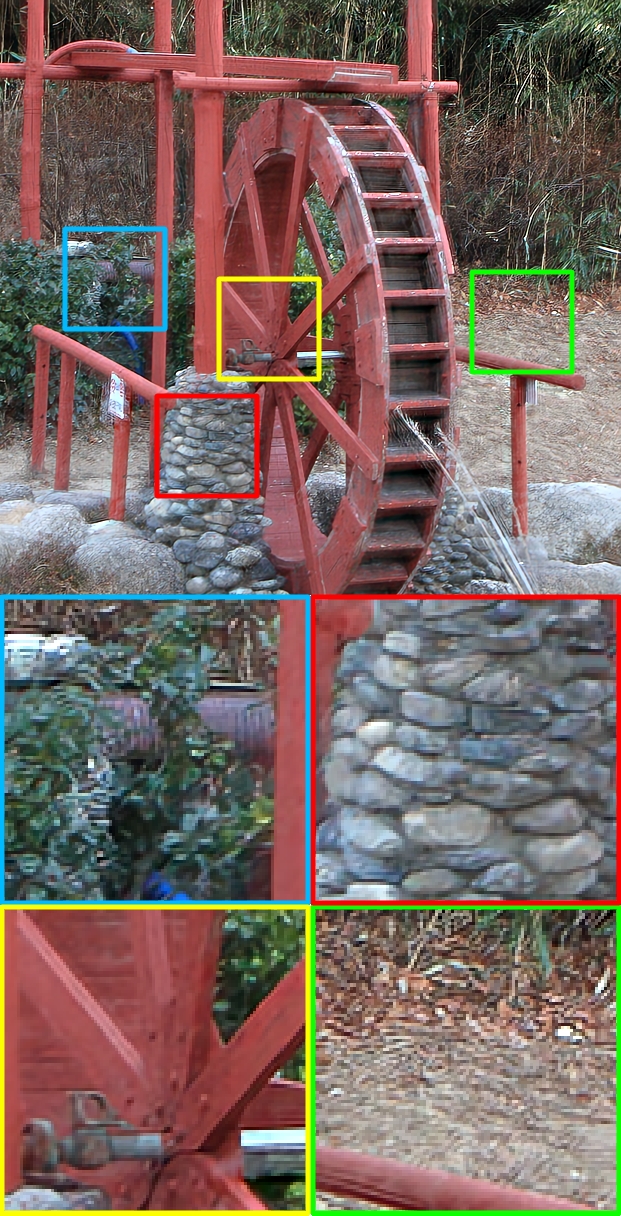}&
    \includegraphics[width=.142\linewidth]{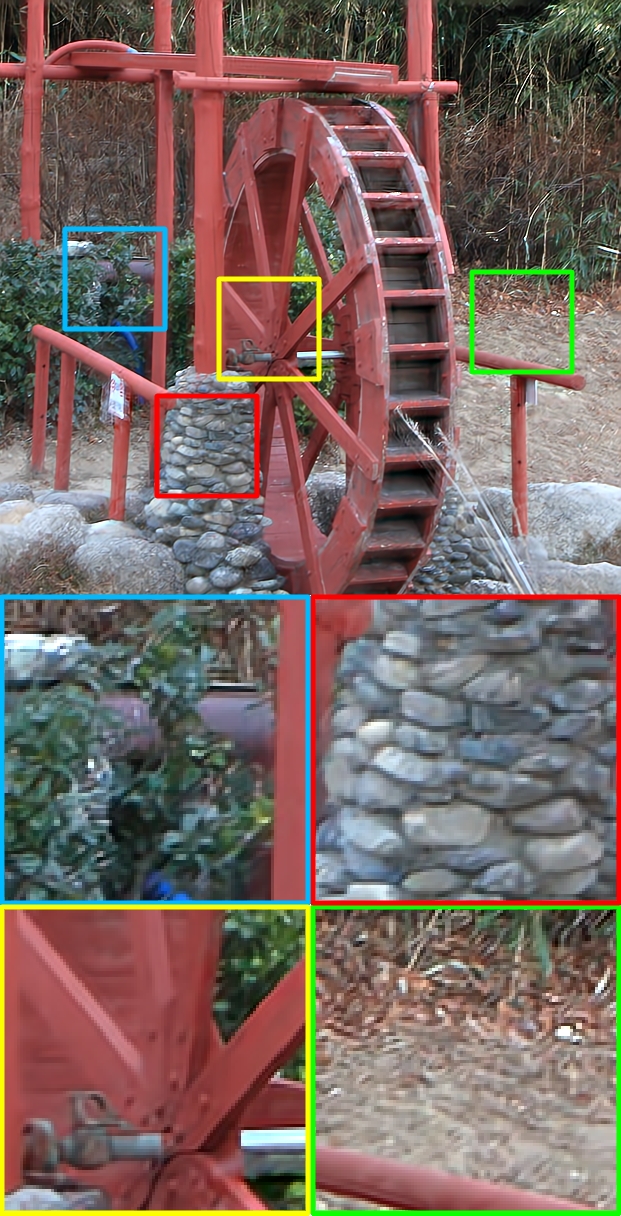}&
    \includegraphics[width=.142\linewidth]{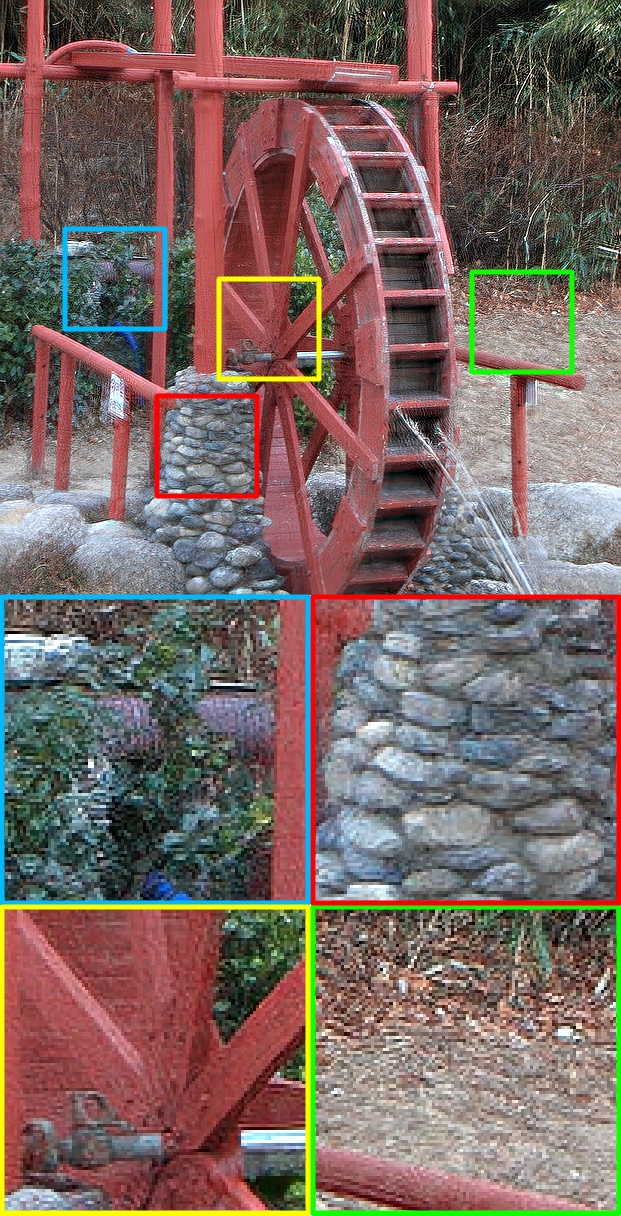}&
    \includegraphics[width=.142\linewidth]{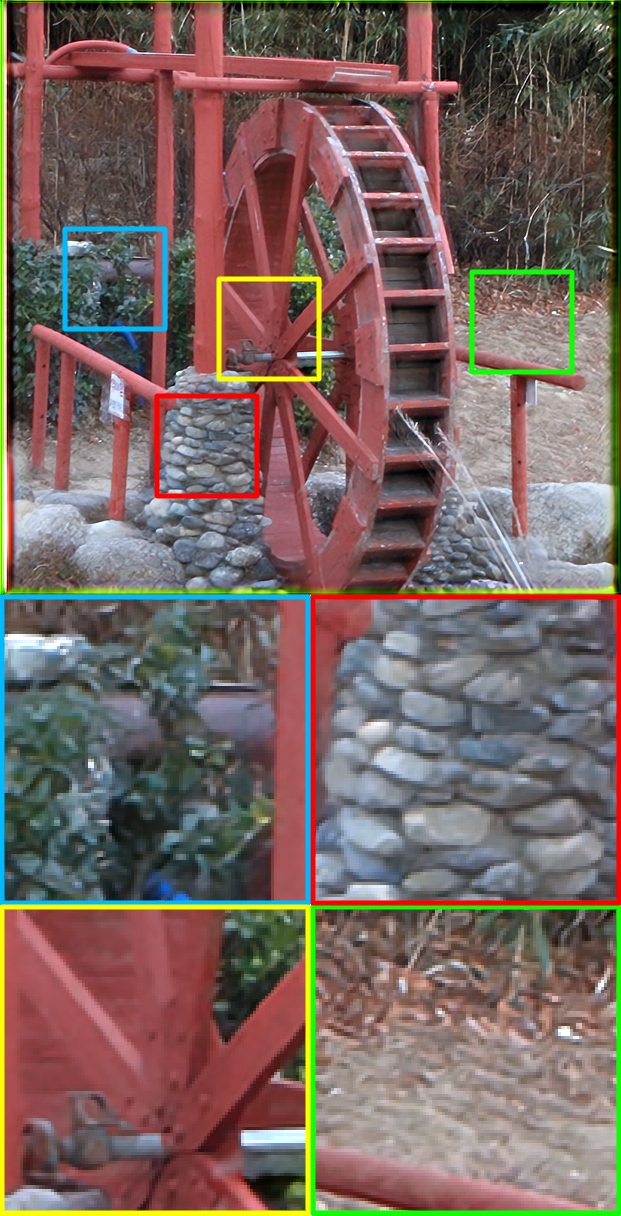}&
    \includegraphics[width=.142\linewidth]{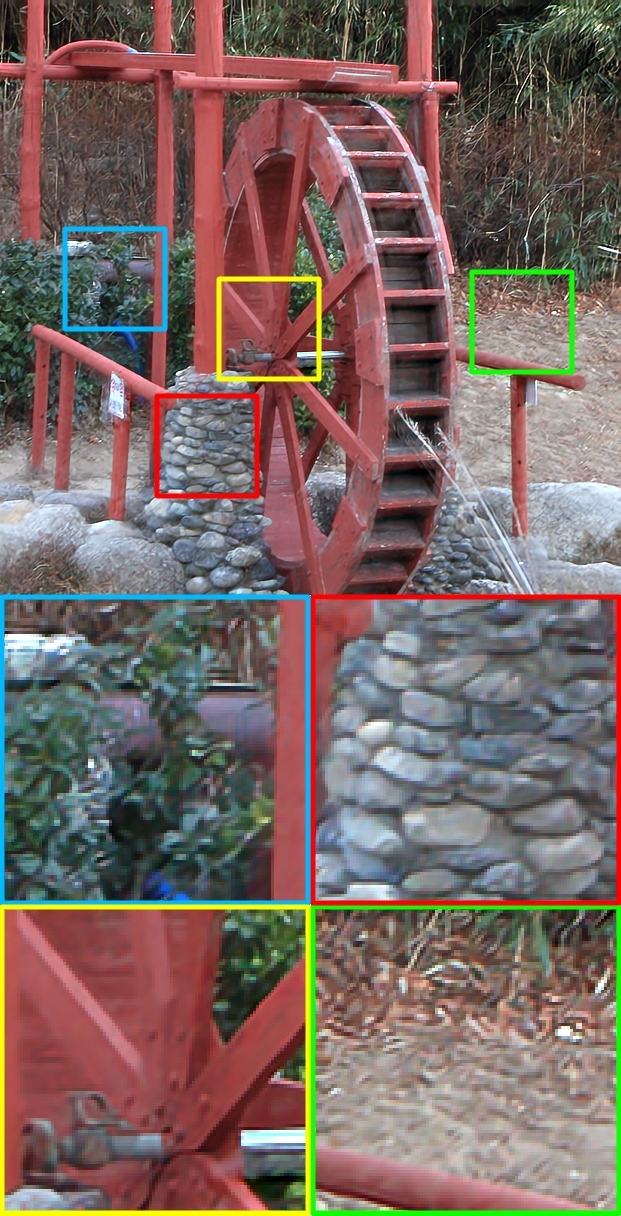}&
    \includegraphics[width=.142\linewidth]{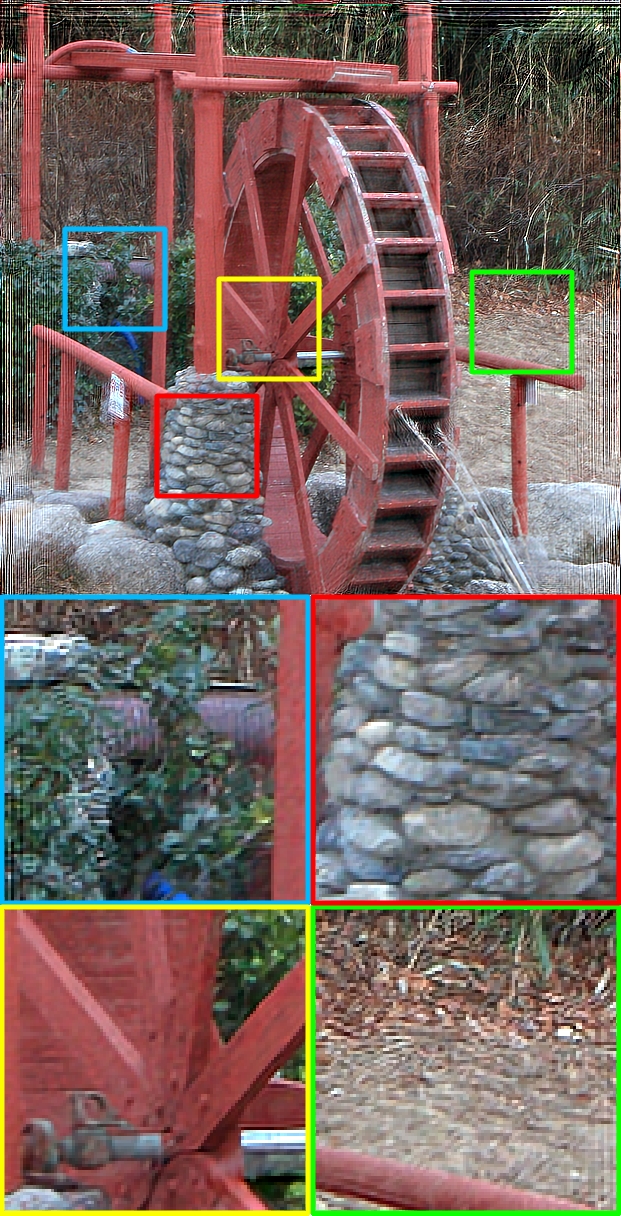}\\
    
    \includegraphics[width=.142\linewidth]{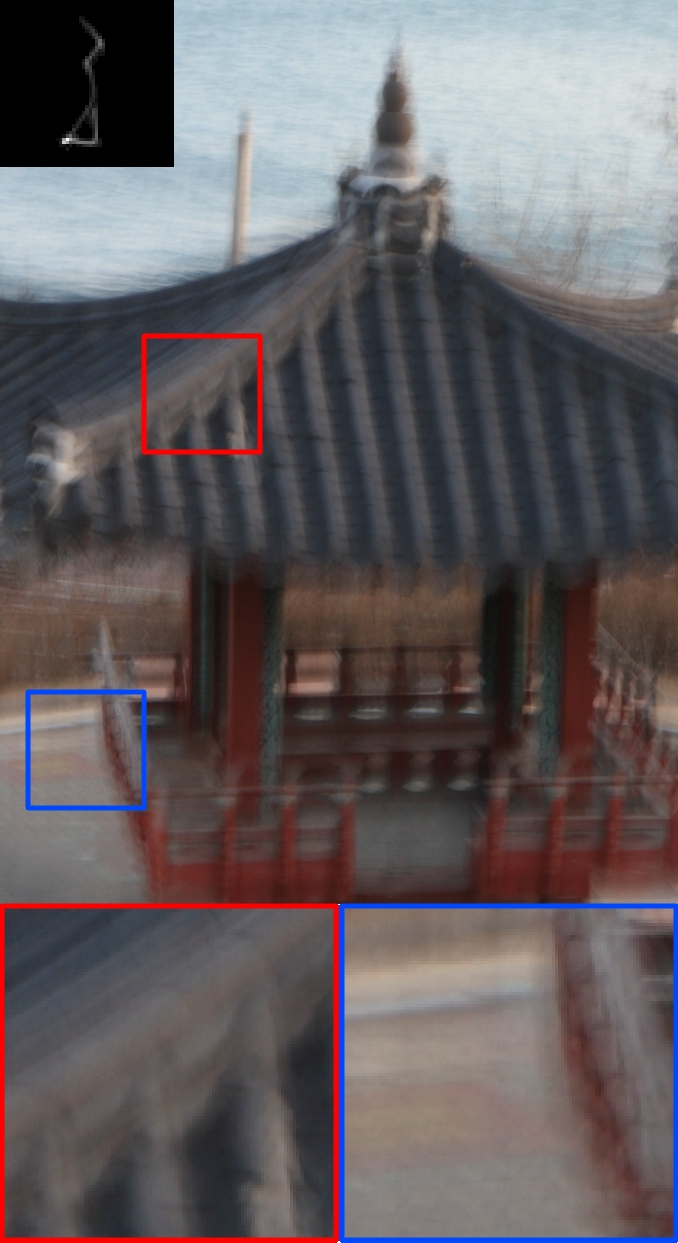}&
    \includegraphics[width=.142\linewidth]{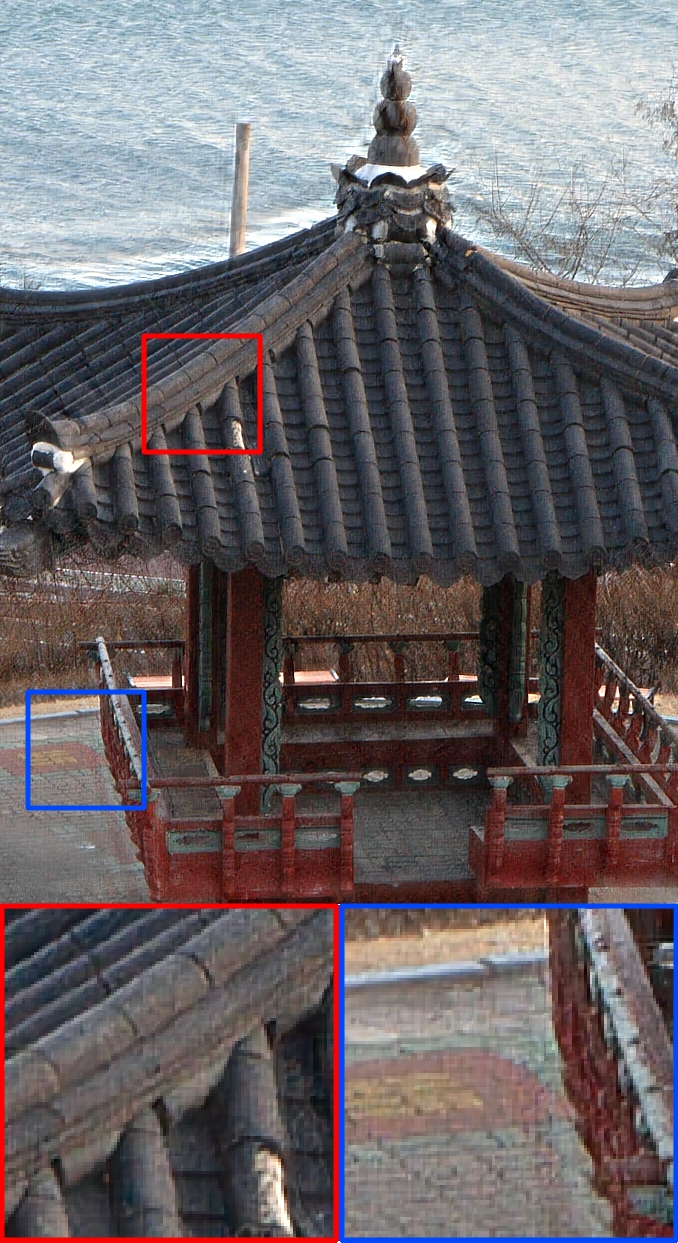}&
    \includegraphics[width=.142\linewidth]{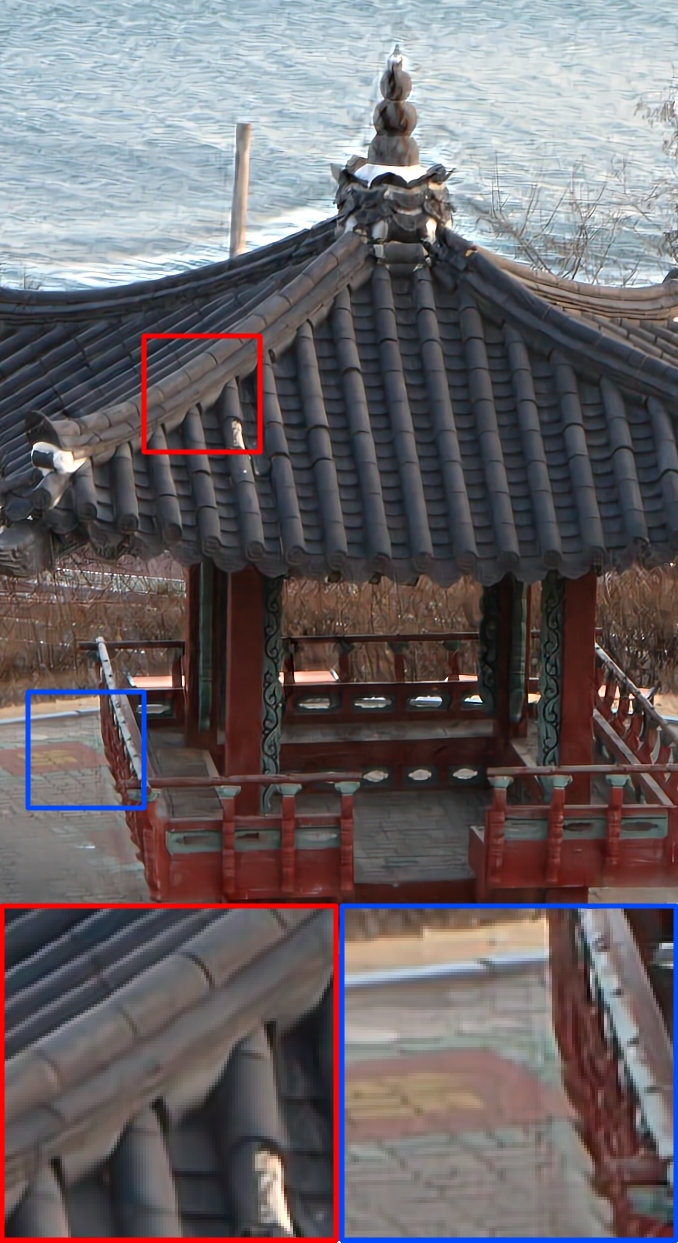}&
    \includegraphics[width=.142\linewidth]{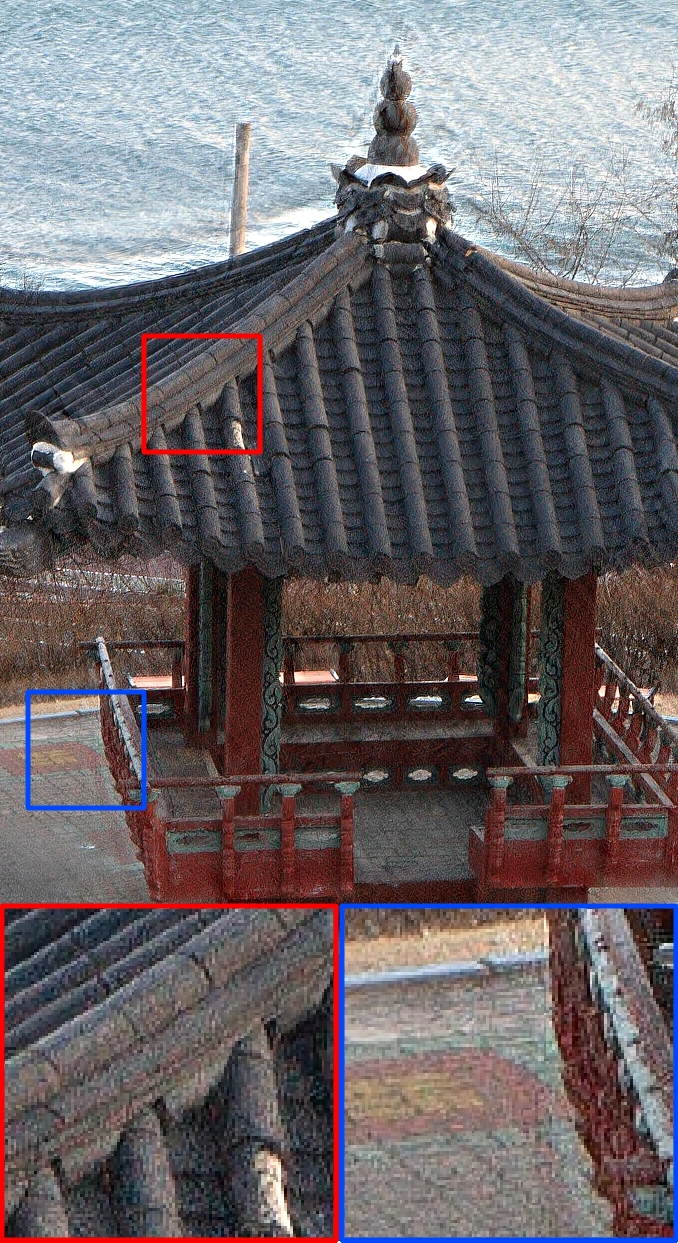}&
    \includegraphics[width=.142\linewidth]{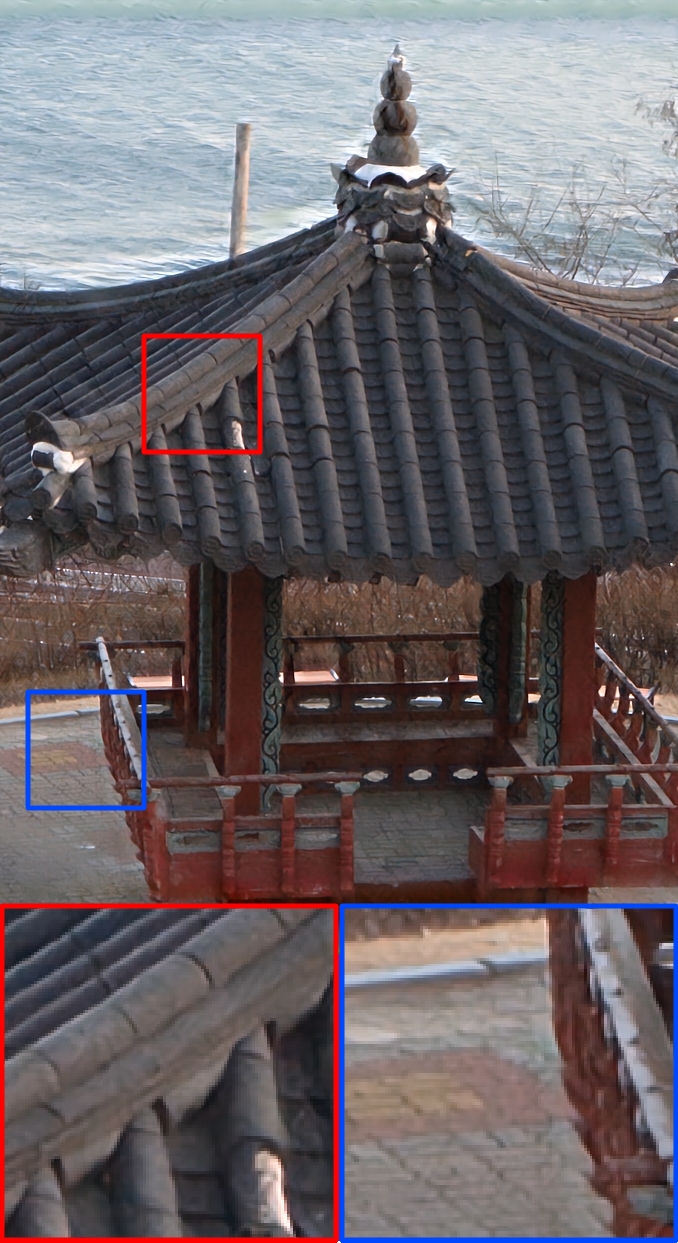}&
    \includegraphics[width=.142\linewidth]{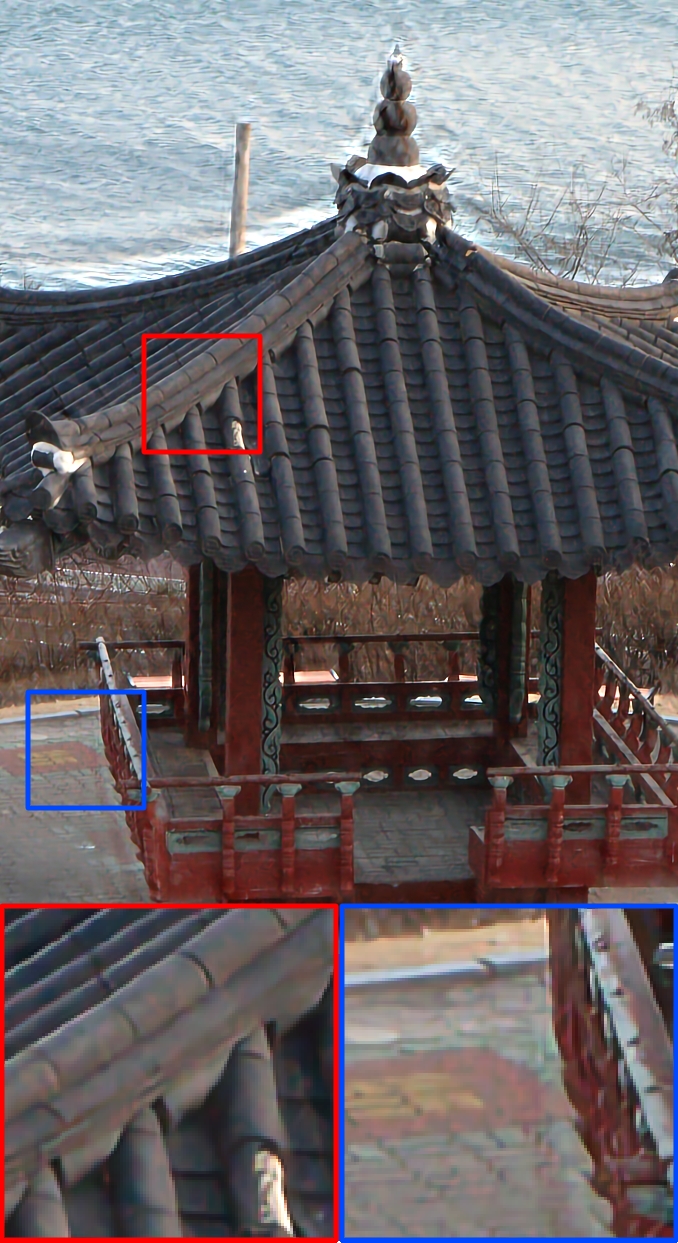}&
    \includegraphics[width=.142\linewidth]{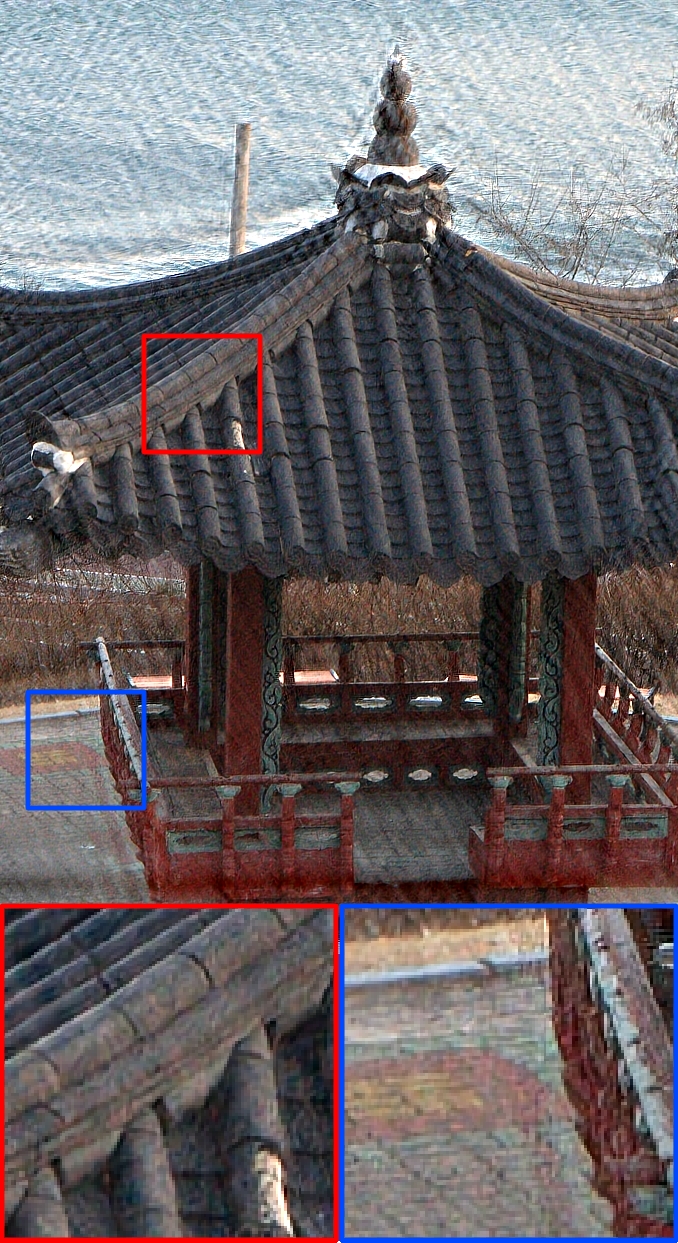}\\
    \small{Input} & \small{RLSDN (ours)} & \small{SVMAP \cite{Dong2021}} & \small{DWDN \cite{Dong2021b}} & \small{RGDN \cite{Gong2020}} & \small{FDN \cite{Kruse2017}} & \small{IRCNN \cite{Zhang2017b}}\\
\end{tabular}
   \caption{Visual comparison of RLSDN with state-of-the art methods on real color blurred images with blur kernels estimated by a third-party estimation method. First row: image and kernel from \cite{Pan2016}, middle row: image and kernel from \cite{Pan2016b}, bottom row: image from \cite{Lai2016}, kernel estimated by \cite{Pan2016}.}
   \label{fig:RealBlurColorComp}
\end{figure*}

\vspace{-0.1cm}
In this section we present more visual comparisons of our proposed models with current state-of-the-art deblurring methods for real examples. We present the most interesting practical case of real image deblurring, when the blur kernel is estimated by a third-party method. To evaluate the performance of RLSDN we use a collection of three images and kernels provided by different methods and we present the results in \cref{fig:RealBlurColorComp}. For a fair comparison, we do not tune the noise standard deviation for all methods requiring it as an additional input (including ours), but instead use the value predicted by the WMAD estimation method~\cite{Donoho1994}. As it can be seen from the presented examples, compared to the rest of the methods our network restores finer details in a more accurate manner without amplifying the noise or exhibiting an over-smoothing effect. 

\section{Impact of the blur kernel}
\begin{figure*}[!t]
\centering
\begin{tabular}{@{} c @{ } c @{ } c @{ }}
    \includegraphics[width=0.3346\linewidth]{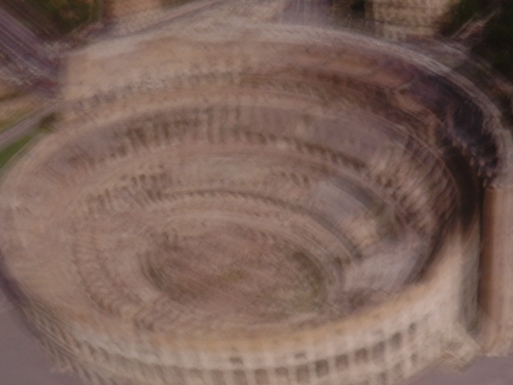}&
    \includegraphics[width=0.3346\linewidth]{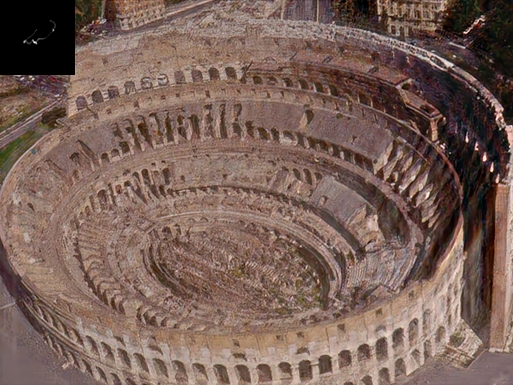}& 
    \includegraphics[width=0.3346\linewidth]{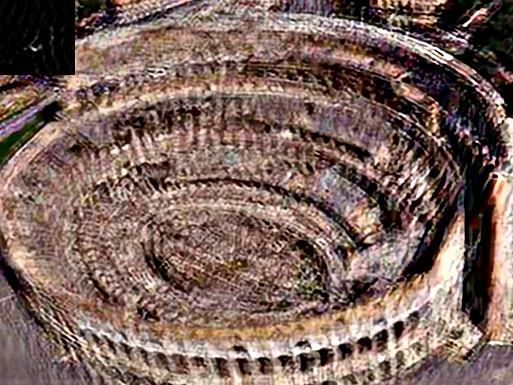}\\
    \small{(a) Input} & \small{(b) RLSDN (ours), kernel from \cite{Wen2021}} & \small{(c) RLSDN (ours), kernel from \cite{Perrone2014}}\\
    
    
\end{tabular}
   \caption{Performance of our method on a real image (``Roma" from \cite{Lai2016}) suffering from extensive blur. Two existing kernel estimation methods have been used in order to estimate the blur kernel. From the result depicted in (b) it is clear, that our network is capable of producing deblurring results of high quality when an accurate enough blur kernel is provided. However, when the kernel estimation fails, then the quality of the reconstruction result can be very poor as shown in (c).}
   \label{fig:bad_cases}
\end{figure*}

\vspace{-0.1cm}
All the above results suggest that our proposed network leads to very competitive performance for different scenarios. However, a limiting factor to our approach, similarly to the rest of non-blind deconvolution methods, is that our network relies on a third-party blur kernel estimation method. Consequently, when the blur kernel is poorly estimated this has immediate effect in the reconstruction result. This is depicted in \cref{fig:bad_cases}, where we observe that when the estimated kernel is relatively accurate our network is able to produce a high quality image reconstruction, while when the used blur kernel is inaccurate the quality of the deblurred image can drop significantly.

\newpage
{\small
\bibliographystyle{ieee_fullname}
\bibliography{main}

\begin{thebibliography}{10}\itemsep=-1pt

\bibitem{Agustsson2017}
Eirikur Agustsson and Radu Timofte.
\newblock Ntire 2017 challenge on single image super-resolution: Dataset and
  study.
\newblock In {\em The IEEE Conference on Computer Vision and Pattern
  Recognition (CVPR) Workshops}, July 2017.

\bibitem{Babacan2012}
D. Babacan, M.~N. Molina, R.~Do, and A.~K. Katsaggelos.
\newblock Bayesian blind deconvolution with general sparse image priors.
\newblock In {\em European Conference on Computer Vision}, pages 341--355,
  2012.

\bibitem{Bai2019}
Shaojie Bai, J.~Zico Kolter, and Vladlen Koltun.
\newblock Deep equilibrium models.
\newblock In {\em Advances in Neural Information Processing Systems},
  volume~32, 2019.

\bibitem{Beck2009}
A. Beck and M. Teboulle.
\newblock A fast iterative shrinkage-thresholding algorithm for linear inverse
  problems.
\newblock {\em SIAM J. Imaging Sci.}, 2:183--202, 2009.

\bibitem{Bertero1998}
Mario Bertero and Patrizia Boccacci.
\newblock {\em Introduction to Inverse Problems in Imaging}.
\newblock IOP Publishing, 1998.

\bibitem{Boracchi2012}
Giacomo Boracchi and Alessandro Foi.
\newblock Modeling the performance of image restoration from motion blur.
\newblock {\em IEEE Transactions on Image Processing}, 21(8):3502--3517, 2012.

\bibitem{Boyd2011}
S. Boyd, N. Parikh, E. Chu, B. Peleato, and J. Eckstein.
\newblock {\em Distributed Optimization and Statistical Learning via the
  Alternating Direction Method of Multipliers}.
\newblock Now Publishers, 2011.

\bibitem{Candes2008}
Emmanuel~J. Candes and Michael~B. Wakin.
\newblock An introduction to compressive sampling.
\newblock {\em IEEE Signal Processing Magazine}, 25(2):21--30, 2008.

\bibitem{Cho2009}
Sunghyun Cho and Seungyong Lee.
\newblock Fast motion deblurring.
\newblock In {\em ACM SIGGRAPH Asia 2009 papers}, pages 1--8. 2009.

\bibitem{Cho2011}
Sunghyun Cho, Jue Wang, and Seungyong Lee.
\newblock Handling outliers in non-blind image deconvolution.
\newblock In {\em 2011 International Conference on Computer Vision}, pages
  495--502. IEEE, 2011.

\bibitem{Daubechies2010}
Ingrid Daubechies, Ronald DeVore, Massimo Fornasier, and C~Sinan
  G{\"u}nt{\"u}rk.
\newblock Iteratively reweighted least squares minimization for sparse
  recovery.
\newblock {\em Communications on Pure and Applied Mathematics}, 63(1):1--38,
  2010.

\bibitem{Dong2017}
Jiangxin Dong, Jinshan Pan, Zhixun Su, and Ming-Hsuan Yang.
\newblock Blind image deblurring with outlier handling.
\newblock In {\em Proceedings of the IEEE International Conference on Computer
  Vision}, pages 2478--2486, 2017.

\bibitem{Dong2021b}
Jiangxin Dong, Stefan Roth, and Bernt Schiele.
\newblock Deep wiener deconvolution: Wiener meets deep learning for image
  deblurring.
\newblock {\em arXiv preprint arXiv:2103.09962}, 2021.

\bibitem{Dong2021}
Jiangxin Dong, Stefan Roth, and Bernt Schiele.
\newblock Learning spatially-variant map models for non-blind image deblurring.
\newblock In {\em Proceedings of the IEEE/CVF Conference on Computer Vision and
  Pattern Recognition (CVPR)}, pages 4886--4895, June 2021.

\bibitem{Donoho2006}
David~L Donoho.
\newblock Compressed sensing.
\newblock {\em IEEE Transactions on information theory}, 52(4):1289--1306,
  2006.

\bibitem{Donoho1994}
David~L. Donoho and Iain~M. Johnstone.
\newblock Ideal spatial adaptation by wavelet shrinkage.
\newblock {\em Biometrika}, 81(3):425--455, 1994.

\bibitem{Fergus2006}
Rob Fergus, Barun Singh, Aaron Hertzmann, Sam~T Roweis, and William~T Freeman.
\newblock Removing camera shake from a single photograph.
\newblock In {\em ACM SIGGRAPH}, pages 787--794. 2006.

\bibitem{Figueiredo2007}
M.A.T. Figueiredo, J.M. Bioucas-Dias, and R.D. Nowak.
\newblock Majorization--minimization algorithms for wavelet-based image
  restoration.
\newblock {\em {IEEE} Trans. Image Process.}, 16:2980--2991, 2007.

\bibitem{Goldstein2009}
T. Goldstein and S. Osher.
\newblock The split {Bregman} method for ${L}_1$-regularized problems.
\newblock {\em SIAM J. Imaging Sci.}, 2:323--343, 2009.

\bibitem{Gong2020}
Dong Gong, Zhen Zhang, Qinfeng Shi, Anton van~den Hengel, Chunhua Shen, and
  Yanning Zhang.
\newblock Learning deep gradient descent optimization for image deconvolution.
\newblock {\em IEEE transactions on neural networks and learning systems},
  31(12):5468--5482, 2020.

\bibitem{Hansen2006}
P.~C. Hansen, J.~G. Nagy, and D.~P. O'Leary.
\newblock {\em Deblurring Images: Matrices, Spectra, and Filtering}.
\newblock SIAM, 2006.

\bibitem{He2016}
Kaiming He, Xiangyu Zhang, Shaoqing Ren, and Jian Sun.
\newblock Deep residual learning for image recognition.
\newblock In {\em Proc. IEEE Int. Conf. Computer Vision and Pattern
  Recognition}, 2016.

\bibitem{Kay1993}
Steven~M Kay.
\newblock {\em Fundamentals of statistical signal processing: estimation
  theory}.
\newblock Prentice-Hall, Inc., 1993.

\bibitem{Keshavan2010}
Raghunandan~H Keshavan, Andrea Montanari, and Sewoong Oh.
\newblock Matrix completion from a few entries.
\newblock {\em IEEE transactions on information theory}, 56(6):2980--2998,
  2010.

\bibitem{Kingma2014}
Diederik Kingma and Jimmy Ba.
\newblock Adam: A method for stochastic optimization.
\newblock {\em arXiv preprint arXiv:1412.6980}, 2014.

\bibitem{Kohler2012}
Rolf K{\"o}hler, Michael Hirsch, Betty Mohler, Bernhard Sch{\"o}lkopf, and
  Stefan Harmeling.
\newblock Recording and playback of camera shake: Benchmarking blind
  deconvolution with a real-world database.
\newblock In {\em Computer Vision -- ECCV 2012}, pages 27--40, 2012.

\bibitem{Kokkinos2019}
Filippos Kokkinos and Stamatios Lefkimmiatis.
\newblock Iterative joint image demosaicking and denoising using a residual
  denoising network.
\newblock {\em IEEE Transactions on Image Processing}, 28(8):4177--4188, 2019.

\bibitem{Krishnan2011}
Dilip Krishnan, Terence Tay, and Rob Fergus.
\newblock Blind deconvolution using a normalized sparsity measure.
\newblock In {\em Proc. IEEE Int. Conf. Computer Vision and Pattern
  Recognition}, pages 233--240. IEEE, 2011.

\bibitem{Kruse2017}
Jakob Kruse, Carsten Rother, and Uwe Schmidt.
\newblock Learning to push the limits of efficient fft-based image
  deconvolution.
\newblock In {\em Proceedings of the IEEE International Conference on Computer
  Vision}, pages 4586--4594, 2017.

\bibitem{Lai2016}
Wei-Sheng Lai, Jia-Bin Huang, Zhe Hu, Narendra Ahuja, and Ming-Hsuan Yang.
\newblock A comparative study for single image blind deblurring.
\newblock In {\em IEEE Conf. Comput. Vision and Patt. Recogn. {(CVPR)}}, pages
  1701--1709. IEEE, 2016.

\bibitem{Laurent2018}
Thomas Laurent and James Brecht.
\newblock Deep linear networks with arbitrary loss: All local minima are
  global.
\newblock In {\em International conference on machine learning}, pages
  2902--2907. PMLR, 2018.

\bibitem{Lefkimmiatis2012J}
S. Lefkimmiatis, A. Bourquard, and M. Unser.
\newblock Hessian-based norm regularization for image restoration with
  biomedical applications.
\newblock {\em IEEE Trans. Image Process.}, 21(3):983--995, 2012.

\bibitem{Lefkimmiatis2015J}
S. Lefkimmiatis, A. Roussos, P. Maragos, and M. Unser.
\newblock {Structure} tensor total variation.
\newblock {\em SIAM J. Imaging Sci.}, 8:1090--1122, 2015.

\bibitem{Lefkimmiatis2013J}
S. Lefkimmiatis, J. Ward, and M. Unser.
\newblock {Hessian} {Schatten}-norm regularization for linear inverse problems.
\newblock {\em IEEE Trans. Image Process.}, 22(5):1873--1888, 2013.

\bibitem{Levin2009}
A. Levin, Y. Weiss, F. Durand, and W.~T. Freeman.
\newblock Understanding and evaluating blind deconvolution algorithms.
\newblock In {\em IEEE Conf. Comput. Vision and Patt. Recogn. {(CVPR)}}, pages
  1964--1971, 2009.

\bibitem{Levin2011B}
Anat Levin, Yair Weiss, Fredo Durand, and William~T Freeman.
\newblock Efficient marginal likelihood optimization in blind deconvolution.
\newblock In {\em CVPR 2011}, pages 2657--2664. IEEE, 2011.

\bibitem{Magnus1999}
Jan~R Magnus and Heinz Neudecker.
\newblock {\em Matrix differential calculus with applications in statistics and
  econometrics}.
\newblock John Wiley \& Sons, 1999.

\bibitem{Nikolova2005}
Mila Nikolova and Michael~K Ng.
\newblock Analysis of half-quadratic minimization methods for signal and image
  recovery.
\newblock {\em SIAM Journal on Scientific computing}, 27(3):937--966, 2005.

\bibitem{Pan2016b}
Jinshan Pan, Zhouchen Lin, Zhixun Su, and Ming-Hsuan Yang.
\newblock Robust kernel estimation with outliers handling for image deblurring.
\newblock In {\em 2016 IEEE Conference on Computer Vision and Pattern
  Recognition (CVPR)}, pages 2800--2808, 2016.

\bibitem{Pan2016}
Jinshan Pan, Deqing Sun, Hanspeter Pfister, and Ming-Hsuan Yang.
\newblock Blind image deblurring using dark channel prior.
\newblock In {\em 2016 IEEE Conference on Computer Vision and Pattern
  Recognition (CVPR)}, pages 1628--1636, 2016.

\bibitem{Perrone2014}
Daniele Perrone and Paolo Favaro.
\newblock Total variation blind deconvolution: The devil is in the details.
\newblock In {\em 2014 IEEE Conference on Computer Vision and Pattern
  Recognition}, pages 2909--2916, 2014.

\bibitem{Ren2020}
Dongwei Ren, Kai Zhang, Qilong Wang, Qinghua Hu, and Wangmeng Zuo.
\newblock Neural blind deconvolution using deep priors.
\newblock In {\em Proceedings of the IEEE/CVF Conference on Computer Vision and
  Pattern Recognition}, pages 3341--3350, 2020.

\bibitem{Robinson2008}
A.~J. Robinson and Frank Fallside.
\newblock The {U}tility {D}riven {D}ynamic {E}rror {P}ropagation {N}etwork.
\newblock Technical Report CUED/F-INFENG/TR.1, Engineering Department,
  Cambridge University, Cambridge, UK, 1987.

\bibitem{Roth2009}
Stefan Roth and Michael~J Black.
\newblock Fields of experts.
\newblock {\em International Journal of Computer Vision}, 82(2):205--229, 2009.

\bibitem{Rudin1992}
L. Rudin, S. Osher, and E. Fatemi.
\newblock Nonlinear total variation based noise removal algorithms.
\newblock {\em Physica D}, 60:259--268, 1992.

\bibitem{Shewchuk1994}
J.~R. Shewchuk.
\newblock An introduction to the conjugate gradient method without the
  agonizing pain, 1994.

\bibitem{Sun2013}
Libin Sun, Sunghyun Cho, Jue Wang, and James Hays.
\newblock Edge-based blur kernel estimation using patch priors.
\newblock In {\em Proc. IEEE International Conference on Computational
  Photography}, 2013.

\bibitem{Sun2012}
Libin Sun and James Hays.
\newblock Super-resolution from internet-scale scene matching.
\newblock In {\em Proceedings of the {IEEE} Conf. on International Conference
  on Computational Photography ({ICCP})}, 2012.

\bibitem{Tran2021}
Phong Tran, Anh Tran, Quynh Phung, and Minh Hoai.
\newblock Explore image deblurring via encoded blur kernel space.
\newblock In {\em Proceedings of the {IEEE} Conference on Computer Vision and
  Pattern Recognition (CVPR)}, 2021.

\bibitem{Wang2018}
Xintao Wang, Ke Yu, Shixiang Wu, Jinjin Gu, Yihao Liu, Chao Dong, Yu Qiao, and
  Chen Change~Loy.
\newblock Esrgan: Enhanced super-resolution generative adversarial networks.
\newblock In {\em Proceedings of the European conference on computer vision
  (ECCV) workshops}, pages 0--0, 2018.

\bibitem{Wen2021}
Fei Wen, Rendong Ying, Yipeng Liu, Peilin Liu, and Trieu-Kien Truong.
\newblock A simple local minimal intensity prior and an improved algorithm for
  blind image deblurring.
\newblock {\em IEEE Transactions on Circuits and Systems for Video Technology},
  31(8):2923--2937, 2021.

\bibitem{Whyte2014}
Oliver Whyte, Josef Sivic, and Andrew Zisserman.
\newblock Deblurring shaken and partially saturated images.
\newblock {\em International journal of computer vision}, 110(2):185--201,
  2014.

\bibitem{Xu2013}
Li Xu, Shicheng Zheng, and Jiaya Jia.
\newblock Unnatural l0 sparse representation for natural image deblurring.
\newblock In {\em Proc. IEEE Int. Conf. Computer Vision and Pattern
  Recognition}, pages 1107--1114, 2013.

\bibitem{Xu2017}
Xiangyu Xu, Jinshan Pan, Yu-Jin Zhang, and Ming-Hsuan Yang.
\newblock Motion blur kernel estimation via deep learning.
\newblock {\em IEEE Transactions on Image Processing}, 27(1):194--205, 2017.

\bibitem{Zhang2017}
Kai Zhang, Wangmeng Zuo, Yunjin Chen, Deyu Meng, and Lei Zhang.
\newblock Beyond a {Gaussian} denoiser: Residual learning of deep {CNN} for
  image denoising.
\newblock {\em IEEE Trans. Image Process.}, 2017.

\bibitem{Zhang2017b}
Kai Zhang, Wangmeng Zuo, Shuhang Gu, and Lei Zhang.
\newblock Learning deep {CNN} denoiser prior for image restoration.
\newblock In {\em Proc. IEEE Int. Conf. Computer Vision and Pattern
  Recognition}, July 2017.

\end{thebibliography}
}

\end{document}